\documentclass[referee,sn-standardnature,pdflatex]{sn-jnl}
\jyear{2022}

\usepackage{graphicx}
\graphicspath{ {Figures/} }
\usepackage{enumerate}
\usepackage{threeparttable} 
\usepackage{dcolumn}
\usepackage{bm}
\usepackage{multirow} 
\usepackage{comment}
\usepackage{subcaption}
\usepackage[version=4]{mhchem} 
\usepackage{parskip} 
\usepackage[capitalise]{cleveref}
\usepackage{color}
\usepackage{siunitx}
\setcitestyle{super}
\usepackage[numbers]{natbib}
\usepackage{etoolbox}
\usepackage{pdfpages}

\usepackage{pdftexcmds}
\makeatletter
\newcommand\kv[2]{%
  \ifnum\pdf@strcmp{\unexpanded{#1}}{V}=0 %
     \expandafter\@firstoftwo
  \else
    \expandafter\@secondoftwo
  \fi
    {\textit{#1}$_{\rm#2}$}
    {#1$_{\rm#2}$}%
}
\makeatother

\makeatletter
\newcommand\kvc[3]{%
  \ifnum\pdf@strcmp{\unexpanded{#1}}{V}=0 %
     \expandafter\@firstoftwo
  \else
    \expandafter\@secondoftwo
  \fi
    {\textit{#1}$_{\rm #2}^{#3}$}
    {#1$_{\rm#2}^{#3}$}
}
\makeatother

\makeatletter
\patchcmd{\ps@headings}
{\hbox to \hsize{\hfill Springer Nature 2021 \LaTeX\ template\hfill}}
{\hbox to \hsize{}}
{}
{}
\patchcmd{\ps@headings}
{\hbox to \hsize{\hfill Springer Nature 2021 \LaTeX\ template\hfill}}
{\hbox to \hsize{}}
{}
{}
\patchcmd{\ps@titlepage}
{\hbox to \hsize{\hfill Springer Nature 2021 \LaTeX\ template\hfill}}
{\hbox to \hsize{}}
{}
{}
\makeatother

\begin{document}

\title{Identifying the ground state structures of point defects in solids}

\author[1]{\fnm{Irea} \sur{Mosquera-Lois}}
\equalcont{These authors contributed equally to this work.}

\author*[1,2]{\fnm{Seán R.} \sur{Kavanagh}}\email{sean.kavanagh.19@ucl.ac.uk}
\equalcont{These authors contributed equally to this work.}

\author[2]{\fnm{Aron} \sur{Walsh}}

\author*[1]{\fnm{David O.} \sur{Scanlon}}\email{d.scanlon@ucl.ac.uk}

\affil*[1]{\orgdiv{Thomas Young Centre and Department of Chemistry}, \orgname{University College London}, \orgaddress{\street{20
Gordon Street}, \city{London}, \postcode{WC1H 0AJ}, \country{UK}}}

\affil*[2]{\orgdiv{Thomas Young Centre and Department of Materials}, \orgname{Imperial College London}, \orgaddress{\street{Exhibition Road}, \city{London}, \postcode{SW7 2AZ}, \country{UK}}}

\abstract{Point defects are a universal feature of crystalline materials. Their identification is often addressed by combining experimental measurements with theoretical models. The standard approach of simulating defects is, however, prone to missing the ground state atomic configurations associated with energy-lowering reconstructions from the idealised crystallographic environment. Missed ground states compromise the accuracy of calculated properties. To address this issue, we report an approach to efficiently navigate the defect configurational landscape using targeted bond distortions and rattling. Application of our workflow to a range of materials (CdTe, GaAs, \ce{Sb2S3}, \ce{Sb2Se3}, \ce{CeO2}, \ce{In2O3}, \ce{ZnO}, anatase-\ce{TiO2}) reveals symmetry breaking in each host crystal that is not found via conventional local minimisation techniques. The point defect distortions are classified by the associated physico-chemical factors. We demonstrate the impact of these defect distortions on derived properties, including formation energies, concentrations and charge transition levels. Our work presents a step forward for quantitative modelling of imperfect solids.}

\maketitle

\section{\label{sec:level1}Introduction\protect}
\begin{figure}
    \centering
    \includegraphics[width=0.6\linewidth]{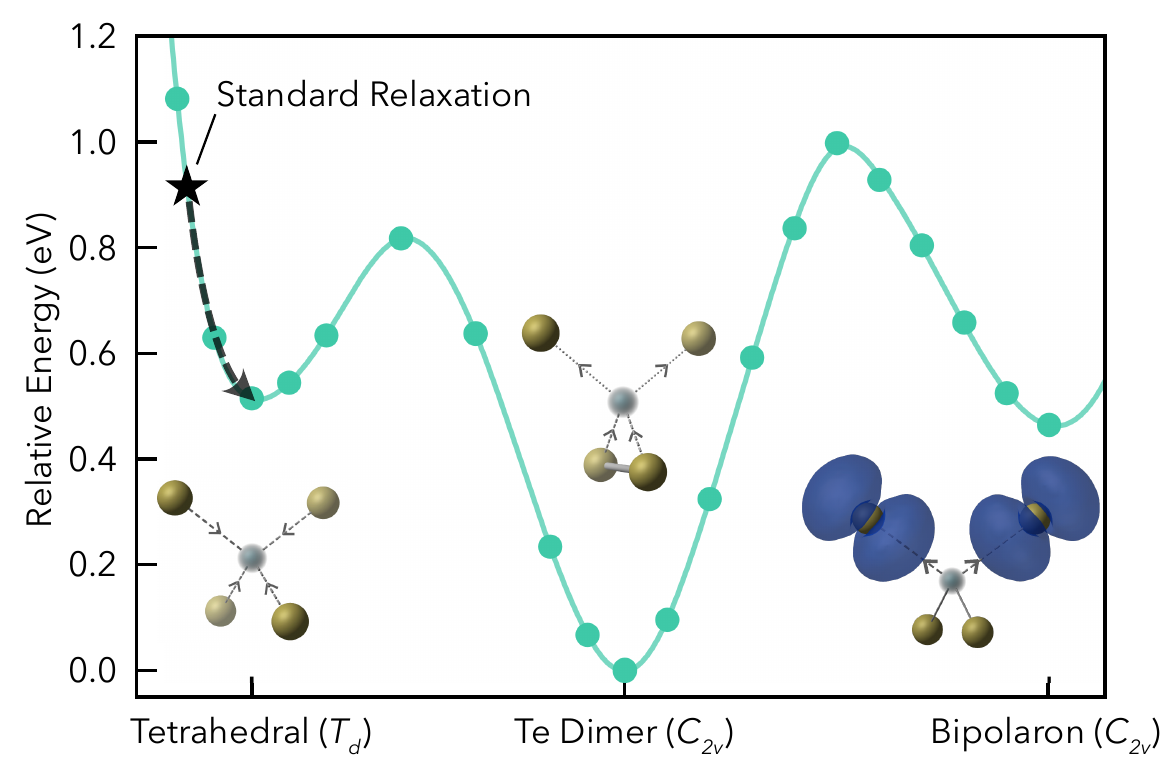} 
    \caption{Potential energy surface for the neutral cadmium vacancy in \ce{CdTe}, illustrating the global minimum (Te dimer) and local minima (Bipolaron and Tetrahedral configurations). A standard optimisation from the initial, high-symmetry configuration (black star) gets trapped in the metastable Tetrahedral configuration. Adapted from \textit{Matter}, I. Mosquera-Lois and S.R. Kavanagh, \textit{In Search of Hidden Defects} (2021) with permission from Elsevier.}
    \label{fig:neb_vac_1_Cd_0}
\end{figure}

Defects control the properties and performance of most functional materials and devices. 
Unravelling the identity and impact of these imperfections is, however, a challenging task. Their dilute concentrations hinder experimental identification, which is often tackled by combining characterisation measurements with \textit{ab-initio} techniques. 
The standard modelling approach, based on local optimisation of a defect containing crystal, is prone to miss the true ground state atomic arrangement, however. 
The chosen initial configuration, which is often initiated as a vacancy/substitution/interstitial on a known crystal site (Wyckoff position) and all other atoms retaining their typical lattice positions, may lie within a local minimum or on a saddle point of the potential energy surface (PES), trapping a gradient-based optimisation algorithm in an unstable or metastable arrangement\cite{Arrigoni_evolutionary_2021,Ong_2012,Evarestov_2017,Lany_dimer_2004,Lindstrom_2015,Sokol_2010,Osterbacka_2022,Krajewska_2021,Mosquera_2021} as illustrated in \cref{fig:neb_vac_1_Cd_0}. 
By yielding incorrect geometries, the predicted defect properties, such as equilibrium concentrations, charge transition levels and recombination rates, are rendered inaccurate. This behaviour can severely impact theoretical predictions of material performance, including photovoltaic efficiency\cite{Kavanagh_2021}, catalytic activity\cite{Kehoe_2011}, absorption spectra\cite{Lany_2004} and carrier doping\cite{Goyal_2020}, highlighting the pressing requirement for improved structure prediction techniques for defects in solids.
Several approaches have recently been devised to navigate the defect configurational landscape. 
Arrigoni and Madsen\cite{Arrigoni_evolutionary_2021} used an evolutionary algorithm enhanced with a machine learning model to explore the defect PES and identify low energy structural configurations. While its robust performance makes it suitable for in-depth studies of specific defects, its complexity and computational cost hinders its applicability in standard defect studies, where all intrinsic defects in all plausible charge states (and relevant extrinsic defects) are modelled. 
On the other hand, Pickard and Needs\cite{Pickard_2006} employed random sampling of the PES.
Essentially, they remove the defect atom and its nearest neighbours and reintroduce them at random positions within a 5 \AA~cube centred on the defect, thus ensuring sampling of an important region of the PES. Despite successfully identifying the ground state structures for several systems\cite{Morris_2008,Morris_2009,Morris_2011,Pickard_2006,Mulroue_2011}, constrained random sampling in a high-dimensional space lowers efficiency and increases computational cost. 

To improve PES sampling efficiency, domain knowledge can be employed to bias the search towards energy-lowering structural distortions. 
Accordingly, we exploit the localised nature of defect distortions, as well as the key role of the defect valence electrons, to guide the exploration of the PES. 
By combining these biases with lessons learnt from crystal structure prediction, we develop a practical and robust method to identify the defect ground and low-energy metastable states. Its application to eight host materials (CdTe, GaAs, \ce{Sb2S3}, \ce{Sb2Se3}, \ce{CeO2}, \ce{In2O3}, \ce{ZnO}, \ce{TiO2}) reveals a myriad of energy-lowering defect distortions, which are missed when relaxing the ideal high-symmetry defect configurations. 
Notably, energy-lowering distortions that are missed by standard geometry relaxation were found in each system investigated.
Classifying these atomic rearrangements by distortion motif, we outline the main physico-chemical factors that underlie defect reconstructions. Moreover, we demonstrate the strong effect on defect formation energies and their charge transition levels, illustrating the
importance of exploring the defect configurational space for accurate predictions of defect properties. 
Additionally, while defect properties are often determined by the ground state structure, there are cases when low energy metastable structures can significantly affect behaviour\cite{Kavanagh_2021,Coutinho_2020,kavanagh2022impact}. Accordingly, we tested our approach for low-energy metastable configurations by considering the DX centres in GaAs, which have been extensively studied due to their anomalous physical properties and technological importance. By finding the vast majority ($>90$\%) of configurations identified by previous investigations, we demonstrate the applicability of our approach to locate both ground state and low-energy metastable defect structures, constituting an affordable and effective tool to explore the defect configurational landscape.

\section{Methodology}
\subsection{\label{sec:level2} Structure search strategy}
{I. \emph{Local bond distortion}}\\
The role of valence electrons in defect reconstructions has been demonstrated by the defect-molecule model developed by Watkins\cite{Watkins_1983,Watkins_1993,Watkins_1996,Watkins_1997,Watkins_2000,Watkins_2001} and Coulson\cite{Coulson_1957}. 
It successfully explained the Jahn-Teller distortions observed for vacancy centres in silicon and diamond, and continues to be applied to rationalise defect reconstructions\cite{El-Maghraby_1998,Stoneham_2001_vacancies,Lindstrom_2015,Carvalho_2010,Lanoo_1981,Kavanagh_2021,Lany_2004,Lany_2005,Chanier_2008,Schultz_2009,Feichtinger_1986}. 
This highlights the suitability and utility of defect valence electrons as an approximate indicator of likely defect distortions. As such, our method incorporates this feature, as well as the localised nature of defect reconstructions, to generate a reasonable set of chemically-guided distortions to sample the PES. 
Specifically, we use the number of missing/additional valence electrons of the defect species to determine the number of nearest neighbour atoms to distort. To generate a set of trial structures, the initial defect-neighbour distances $d_0$ are distorted by varying amounts, with the range set by the minimum and maximum distortion factors $k_{min}$ and $k_{max}$, such that $d_{min} = k_{min}d_0$ and $d_{max} = k_{max}d_0$, and the increment parameter $\delta$ which determines the sampling density of the trial distortion mesh. 
Thus, the distorted defect-neighbour distances $d$ are given by 
\begin{equation}
    d_n =  k_{n}d_0
\end{equation}
\begin{equation}
    k_{n} = k_{min} + n\delta \quad \textrm{with} \quad n=0,1,2,...,\Big(\frac{k_{max}-k_{min}}{\delta}\Big)
\end{equation}
While these parameters may be varied, in our work we found a distortion range of $\pm60\%$ (i.e. $k_{min} = 0.4$ and $k_{max} = 1.6$) and increment of $10\%$ ($\delta = 0.1$) to be an optimal choice, with a sufficiently wide and dense distortion mesh to identify all known energy-lowering distortions (see SI Section III for more details). 
Similar bond distortions have previously been applied by Pham and Deskins\cite{Pham_2020} to locate the ground state of small polarons in three metal oxides. Analogously, these initial distortions aim to escape the local energy basin in which the ideal, undistorted structure may be trapped.  
Further, by only distorting the defect nearest neighbours --- and thus restricting to a lower-dimensional subspace, the method aims to sample the key regions of the PES that may comprise energy-lowering reconstructions, such as Jahn-Teller effects or rebonding between under-coordinated atoms. 

\begin{figure}
    \includegraphics[width=\textwidth]{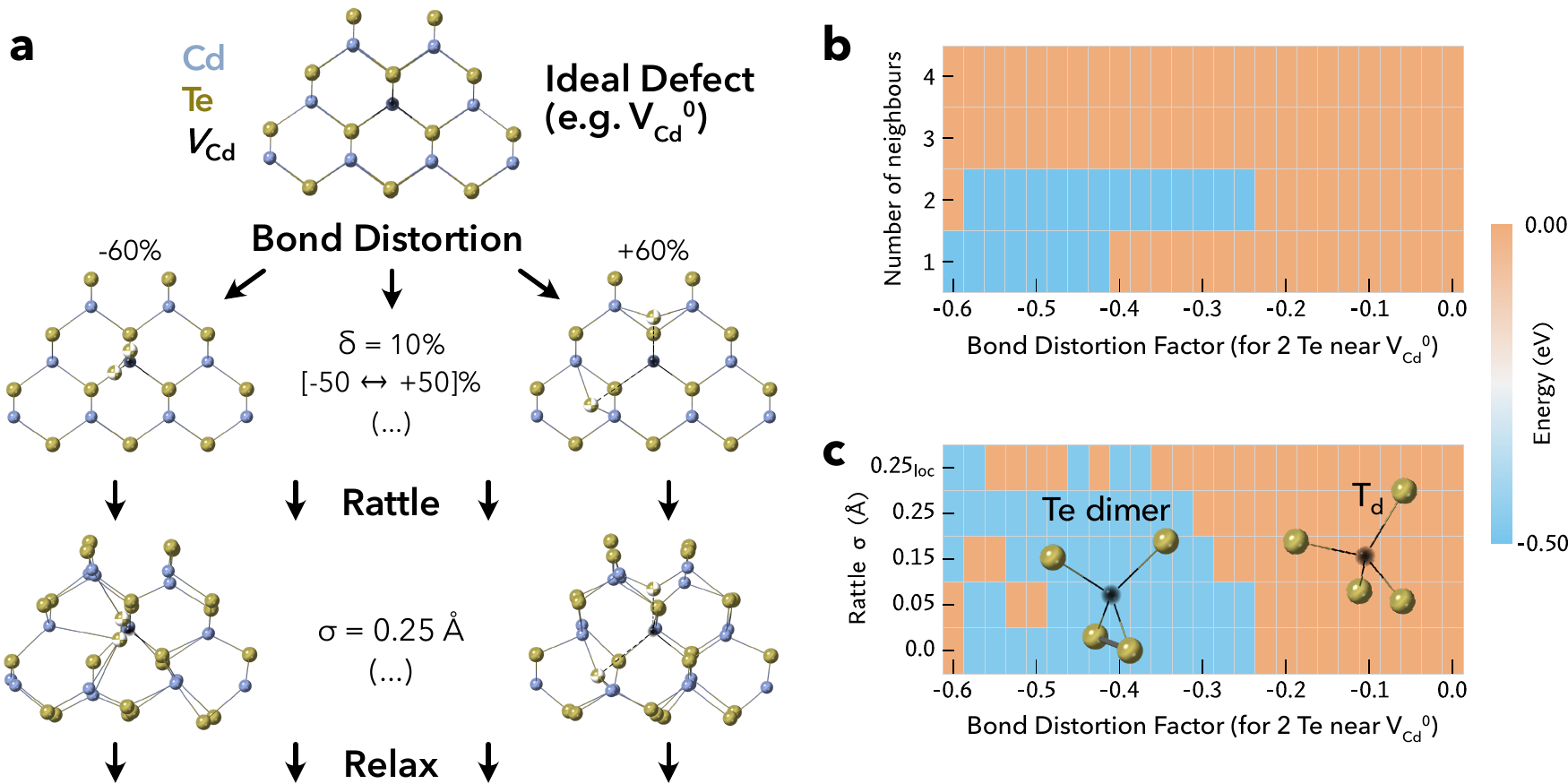}
    \caption{\textbf{(a)} Schematic representation of our approach, as described in the Methods section, using the neutral cadmium vacancy (\kvc{V}{Cd}{0}\!\!) in CdTe as an example. Cd in blue, Te in gold and \kv{V}{Cd} in black, with the distorted Te neighbours shown in a gold/white. \textbf{(b)} Relative energy of relaxed structures for \kvc{V}{Cd}{0}\!\! in CdTe, for different numbers of distorted defect neighbours and \textbf{(c)} rattle standard deviations. A localised rattle (where only atoms within 5~\AA~to the defect are distorted) is also shown ($\sigma=0.25_{loc}$ \AA). The widest interval of bond distortions leading to the ground state is obtained by distorting two nearest neighbours.
    Only negative bond distortions are shown since all positive ones yield the metastable `Tetrahedral' configuration.}
    \label{fig:BDM_parameters_V_Cd}
\end{figure}

The utility of electron count as an indicator for the number of defect-neighbour bonds to distort is demonstrated for the neutral cadmium vacancy in CdTe in \cref{fig:BDM_parameters_V_Cd} (a). 
Here, bond distortions were applied to differing numbers of defect neighbours. Best performance (i.e. widest range of distortions yielding the ground state) was obtained when distorting two neighbours, supporting the feature choice given that the vacancy site lacks two valence electrons relative to the original Cd atom.\\ 

{II. \emph{Atomic rattle}}\\
Following the nearest neighbour distortion, the method applies random perturbations to the coordinates $r$ of all atoms $N$ in the supercell. This aids to escape from any metastable symmetry trapping and locate small, symmetry-breaking distortions (i.e. nearby lower energy basins). 
This step has proven useful in other structure-searching approaches, such as the investigation of symmetry breaking distortions in perovskites\cite{Wang_2021} or point defects\cite{Huang_2022,kavanagh2022impact}.
The magnitudes of these randomly-directed atomic displacements $d$ are themselves randomly selected from a normal distribution of chosen standard deviation $\sigma$ (\cref{eq:rattle}).
\begin{equation}\label{eq:rattle}
\begin{aligned}
    r'_i = r_i + d \quad 
    & \textrm{with} \quad i = 0,1,...,3N
    & \textrm{and} \quad d 
    \leftarrow \frac{1}{\sigma \sqrt{2\pi}}
    \exp \big( -\frac{d^2}{2\sigma^2}  \big) 
\end{aligned}
\end{equation}
To determine the optimum magnitude of the perturbation, several standard deviations were tested for a series of defects -- shown in the SI Section III and exemplified in \cref{fig:BDM_parameters_V_Cd} (b) for \kvc{V}{Cd}{0} in CdTe. In general, we found values between 0.05 - \SI{0.25}{\AA} to give good performance, and adopted \SI{0.25}{\AA} since it was the only distortion capable of finding the ground state of $\rm S_i^{-1}$ in \ce{Sb2S3} (SI Fig. S3). 

Additionally, we also investigated a `localised' rattle, where the perturbations were only applied to atoms located within 5~\AA~of the defect, as recently employed in other defect studies\cite{Gake_2021,Kumagai_2021,Huang_2022}.
However, this yielded inferior performance for several defects (less bond distortions leading to the ground state) without significant reduction in the number of ionic relaxation steps (SI Fig. S4). While the final structure is still only locally deformed, it appears that the external potential induced by the long-range symmetry of the surrounding host crystal biases the initial forces toward retaining the high-symmetry metastable defect structure. Therefore rattling was applied to all atoms in the simulation supercell.

\subsection{Computational details}
The bond distortion and rattling procedure for defect structure generation has been implemented in Python\cite{Ong_pymatgen_2013,Bahn_ase_2002}. 
The underlying total energy and force calculations were performed with plane-wave Density Functional Theory (DFT) within VASP\cite{Kresse_1993,Kresse_1994}, using the projector augmented wave method.\cite{Kresse_1996}. 
All calculations were spin-polarised and based on the HSE screened hybrid exchange-correlation functional\cite{Heyd_2003}. 
When reported in previous studies, the bandgap-corrected Hartree-Fock exchange fraction ($\alpha$) was used, which is common practice in the field\cite{Alkauskas_2011,Alkauskas_2014,Lyons_2010,Deak_2005}. 
Otherwise HSE06 ($\alpha = 25\%$) was employed. The basis set energy cut-off and $k$-point grid were converged to 3 meV/atom in each case. The converged parameters, valence electron configurations and fractions of Hartree-Fock exchange used for each system are tabulated in the SI Table I.
\\
The conventional supercell approach for modelling defects in periodic solids\cite{Huang_2021} was used. 
To reduce periodic image interactions, supercell dimensions of at least 10 \AA~in each direction\cite{Zunger_2008_assessment} were employed (SI Table I). 
The use of a large supercell justifies performing the PES exploration with $\Gamma$-point-only reciprocal space sampling, thus increasing speed while retaining qualitative accuracy. This makes the cost of these test geometry optimisations very small compared to the total cost of the fully-converged defect calculations. 
The $\Gamma$-point approximation was further validated by optimising the defect ground state and metastable geometry (obtained by relaxing the high-symmetry, undistorted structure) with denser reciprocal space sampling (converged within 3 meV/atom). For materials containing heavy-atom elements (CdTe, \ce{Sb2S3}, \ce{Sb2Se3}, \ce{CeO2}, \ce{In2O3}), relativistic effects were accounted for by including spin-orbit coupling (SOC) interactions. To calculate the defect formation energy, we followed the approach described by Freysoldt \emph{et al}\cite{Freysoldt_2014}, using the charge correction developed by Kumagai and Oba\cite{Kumagai_2014}. Madelung energies were calculated using Mulliken charges with the LOBSTER package\cite{ertural_development_2019,dronskowski_crystal_1993,deringer_crystal_2011,maintz_analytic_2013,nelson_lobster_2020}. To generate and analyse the defect distortions described in this work, we developed the \texttt{ShakeNBreak} python and command-line package (\url{https://github.com/SMTG-UCL/ShakeNBreak}) which will be described in a forthcoming publication.
\subsection{Defect suite}
To test our method we chose a set of eight host crystals that provide diversity in crystal symmetry, coordination environments and electronic structure. CdTe and GaAs are tetrahedral semiconductors that adopt the zincblende structure (space group $F\bar43m$). \ce{Sb2Se3} and \ce{Sb2S3} are layered materials (space group $Pnma$), composed of one-dimensional [\ce{Sb4X6}]$_n$ ribbons with covalent metal-chalcogen bonds within each ribbon and van der Waals interactions between ribbons\cite{Guo_2018,Wang_2022,Caruso_2015,Song_2017,Yang_2018,Gusmao_2019}. Beyond these covalently-bonded crystals, we also studied four metal oxides (\ce{In2O3}, \ce{ZnO}, \ce{CeO2}, \ce{TiO2}). \ce{ZnO} crystallises in the hexagonal wurtzite structure (space group $P6_3mc$) with tetrahedral coordination, while \ce{In2O3} adopts a bixbyite structure (space group $Ia3$). \ce{CeO2} crystallises in the fluorite structure with a face-centred cubic unit cell (space group $Fm\bar3m$). The anatase phase of \ce{TiO2} was considered, which adopts a tetragonal structure composed of distorted \ce{TiO6} octahedra (space group $I4_1/amd$). For the covalently-bonded systems, all intrinsic defects were studied, while for the oxides we focused on defects where reconstructions had been reported or were likely (oxygen interstitials in \ce{ZnO}, \ce{In2O3} and a-\ce{TiO2} and dopant defects in \ce{CeO2}). For the materials that can host defects in several symmetry inequivalent sites, the lowest energy one was selected if the energy difference with the other sites was higher than 1 eV, otherwise all symmetry inequivalent sites were considered.

\section{Results}
\begin{figure*}[ht]
    \centering
    \includegraphics[width=0.95\textwidth]{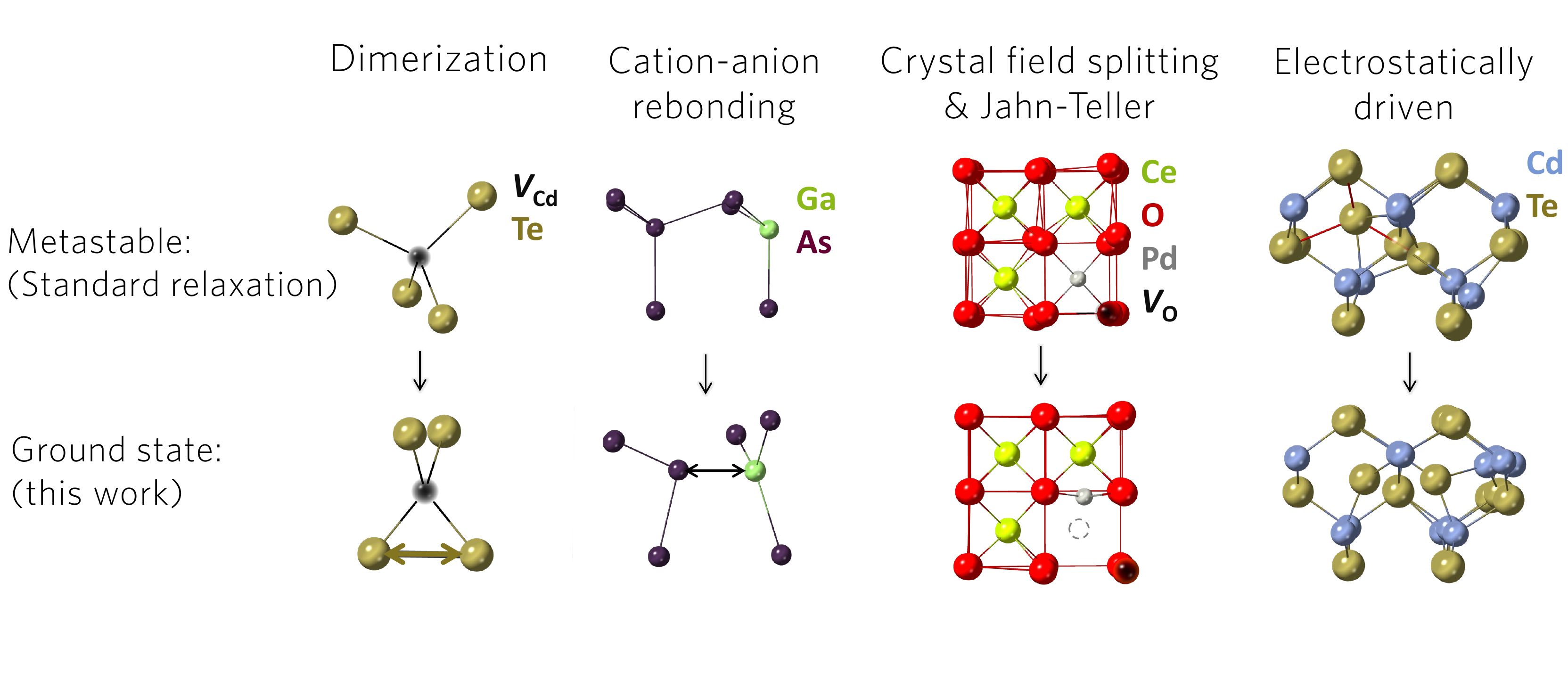}
    \caption{Types of symmetry-breaking reconstructions at point defect sites identified by our method.}
    \label{fig:distortion_rationales}
\end{figure*}

The application of our method to a series of semiconductors revealed a range of defect reconstructions, which are missed when performing a geometry optimisation of the high symmetry configuration. We categorise these reconstructions into distinct motifs according to the chemical origins of the energy-lowering distortion, as exemplified in  \cref{fig:distortion_rationales} and discussed hereafter. 

\subsection{Rebonding}
For covalently-bonded materials, defect reconstructions identified by our method often entailed a change in the bonding arrangement at the defect site (`rebonding'), such as dimer formation or replacing cation-cation/anion-anion bonds with more favourable cation-anion ones.\\
\emph{Dimerisation.} For many vacancies and interstitials, symmetry breaking was found to produce dimer bonds between under-coordinated atoms. This novel bond formation results in these distortions being highly-favourable, with energy decreases from \SI{0.4}{eV} to over \SI{2.5}{eV} (\cref{tab:dimer}), between the ground state and metastable structure obtained from standard relaxations. 
As expected given the large energy differences, the metastable and ground state configurations also display significant structural differences, which we quantify here by summing the atomic displacements between structures which are above a 0.1 \AA~threshold. The large structural rearrangements for dimer reconstructions are demonstrated by displacement values ranging between 2.0 and 24.3 \AA, with the distortion mainly localised to the dimer-forming atoms (SI Figs.~S8, S9, S10). For vacancies, the distortion typically entails two of the under-coordinated defect neighbours displacing toward each other to form a bond, while for interstitials it is the additional atom which displaces towards a nearby neighbour.

\begin{table}[ht]
    \centering
        \caption{Dimerisation. Energy difference and summed atomic displacements ($\Sigma$Disp.) between the ground state identified by our method and the metastable configuration found when relaxing the ideal, undistorted defect structure (i.e. $\rm \Delta E = E_{\rm ground} - E_{\rm metastable}$).}
		\label{tab:dimer}
		\setlength{\tabcolsep}{8pt} 
		\renewcommand{\arraystretch}{1.30} 
		\begin{tabular}{ccccc}
			\hline 
			\multicolumn{5}{c}{Cation-cation}\\
			\hline
			Defect & Material & Charge &  $\Delta$E~(eV) & $\Sigma$Disp.~(\AA) \\
			\hline
			\kv{V}{Te} & CdTe & 0 & -0.42 & 6.0 \\ 
			\hline
			{\ce{Sb_i}} & {\ce{Sb2S3}} & -3 & -0.46 &  12.8 \\ 
			\hline
			\multicolumn{5}{c}{Anion-anion}\\
			\hline
			Defect & Material & Charge & $\Delta $E~(eV) & $\Sigma$Disp.~(\AA) \\
			\hline
			\kv{V}{Cd} & CdTe & 0 & -0.50 & 6.1 \\
			\hline			  
			\multirow{4}{*}{\kv{V}{Sb,1}} &
			\multirow{4}{*}{\ce{Sb2Se3}}
			&-1 & -0.66 & 8.2\\ 
			& & 0 & -0.78 & 8.7\\ 
			& & +1 & -0.84 & 8.9\\ 
			& & +2 & -2.02 & 12.8\\ 
			\hline			
			\multirow{4}{*}{\kv{V}{Sb,2}} &
			\multirow{4}{*}{\ce{Sb2Se3}}
			&-1 & -0.80 & 7.2\\ 
			& & 0 & -0.91 & 9.0\\ 	
			& & +1 & -2.10 & 14.7\\  
			& & +2 & -2.15 & 16.6\\ 
			\hline		
			\multirow{4}{*}{\kv{V}{Sb,1}} &
			\multirow{4}{*}{\ce{Sb2S3}}
			&-1 & -0.80 & 9.9 \\ 
			& & 0 & -1.03 & 9.0 \\ 
			& & +1 & -2.31 & 11.8 \\  
			& & +2 & -2.03 & 24.3 \\ 
			\hline			
			\multirow{4}{*}{\kv{V}{Sb,2}} &
			\multirow{4}{*}{\ce{Sb2S3}}
			&-1 & -0.99 & 8.9 \\ 
			& & 0 & -1.10 & 12.2 \\ 
			& & +1 & -2.50 & 12.8 \\
			& & +2 & -2.59 & 21.9 \\
			\hline					
			$\rm S_i$& \ce{Sb2S3} & -1 & -0.54 & 9.4 \\
			\hline 
			\multirow{2}{*}{\kv{O}{i}} &
			\multirow{2}{*}{\ce{In2O3}}
			& 0 & -1.47 & 1.6 \\ 
			& & +1 & -2.44 & 3.4 \\  
			\hline
			\multirow{2}{*}{\kv{O}{i}} &
			\multirow{2}{*}{\ce{ZnO}}
			& 0 & -1.87 & 4.3 \\ 
			& & +1 & -2.22 & 3.9  \\
			\hline
			\multirow{2}{*}{\kv{O}{i}} &
			\multirow{2}{*}{a-\ce{TiO2}}
			& 0 & -2.23 & 3.1 \\ 
			& & +1 & -1.99 & 3.2  \\ 
			\hline
		\end{tabular}
\footnotetext[1]{Metastable and ground state structures with bond distance labelling shown in SI Figs.~S8, S9, S10.}
\end{table}

\begin{figure}[h]
    \centering
    \includegraphics[width=1.0\textwidth]{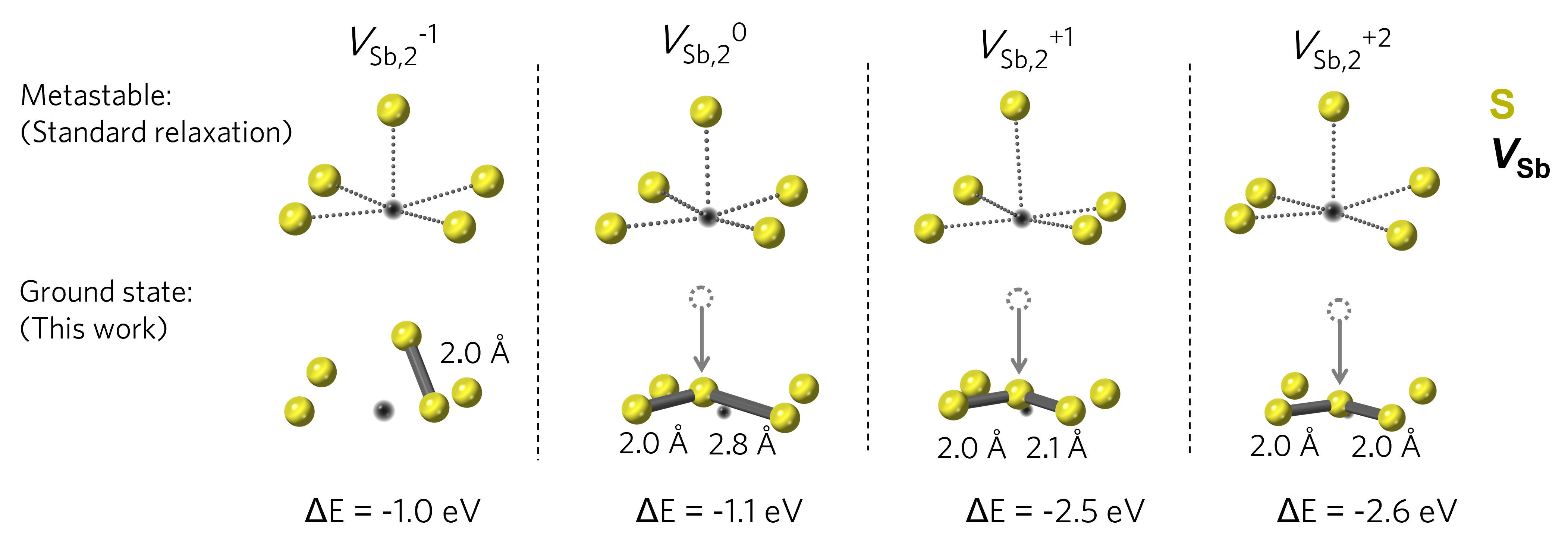}
    \caption{Anion-anion dimer formation undergone by \kvc{V}{Sb,2}{-1,0,+1,+2} in \ce{Sb2S3}, where the subscript differentiates between the two symmetry inequivalent sites of Sb (shown in SI Fig. S7). Similar reconstructions were observed for the other symmetry inequivalent site and in \ce{Sb2Se3}. For each charge state, on the top we show the configuration obtained by relaxing the ideal, high-symmetry geometry and on the bottom the ground state found with our method. For the singly negative charge states, the distortion is driven by the formation of one sulfur dimer while for the neutral and positively charge states a trimer is formed.
    \kv{V}{Sb} in black and S in yellow. Pseudo-bonds from the vacancy position to neighbouring S atoms are shown for the metastable configurations to illustrate the coordination environment and the arrows illustrate the key displacement from the high-symmetry structure.}
    \label{fig:vac_2_Sb}
\end{figure}

Within the cases reported in \cref{tab:dimer}, it is instructive to further consider the antimony vacancies (\kv{V}{Sb}) in the quasi-one dimensional materials \ce{Sb2(S/Se)3}. 
In contrast to other systems, the dimer reconstruction is favourable across several charge states, with its stability increasing with the magnitude of the charge state of the defect. As indicated by Crystal Orbital Hamilton Population analysis (COHP) \cite{Deringer_2011,Dronskowski_1993,Maintz_2013}, from the singly negative to the double positive state, we observe greater net bonding anion-anion interactions (SI: Table 3, Figure 11), thereby leading to stronger bonds and thus a more favourable reconstruction. Furthermore, for the neutral and positive charge states, the under-coordinated vacancy neighbours form \emph{two} anion-anion bonds (yielding a S-S-S trimer), which is achieved by a large displacement of one of the vacancy neighbours towards the other two (\cref{fig:vac_2_Sb}). This remarkable ability to distort likely stems from their soft, quasi-1D structure\cite{Wang_2022}, with van der Waals voids between atomic chains\cite{Zhang_2022} (SI Fig.~S5). 
Despite the key role of anion dimerisation for \kv{V}{Sb}, which significantly impacts its energy and structure, this reconstruction has not been reported by previous studies on \ce{Sb2S3}\cite{Cai_2020} or \ce{Sb2Se3}\cite{Savory_2019,Huang_2019,Liu_2017,Zhao_2021,Tumelero_2016,Stoliaroff_2020}, due to the local minimum trapping which our method aims to tackle. While sulfur dimers were not identified in \ce{Sb2(S/Se)3}, they were reported for \kv{V}{Bi} and $\rm S_i$ in the isostructural material \ce{Bi2S3}\cite{Han_2017}, where their formation stabilises abnormal charge states for both species, rendering them \emph{donor} defects --- contradicting the typical acceptor nature of cation vacancies and anion interstitials. 

Beyond covalently-bonded materials, anion-anion bond formation is also observed for anion interstitials in systems with a stronger ionic character (\ce{In2O3}, \ce{ZnO}, a-\ce{TiO2}). 
Rather than leading to hole localisation in the interstitial atom and/or neighbouring anions (metastable state), the defect can accommodate the charge deficiency by forming oxygen dimers, resulting in highly favourable reconstructions (\cref{tab:dimer}). When only two electrons are missing ($\rm O_i^{0}$), the interstitial atom displaces towards a lattice oxygen forming a peroxide ($\rm O_2^{-2}$, d(O-O)=\SI{1.45}{\AA}, SI Fig.~S10), which is 1.5-2.0 eV more favourable than the metastable double polaron state. If an additional electron is removed (\kvc{O}{i}{+1}\!\!), a hole is trapped in one of the peroxide antibonding $\pi$* orbitals (e.g. $\rm h^+ + O_2^{-2}$, d(O-O)=\SI{1.31}{\AA}, SI Fig.~S12). Remarkably, this peroxide hole trapping yields a stabilisation $>$\SI{2}{eV} compared to the system with 3 holes localised on the defect region, and has previously been reported for \emph{bulk} lithium peroxide (\ce{Li2O2})\cite{Ong_2012}. 

Overall, these anion-anion dimers lead to highly-favourable distortions and stabilise abnormal charge states for anion interstitials and cation vacancies. This unexpected behaviour highlights the key role of exploring the defect configurational space for accurate theoretical predictions of defect properties, while also demonstrating the importance of considering a wide range of charge states, as unforeseen chemical interactions (facilitated by defect reconstructions) may lead to unexpected stabilisations. 
Finally, we note that dimer reconstructions at defects are not uncommon, and have been previously reported for numerous vacancies and interstitials, including \kvc{V}{Se}{0} in ZnSe, \ce{CuInSe2} and \ce{CuGaSe2}\cite{Lany_2004}, \kvc{V}{S}{0} in ZnS\cite{Lany_2004}, \kvc{V}{Ti}{0,-1} and \kvc{V}{Zr}{0} in \ce{CaZrTi2O7} \cite{Mulroue_2011}, \kvc{O}{i}{0} in \ce{In2O3}\cite{Agoston_2009}, \ce{ZnO}\cite{Erhart_2005}, \ce{Al2O3}\cite{Sokol_2010}, \ce{MgO}\cite{Evarestov_1996,Kotomin_1998}, \ce{CdO}\cite{Burbano_2011}, \ce{SnO2}\cite{Scanlon_2012_Onthe,Godinho_2009}, \ce{PbO_2}\cite{Scanlon_2011_Nature}, \ce{CeO2}\cite{Keating_2012}, \ce{BaSnO3}\cite{Scanlon_2013_Defect} and
\ce{In2ZnO4}\cite{Walsh_2009}, $\rm Ag_{i}^{0}$ in AgCl and AgBr\cite{Wilson_2008}, \kvc{V}{I}{-}, \kvc{I}{i}{0},  \kvc{Pb}{i}{0}, $\rm Pb_{CH_3NH_3}^{0}$ and $\rm I_{CH_3NH_3}^{0}$ in \ce{CH3NH3PbI3} \cite{Agiorgousis_2014,Whalley_2017,Whalley_2021,Motti_2019}, $\rm Pb_i$ in \ce{CsPbBr3}\cite{Kang_2017}, \ce{(CH3NH3)3Pb2I7}\cite{Zhao_2016}, 
\ce{(CH3NH3)2Pb(SCN)2I2}\cite{Xiao_2016} and \kv{Sn}{i} in \ce{CH3NH3SnI3}\cite{Meggiolaro_2020}. 

\begin{figure}[ht]
    \centering
    \includegraphics[width=0.5\textwidth]{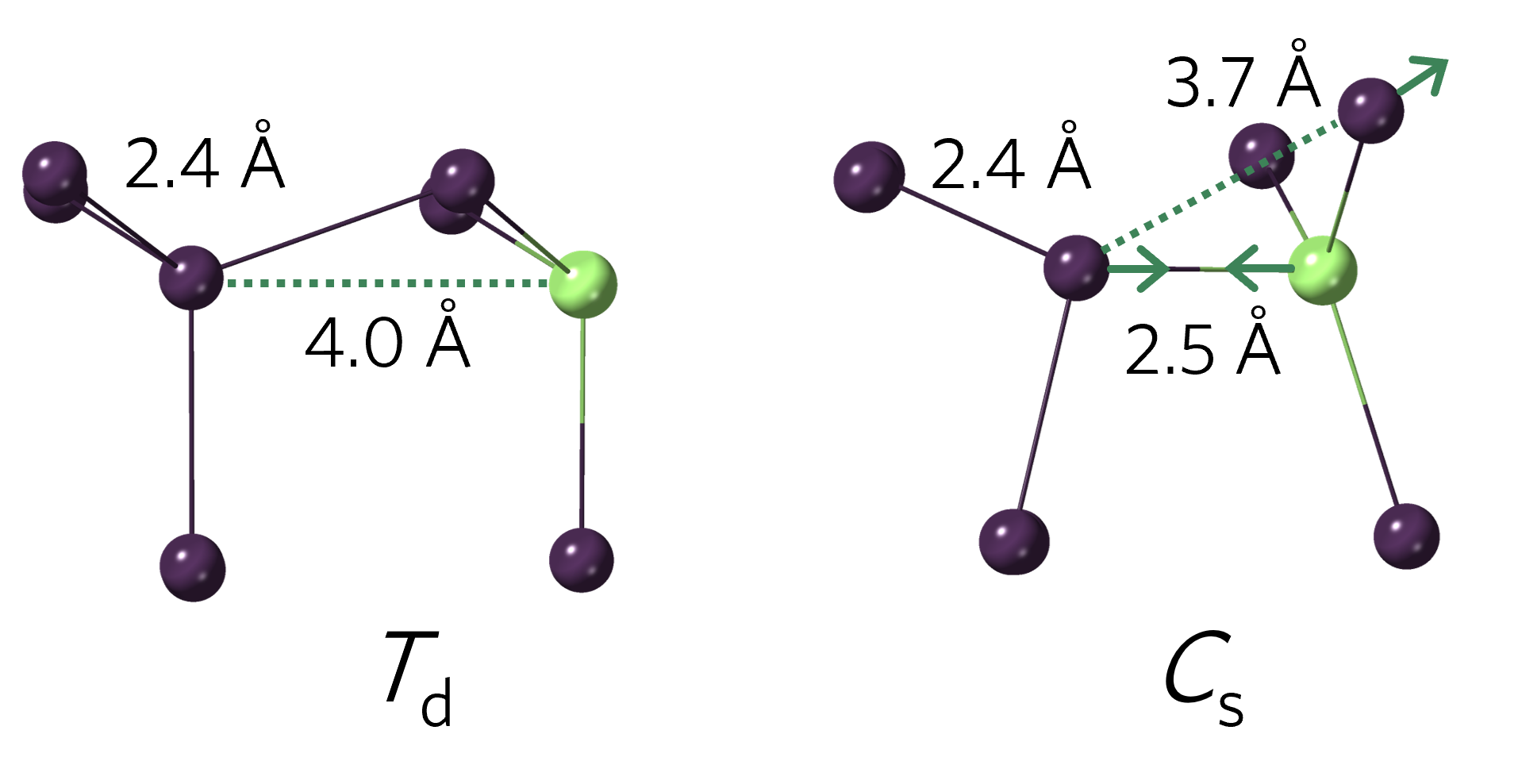}
    \caption{Cation-anion rebonding undergone by $\rm As_{Ga}^{-2}$ in GaAs. On the left is the metastable $T_d$ structure identified by standard defect relaxation, and to the right the ground state $C_s$ symmetry identified by our method. Ga in green and As in purple.}
    \label{fig:as_on_ga_-2}
\end{figure}

\emph{Cation-anion rebonding. }
Beyond dimerisation, rebonding between defect neighbours can also include replacing cation-cation/anion-anion (homoionic) bonds with more favourable cation-anion (heteroionic) bonds. This reconstruction motif is primarily observed for antisite defects, where the antisite may displace from its original position towards an oppositely-charged ion to form a new bond, while breaking a cation-cation/anion-anion bond. This is illustrated in \cref{fig:as_on_ga_-2} for $\rm As_{Ga}^{-2}$ in GaAs and was also found for $\rm Sb_{S}^{+3}$ and $\rm S_{Sb}^{+3}$ in \ce{Sb2S3}. In the former ($\rm As_{Ga}^{-2}$), both the cation and anion displace towards each other (\cref{fig:as_on_ga_-2}), from an original separation of \SI{4.00}{\AA} to \SI{2.49}{\AA} (close to the bulk Ga-As bond length of \SI{2.45}{\AA}), breaking an As-$\rm As_{Ga}$ bond in the process.
Interestingly, this behaviour contrasts with that of $\rm Te_{Cd}^{-2}$ in the isostructural compound semiconductor; CdTe, which displaces to form a Te-Te dimer\cite{Lindstrom_2015} rather than the cation-anion rebonding witnessed here. The differing reconstructions exhibited by these nominally similar anion-on-cation antisite defects -- both in zinc blende compound semiconductors, may be rationalised by the greater bond energy of Te-Te (\SI{260}{kJ/mol}) compared to As-As (\SI{200}{kJ/mol})\cite{liao_2006}, suggesting a greater proclivity toward anion dimerisation.

\begin{table}[ht]
    \begin{center}
        \caption{Cation--anion rebonding. Energy difference and summed atomic displacements ($\Sigma$Disp.) between the ground state identified by our method and the metastable configuration found when relaxing the ideal, undistorted defect structure (i.e. $\rm \Delta E = E_{\rm ground} - E_{\rm metastable}$).}
        \setlength{\tabcolsep}{8pt} 
        \renewcommand{\arraystretch}{1.3} 
		\label{tab:cation_anion}
		\begin{tabular}{ccccc}
			\toprule
        Defect & Material & Charge & $\Delta$E~(eV) & $\Sigma$Disp.~(\AA) \\
			\midrule
			{\ce{Sb_S}} & {\ce{Sb2S3}} 
            & +3 & -0.39 & 15.8\\
            {\ce{S_{Sb}}} & {\ce{Sb2S3}} 
            & +3 & -1.22 & 9.0 \\
            {\ce{S_i}} & {\ce{Sb2S3}} 
            & +3 & -0.45 & 11.5\\
            \ce{As_{Ga}} & GaAs & -2 & -0.37 & 5.2 \\ 
            \ce{Cd_{i}} & CdTe & +2 & -0.46 & 5.8 \\
			\botrule
		\end{tabular}
\footnotetext[1]{Metastable and ground state structures with bond distance labelling shown in SI Fig.~S13}
\end{center}
\end{table}	

Cation-anion rebonding is also found in the antimony antisite in \ce{Sb2S3} ($\rm Sb_{S}^{+3}$), where the high-symmetry metastable structure and the ground state differ by the formation of an additional Sb-S bond (SI Figure S13). 
This extra cation-anion bond has a distance of \SI{2.5}{\AA}, again comparable to the bulk Sb-S bond length of 2.5-2.8 \AA, and lowers the energy by \SI{0.4}{eV} (\cref{tab:cation_anion}). Beyond antisites, this behaviour is also expected for \emph{extrinsic} substitutional defects and dopants, where the same preference for heteroionic bonds can drive reconstructions to distorted, lower-energy arrangements. 

This behaviour is not exclusive to substitution-type defects, however, with similar reconstructions undergone by interstitial defects. For instance, the sulfur interstitial ($\rm S_{i}^{+3}$) in \ce{Sb2S3} displaces to form a $\rm S_i-Sb$ bond which lowers the energy by 0.45 eV (\cref{tab:cation_anion}, SI Fig. S13). 
Similar interactions drive the reconstruction for the cadmium interstitial in CdTe ($\rm Cd_{i}^{+2}$), where $\rm Cd_{i}^{+2}$ moves to a Te (rather than Cd) tetrahedral coordination, again with bond lengths (\SI{2.85}{\AA}) similar to the bulk cation-anion bond length (\SI{2.84}{\AA}) and lowering the energy by \SI{0.46}{eV}. Beyond the explicit formation of these heteroionic bonds, the change in the interstitial position also results in more favourable electrostatic interactions, as reflected by the lower Madelung energy of the ground state ($\Delta E_{\textrm{Madelung}} =$ \SI{-2.0}{eV}).
While this configuration is also found by considering a range of potential interstitial sites without distortion, the identified reconstruction demonstrates the ability of this approach to locate major structural distortions, where both the position of the defect atom and its coordination change significantly (summed atomic displacements of \SI{5.8}{\AA} and $\rm Cd_i^{+2}$ moving by \SI{0.86}{\AA}).

\subsection{Crystal field \& Jahn-Teller}
For defects in materials with a stronger ionic character, crystal field or Jahn-Teller effects can also drive symmetry-breaking reconstructions. In this category, we focus on the distortions introduced by a series of divalent metal dopants (Be, Ni, Pd, Pt, Cu) in ceria (\ce{CeO2}). 
Previous studies had shown these dopants to favour the formation of a charge-compensating oxygen vacancy (\kvc{V}{O}{+2}\!\!) and undergo significant reconstructions, which were only found by testing a range of chemically-guided manual distortions for each dopant\cite{Kehoe_2011}.

Accordingly, we tested our method by applying it to this system, finding it to successfully reproduce all previously-reported ground state structures, as well as identifying a novel lower-energy reconstruction for the Ni dopant. Here, Ni and Pd are found to distort from an originally cubic coordination to a more favourable square planar arrangement (SI Fig.~S14). This is also the case for Pt, however unlike the other dopants this ground state configuration is also found by a standard relaxation. 
For Ni and Pd, crystal field effects drive the displacement of the dopant by 1/4 of a unit cell (\SI{1.2}{\AA}) from the centre of the cube towards one of the faces (\cref{fig:distortion_rationales}, SI Fig.~S14). By lowering the energy of the occupied $d$ orbitals (as depicted in the electron energy diagram of SI Fig.~S15), the reconstruction leads to an overall stabilisation of 0.30 and \SI{0.87}{eV} for Ni and Pd, respectively.
Notably, the nickel distortion was overlooked by the previous study\cite{Kehoe_2011}, likely due to limited exploration of the PES and the shallower and narrower energy basin -- stemming from reduced crystal field stabilisation energy (\cref{tab:crystal_field}) due to weaker hybridisation between the anions and less diffuse 3$d$ orbitals (compared to 4$d$ and 5$d$ for Pd and Pt). \\
Finally, in the case of copper, a $d^9$ metal, it is the Jahn-Teller effect which drives a reconstruction towards a \emph{distorted} square-planar arrangement (\cref{fig:distortion_rationales}). By splitting the partially-occupied electron levels, it lowers the energy of the occupied $d$ orbitals, as shown in the orbital energy diagram of SI Fig.~S16, and leads to an overall stabilisation of 0.67 eV. 

\begin{table}[hb]
    \begin{center}
    \caption{Crystal field and Jahn-Teller driven distortions. Energy difference and summed atomic displacements ($\Sigma$Disp.) between the ground state identified by our method and the metastable configuration found when relaxing the ideal, undistorted defect structure (i.e. $\rm \Delta E = E_{\rm ground} - E_{\rm metastable}$).}
    \setlength{\tabcolsep}{8pt} 
    \renewcommand{\arraystretch}{1.3} 
    \label{tab:crystal_field}
    \begin{tabular}{cccc}
         \toprule 
         Defect & Material &  $\Delta$E~(eV) & $\Sigma$Disp.~(\AA) \\
         \midrule
         $\rm Ni_{Ce}^{-2}$ + \kvc{V}{O}{2+} & \ce{CeO2} & -0.30  &  2.2 \\
         $\rm Pd_{Ce}^{-2}$ + \kvc{V}{O}{2+} & \ce{CeO2} & -0.87  & 5.8 \\  
         $\rm Cu_{Ce}^{-2}$ + \kvc{V}{O}{2+} & \ce{CeO2} & -0.67 &  3.7 \\  
         \botrule
    \end{tabular}
\footnotetext[1]{Metastable and ground state structures with bond distance labelling shown in SI Fig.~S14}
\end{center}
\end{table}

\subsection{Electrostatically-driven}
Finally, for certain interstitials or substitutional defects, symmetry-breaking distortions can yield lower energy structures through optimising electrostatic interactions. Compared to cation-anion rebonding, which is driven by a combination of ionic and covalent bonding interactions, here \emph{no} explicit new bonds are formed, but rather the lattice electrostatic (Madelung) energy is lowered by defect rearrangement. The recurrence of this distortion motif for size-mismatched substitutions is related to their disruption of the nearby lattice framework (and thus long-range electrostatics), making reconstructions to reduce the lattice strain favourable.
For instance, this behaviour was observed for the beryllium dopant in \ce{CeO2}, which also forms a charge-compensating oxygen vacancy, i.e. $\rm Be_{Ce}^{-2}+$\kvc{V}{O}{+2}\!\!\! (as witnessed for other divalent dopants\cite{Kehoe_2011,Hiley_2015,Hegde_2015}). 
Surprisingly, instead of adopting the tetrahedral coordination of its parent oxide, beryllium distorts to a trigonal environment (SI Fig.~S17 (a,b)). This behaviour appears to be driven by the significant size mismatch between beryllium and cerium (with ionic radii of $\rm r_{Be^{+2}}=0.59$~\AA~and  $\rm r_{Ce^{+4}}=1.01$~\AA), requiring significant lattice distortion to attain the optimum Be-O distances in the tetrahedral arrangement. To adopt the tetrahedral coordination, three O atoms significantly displace from their lattice sites towards Be, straining the other Ce-O bonds by 0.2 \AA.
In contrast, the trigonal configuration enables optimum Be-O separation while maintaining near ideal distances for the nearby Ce-O bonds, thereby optimising the overall ionic interactions. This is reflected by the Madelung energy, which is 3.2 eV lower in the ground state configuration, and the overall energy lowering of 0.34 eV. 

Interstitials can also exhibit these electrostatically-driven reconstructions, with their higher mobility allowing for displacement to more favourable Madelung potential sites.
For example, the main difference between the metastable and ground state structures of the tellurium interstitial ($\rm Te_{i}^{-2}$) in cadmium telluride (\ce{CdTe}) lies in its coordination environment. While in the metastable configuration $\rm Te_i$ is tetrahedrally coordinated by other Te anions (distances of \SI{3.4}{\AA}), in the ground state the Te atoms rearrange to a split configuration, both occupying a hexagonal void and minimising interactions with neighbouring Te (distances of 3.5, 3.5, 3.6, \SI{4.0}{\AA}) (SI Fig.~S17 (c,d)). This reduction of unfavourable anion-anion interactions is again witnessed in the Madelung energy, which is 2 eV lower for the ground state, and drives an overall energy lowering of \SI{0.20}{eV}. Although these examples are primarily driven by optimising electrostatic interactions, the different distortion motifs in Fig.~3 are not mutually exclusive. Often several effects can contribute to a certain distortion, as observed for $\rm Cd_i^{+2}$ in CdTe, where the distortion leads to cation-anion rebonding (via explicit replacement of cation-cation bonds with cation-anion) but \emph{also} lowering the lattice electrostatic energy (by surrounding the interstitial with oppositely charged ions). 

\begin{table}[ht]
    \begin{center}
    \caption{Electrostatically-driven reconstructions. Energy difference and summed atomic displacements ($\Sigma$Disp.) between the ground state identified by our method and the metastable configuration found when relaxing the ideal, undistorted defect structure (i.e. $\rm \Delta E = E_{\rm ground} - E_{\rm 
    metastable}$).}
    \setlength{\tabcolsep}{8pt} 
    \renewcommand{\arraystretch}{1.3}
    \label{tab:electrostatic}
    \begin{tabular}{cccc}
         \toprule
         Defect & Material &  $\Delta$E~(eV) & $\Sigma$Disp.~(\AA) \\
         \midrule
         $\rm Te_{i}^{-2}$  & \ce{CdTe} & -0.20  & 13.0  \\ 
         $\rm Cd_{i}^{+2}$ & \ce{CdTe} & -0.46 & 5.8 \\
         $\rm Be_{Ce}^{-2}$ + \kvc{V}{O}{+2} & \ce{CeO2} & -0.34  & 19.7 \\
         \botrule
    \end{tabular}
\footnotetext[1]{Metastable and ground state structures with bond distance labelling shown in SI Figs.~S17}
\end{center}
\end{table}

Beyond these motifs, our approach can also be employed to identify defect bound polarons, as exemplified in the supporting information for \kvc{V}{O}{0} in \ce{CeO2}, \kvc{V}{O}{0} and \kvc{Ti}{i}{0} in rutile \ce{TiO2}, and \kvc{V}{In}{-1} in \ce{In2O3} (SI Section VI). For ceria and rutile, the two electrons donated by the defect localise on lattice cations reducing them to Ce/Ti(III). This leads to many competing states depending on both the defect-polaron and polaron-polaron distances. While the different localisation arrangements are successfully identified by our approach, the complexity of the PES (hosting many competing low energy minima with small energy and structural differences) warrants further study. 

\subsection{Impact on defect properties}
\begin{figure}[ht]
    \centering
    \includegraphics[width=1.0\linewidth]{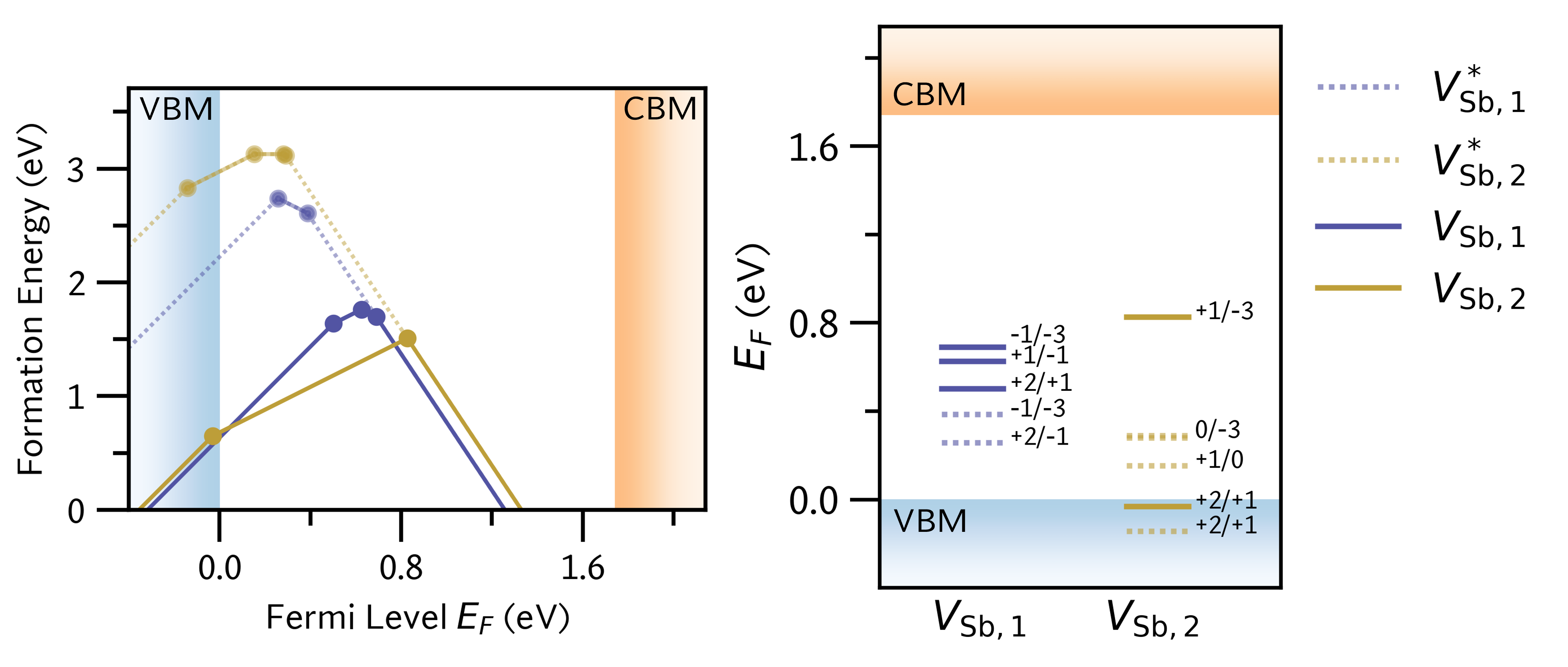}
    \caption{Defect formation energy (left) and vertical energy level (right) diagram for the antimony vacancies in \ce{Sb2S3}, under sulfur rich conditions\cite{lian_revealing_2021} ($\mu_{\rm S}$ = 0, $\mu_{\rm Sb}$=-\SI{0.55}{eV}). 
    The two symmetry inequivalent sites are depicted in purple (\kv{V}{Sb,1}) and gold (\kv{V}{Sb,2}). For each site, the ground state structures are shown with solid lines (\kv{V}{Sb}), while the metastable states are shown with faded dotted lines (\kvc{V}{Sb}{*}). The valence band maximum (VBM) and conduction band minimum (CBM) are shown in blue and orange, respectively.}
    \label{fig:TLD_vac_sb}
\end{figure}

In this section we demonstrate how these reconstructions can affect calculated defect properties. 
We take the charge transition level diagram for antimony vacancies in \ce{Sb2S3} as an exemplar, and compare the formation energies and defect levels calculated for the ground (\kv{V}{Sb}) and metastable states (\kvc{V}{Sb}{*}\!), for both symmetry-inequivalent vacancy positions (\kv{V}{Sb,1} and \kv{V}{Sb,2}; SI Fig. S7). As shown in \cref{fig:TLD_vac_sb,tab:dimer}, the highly favourable dimerisation reconstructions undergone by several charge states (+2, +1, 0, -1) significantly lower the formation energy of \kv{V}{Sb}, by 0.8 -- 2.6 eV depending on charge state. As a result, the predicted concentration under typical growth conditions (T = \SI{550}{K})\cite{Huang_more_2021} increases by $\exp(\frac{\Delta E}{k_BT}) \simeq$ \textbf{21} orders of magnitude for \kvc{V}{Sb,2}{+1} (SI Table IV). Furthermore, the reconstructions also affect the \emph{nature} and \emph{position} of the charge transition levels. The neutral state of \kv{V}{Sb,2} is now predicted to be thermodynamically unfavourable at all Fermi levels, which leads to the disappearance of the (+1/0) and (0/-3) transition levels (\cref{fig:TLD_vac_sb}).  
Similarly, for \kv{V}{Sb,1}, the reconstructions render the singly-positive vacancy stable within the bandgap, resulting in two new transition levels ((+2/+1) and (+1/-1)) (\cref{fig:TLD_vac_sb}, SI Table V).

More significant, however, is the change in the energetic position of the levels. The greater stabilisation of the positive charge states, leads to the levels shifting \emph{deeper} into the bandgap (SI Table V). For instance, the (-1/-3) transition of \kv{V}{Sb,1} shifts 0.3 eV higher in the bandgap, revealing it to be a highly amphoteric defect with strong self-charge-compensating character (\cref{fig:TLD_vac_sb}) -- in fact now aligning with the behaviour of the other intrinsic defects in \ce{Sb2(S/Se)3}\cite{Cai_2020,guo_scalable_2019}. This change in the position of the transition levels can have important consequences when modelling defects in photovoltaic materials, as mid-gap states often act as carrier traps that promote non-radiative electron-hole recombination and decrease performance.
This deep, amphoteric character of the antimony vacancies was missed by previous theoretical studies on \ce{Sb2S3} defects,\cite{Cai_2020,guo_scalable_2019} due to the local minimum trapping which our approach aims to combat, giving significantly underestimated concentrations and shallower transition levels. Overall, this demonstrates the potent sensitivity of predicted concentrations and transitions levels on the underlying defect structures, highlighting the crucial importance of correct defect structure identification for quantitative and reliable defect modelling.

Beyond thermodynamic properties, these reconstructions can also affect calculated rates of non-radiative electron/hole capture. Indeed, charge capture rates depend intimately on the structural PESs of the charge states involved, as well as the position of the transition level. This effect has been recently demonstrated for the neutral cadmium vacancy in CdTe\cite{Kavanagh_2021} as well as for the neutral lead interstitial in \ce{CH3NH3PbI3}\cite{Zhang_2020} (MAPI). For the latter, the timescale of non-radiative charge recombination varies by an order of magnitude depending on whether the Pb interstitial forms a dimer (ground state structure) or remains in a non-bonded position (metastable configuration). 
Moreover, even when defect reconstructions do not lower the energy significantly, this can still result in completely different capture behaviour, as observed for the tellurium interstitial in CdTe\cite{kavanagh2022impact} and the sulfur substitution in silicon\cite{Cai_2021}. For $\rm S_{Si}$, a small structural distortion (with a negligible energy difference) significantly affects its capture behaviour, giving a capture rate now in agreement with experiment\cite{Cai_2021} and once again highlighting the critical role of exploring the defect configurational landscape.

\subsection{Locating metastable structures}
Thus far, we have focused on atomic reconstructions that lower the energy and lead to a new \emph{ground state} structure, which is missed with a standard relaxation from the high symmetry geometry. 
While defect properties such as doping are often determined by the ground state configuration, low energy metastable structures can also be important to performance in device applications. For instance, they often represent transition states in the ion migration process\cite{Krasikov_2018} (crucial for battery materials and semiconductor doping), and can impact non-radiative carrier recombination in solar cells and LEDs, behaving as intermediate species to the charge trapping process\cite{Kavanagh_2021,Yang_2016,kavanagh2022impact}. They can produce anomalous properties, such as persistent photo-conductivity and photo-induced capacitance quenching in semiconductors\cite{Coutinho_2020}. 
Accordingly, we investigated the ability of our approach to locate relevant low-energy metastable configurations by considering the DX centres in GaAs (\kvc{Si}{Ga}{-1}, \kvc{Sn}{Ga}{-1}, \kvc{S}{As}{-1}, \kvc{Te}{As}{-1}). 

\begin{figure}[ht]
    \centering
    \includegraphics[width=0.9\textwidth]{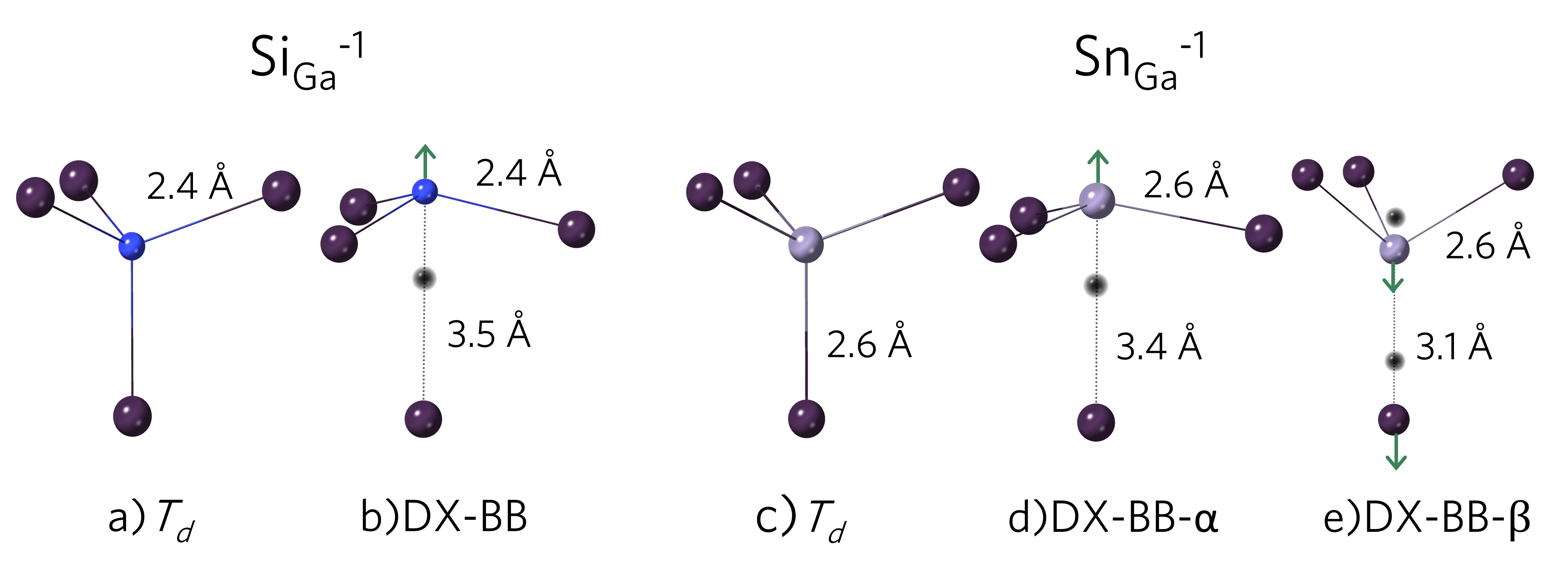}
    \caption{Ground state ($T_d$) and low energy metastable configurations identified for \kvc{Si}{Ga}{-1} and \kvc{Sn}{Ga}{-1} in GaAs. The DX-BB (DX-BB-$\alpha$) configuration consists of a $C_{3\textrm{v}}$ distortion, where Si (Sn) displaces away from the original Ga position, breaking a dopant-As bond.  In the DX-BB-$\beta$ configuration, a Sn-As bond is also broken, but in this case As undergoes the largest displacement. 
    Si in blue, Sn in grey and As in purple. The arrows illustrate how the atoms reconstruct from the $T_d$ configuration and the black spheres depict the original position of the displaced atoms in the $T_d$ structure. }
    \label{fig:group_IV_dopants}
\end{figure}

For Si, Sn and Te impurities, our method successfully identifies all low-energy metastable structures reported by previous studies,\cite{MaoHua_2005,Kim_2019,Dobaczewski_1995,Yamaguchi_1991,Li_2005,Saito_1992,Kundu_2019} while for S, it identifies two of the three structures previously found.\cite{Du_dx_2005,Kundu_2019}  
For instance, in the case of Si, we correctly identify the broken bond configuration (DX-BB)\cite{MaoHua_2005,Kim_2019,Dobaczewski_1995,Yamaguchi_1991,Li_2005,Saito_1992}, which entails a $C_{3\textrm{v}}$ Jahn-Teller distortion with the dopant displacing along the [111] direction thereby breaking a Si-As bond (\cref{fig:group_IV_dopants} (b)). We find it to lie 0.34 eV higher in energy than the ground state $T_d$ configuration (\cref{tab:DX_centres}), in agreement with earlier local DFT (LDA) calculations which found it \SI{0.5}{eV} higher in energy\cite{MaoHua_2005,Li_2005}. 
A similar bond-breaking $C_{3\textrm{v}}$ distortion is also found for the other group IV dopant (Sn), though now with two possible low energy configurations (DX-BB-$\alpha$ and DX-BB-$\beta$). While the former corresponds to the conventional DX behaviour with major off-centring of the dopant atom, in the latter it is mainly a neighbouring As which displaces to an interstitial position (\cref{fig:group_IV_dopants} (d, e)). We find the $\alpha$ configuration to lie slightly lower in energy than $\beta$ (\SI{0.26}{eV} vs \SI{0.41}{eV} above the $T_d$ ground state) (\cref{tab:DX_centres}), agreeing with previous local DFT calculations which found DX-$\alpha$ to be \SI{0.4}{eV} higher in energy than $T_d$\cite{MaoHua_2005}. 

\begin{figure}[ht]
    \centering
    \includegraphics[width=0.9\textwidth]{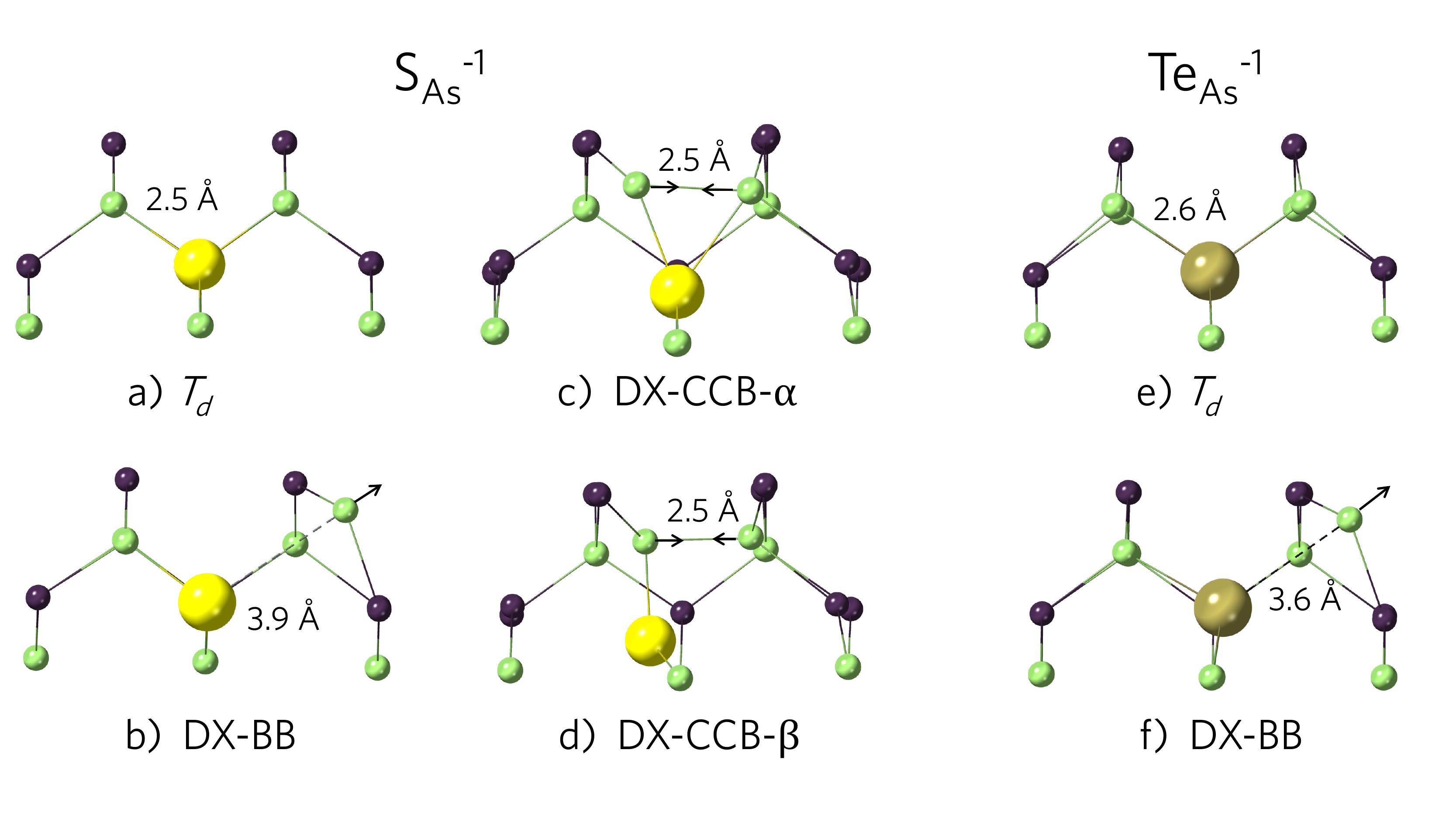}
    \caption{Ground state ($T_d$) and low-energy metastable configurations (DX-BB, DX-CCB-$\alpha$ and DX-CCB-$\beta$) identified for \kvc{S}{As}{-1} and \kvc{Te}{As}{-1} in GaAs.  In the DX-BB configuration, a dopant-Ga bond is broken,  
    while the DX-CCB configurations correspond to cation-cation bond (CCB) formation. 
    S in yellow, Te in gold, Ga in green and As in purple. 
    }
    \label{fig:group_vi_dopants}
\end{figure}

Regarding the sulfur dopant, there are several metastable configurations. Similar to the other species, it can adopt a broken bond arrangement (DX-BB, \cref{fig:group_vi_dopants} (b)), where one of the dopant neighbours displaces off-site and breaks a S-Ga bond. In addition, it can adopt two configurations with cation-cation bonding (CCB) -- corresponding to dimerisation reconstructions. As depicted in \cref{fig:group_vi_dopants}, two of the neighbouring Ga can displace towards each other and form a Ga-Ga bond (\SI{2.5}{\AA}), with the dopant remaining either in the original position (DX-CCB-$\alpha$) or displacing slightly off-site (DX-CCB-$\beta$). The dimer (DX-CCB) configurations are found to be more favourable than the broken bond (DX-BB) one, with energies of \SI{0.30}{eV} (DX-CCB-$\alpha$, DX-CCB-$\beta$) and \SI{0.44}{eV} (DX-BB) relative to the ground state $T_d$ structure. Again, these results agree with previous local DFT (LDA) calculations, which found relative energies of \SI{0.5}{eV} for DX-CCB ($\alpha$ \& $\beta$) and \SI{0.6}{eV} for DX-BB, relative to $T_d$\cite{Du_dx_2005,Kundu_2019}. 
We note that our method did not locate the DX-CCB-$\beta$ arrangement, likely due to a soft PES (with an energy barrier of only \SI{25}{meV} between alpha and beta, SI Section VIII), the close similarity of structure and energies for DX-CCB-$\alpha/\beta$, and the bias toward \emph{ground state} configurations. This highlights a limitation of our approach, where identifying several \emph{metastable} defect structures on a complex PES may require a more exhaustive exploration (e.g. distorting different number of neighbours or using a denser grid of distortions). 

Finally, for \kvc{Te}{As}{-1}, we identify DX-BB as the lowest energy metastable structure (\cref{fig:group_vi_dopants} (f)), in agreement with previous theoretical\cite{MaoHua_2005,Kundu_2019} and experimental\cite{Dobaczewski_1995} studies. 
We calculate an energy difference of \SI{0.25}{eV} between the metastable $C_{3\textrm{v}}$ and ground state $T_d$ configurations, similar to the value of \SI{0.38}{eV} obtained with local DFT\cite{MaoHua_2005}. Overall, these results demonstrate the ability of the method to identify low-energy metastable configurations at negligible additional cost.  

\begin{table}[ht]
    \centering
    \caption{Energy difference and summed atomic displacements ($\Sigma$Disp.) between the identified metastable configurations and the ground state $T_d$ for the DX centres in GaAs. For $\rm S_{As}^{-1}$, we include a configuration not found by our method (DX-CCB-$\beta$*) but reported in a previous study\cite{Kundu_2019}.}
    \label{tab:DX_centres}
    \setlength{\tabcolsep}{7pt} 
    \renewcommand{\arraystretch}{1.4}
    \begin{tabular}{ccccc}
         \toprule 
         Defect & Ground state & Metastable & $\Delta$E~(eV) & $\Sigma$Disp.~(\AA) \\
         \midrule 
         $\rm Si_{Ga}^{-1}$  & $T_d$ & DX-BB & 0.34 & 2.0 \\ 
         \hline
         \multirow{2}{*}{$\rm Sn_{Ga}^{-1}$}  & \multirow{2}{*}{$T_d$} 
         & DX-BB-$\alpha$ & 0.26 & 2.7 \\ %
         & & DX-BB-$\beta$ & 0.41 & 2.0 \\
         \hline
         \multirow{3}{*}{$\rm S_{As}^{-1}$}
         & \multirow{3}{*}{$T_d$} & DX-BB & 0.44 & 1.4 \\
         & & DX-CCB-$\alpha$ & 0.30 & 4.6 \\
         & & DX-CCB-$\beta$* & 0.30 &  5.5 \\ 
         \hline
         $\rm Te_{As}^{-1}$  & $T_d$ & DX-BB & 0.25 & 1.9\\ 
         \botrule 
    \end{tabular}
\end{table}

\section{Conclusions}
In summary, we present a method to explore the configurational space of point defects in solids and identify ground state and low-energy metastable structures. Based on the defects tested, if computational resources only allow limited calculations to be performed, we suggest structural perturbations of $\pm30\%$ using our method provides a higher chance of finding the true ground state defect configuration over starting from the unperturbed bulk structure. Beyond its simplicity and automated application, the low computational cost and the lack of system-dependent parameters make it more practical than current alternative structure-searching methods (comparisons are provided in SI Section IV). The applicability to both standard and high-throughput defect investigations is demonstrated by the range of materials studied herein.

By exploring a variety of defects and materials, the key physico-chemical factors that drive defect reconstructions were also highlighted. 
For systems with mixed ionic-covalent bonding (such as those involving the Ge, Sn, Sb, Bi cations), we expect `rebonding' reconstructions to prevail, with the defect neighbours displacing to yield novel homoionic (dimer) or heteroionic (cation-anion) bonds. Furthermore, if the crystals also display low symmetry (e.g. one-dimensional connectivity in SbSeI, SbSI, and SbSBr), these reconstructions will likely be more significant and prevalent, as exemplified by the antimony chalcogenides (\ce{Sb2(S/Se)3}).
Here, its open and flexible structure hosts several coordination environments, with the empty spaces between the chains enabling atomic displacements without major strain on neighbouring sites, yielding many energy-lowering reconstructions away from the high-symmetry local minimum.
Notably, through the introduction of new bonds, these reconstructions can stabilise unexpected charge states, highlighting the impact on the \emph{qualitative} behaviour of defects. 
For crystals with stronger ionic character, distortions can either be driven by novel homoionic bonds (peroxides in \ce{In2O3}, \ce{ZnO} and \ce{TiO2}) or Jahn-Teller/crystal field stabilisation effects (\ce{CeO2}). Finally, optimising ionic interactions can also result in energy lowering rearrangements, especially for cases of significant size mismatch between dopant and host atoms, or for interstitials, which distort to minimise Coulombic repulsion with neighbouring ions. 

Regarding future improvements, the approach could be enhanced with a machine learning model, which could identify subsets of likely reconstructions to improve sampling efficiency. Similarly, local symmetry analysis could be incorporated to suggest potential subgroup distortions from the initial point group symmetry (e.g. $T_d \rightarrow C_{3\textrm{v}}$, $C_{2\textrm{v}}$, $D_{2\textrm{d}}$ ...) and then generate distorted structures based on this. This feature would likely be ill-suited for low-symmetry materials such as \ce{Sb2(S/Se)3}, however, as the wide range of symmetry-inequivalent distortions could become prohibitively expensive to compute with accurate levels of theory. 

Overall, our results demonstrate the prevalence of energy-lowering reconstructions for defects, that are often missed through local optimisation from a high symmetry configuration. The major structural and energetic differences highlight the crucial impact on defect properties. 
Beyond thermodynamic properties, the modification of the structural PES will completely change recombination, absorption and luminescence behaviour. Consequently, navigating the configurational landscape is key for accurate predictions of defect properties and their impact on materials performance, ranging from energy technologies (thermoelectric/photovoltaic efficiency, battery conductivity and catalytic activity) to defect-enabled applications such as lasing and quantum computing.

\backmatter
{\bf Supplementary information.}
Computational details; Parameter optimisation of the defect structure searching approach; Comparison with other structure searching approaches; Figures of the metastable and ground state structures for the identified energy lowering distortions; Application of the method to defect bound polarons; Concentrations and charge transition levels predicted for the ground state and metastable structures of antimony vacancies in \ce{Sb2S3}; Energy barrier between the DX-CCB-$\alpha$ and DX-CCB-$\beta$ configurations of $\rm S_{As}^{-1}$ in GaAs.

\section*{Declarations}
{\bf Data \& code availability.}
The code used to generate and analyse defect distortions is available from \url{https://github.com/SMTG-UCL/ShakeNBreak}.
The identified ground state and metastable structures are available from the Zenodo repository with DOI 10.5281/zenodo.6558244.

{\bf Author contributions.}
Conceptualisation \& Project Administration: Seán R. Kavanagh, David O.Scanlon. Investigation and methodology: Irea Mosquera-Lois, Seán R. Kavanagh. Supervision: Seán R. Kavanagh, Aron Walsh, David O. Scanlon. Writing - original draft: Irea Mosquera-Lois. Writing - review \& editing: All authors. Resources and funding acquisition: Aron Walsh, David O. Scanlon. These author contributions are defined according to the CRediT contributor roles taxonomy.

{\bf Acknowledgments.}
I.M.L. thanks La Caixa Foundation for funding a postgraduate scholarship (ID 100010434,  fellowship code LCF/BQ/EU20/11810070). 
S.R.K. acknowledges the EPSRC Centre for Doctoral Training in the Advanced Characterisation of Materials (CDT-ACM)(EP/S023259/1) for funding a PhD studentship. DOS acknowledges support from the EPSRC (EP/N01572X/1) and from the European Research Council, ERC (Grant No. 758345). 
Via membership of the
UK’s HEC Materials Chemistry Consortium, which is funded
by the EPSRC (EP/L000202, EP/R029431, EP/T022213),
this work used the UK Materials and
Molecular Modelling (MMM) Hub (Thomas EP/P020194
and Young EP/T022213).

{\bf Competing interests.}
The authors declare no competing interests.

\bibliography{BDM}

\begin{thebibliography}{100}
\expandafter\ifx\csname url\endcsname\relax
  \def\url#1{\burl{#1}}\fi
\expandafter\ifx\csname urlprefix\endcsname\relax\def\urlprefix{URL }\fi
\providecommand{\bibinfo}[2]{#2}
\providecommand{\eprint}[2][]{\url{#2}}
\providecommand{\doi}[1]{\url{https://doi.org/#1}}
\bibcommenthead

\bibitem{Arrigoni_evolutionary_2021}
\bibinfo{author}{Arrigoni, M.} \& \bibinfo{author}{Madsen, G. K.~H.}
\newblock \bibinfo{title}{Evolutionary computing and machine learning for
  discovering of low-energy defect configurations}.
\newblock \emph{\bibinfo{journal}{npj Comp Mater}}
  \textbf{\bibinfo{volume}{7}}~(1) (\bibinfo{year}{2021}) .

\bibitem{Ong_2012}
\bibinfo{author}{Ong, S.~P.}, \bibinfo{author}{Mo, Y.} \&
  \bibinfo{author}{Ceder, G.}
\newblock \bibinfo{title}{Low hole polaron migration barrier in lithium
  peroxide}.
\newblock \emph{\bibinfo{journal}{Phys. Rev. B}} \textbf{\bibinfo{volume}{85}},
  \bibinfo{pages}{081105} (\bibinfo{year}{2012}) .

\bibitem{Evarestov_2017}
\bibinfo{author}{Evarestov, R.~A.} \emph{et~al.}
\newblock \bibinfo{title}{{Use of site symmetry in supercell models of
  defective crystals: polarons in $\rm CeO_2$}}.
\newblock \emph{\bibinfo{journal}{Phys. Chem. Chem. Phys.}}
  \textbf{\bibinfo{volume}{19}}, \bibinfo{pages}{8340--8348}
  (\bibinfo{year}{2017}) .

\bibitem{Lany_dimer_2004}
\bibinfo{author}{Lany, S.} \& \bibinfo{author}{Zunger, A.}
\newblock \bibinfo{title}{{Metal-Dimer Atomic Reconstruction Leading to Deep
  Donor States of the Anion Vacancy in II-VI and Chalcopyrite Semiconductors}}.
\newblock \emph{\bibinfo{journal}{Phys. Rev. Lett.}}
  \textbf{\bibinfo{volume}{93}}, \bibinfo{pages}{156404} (\bibinfo{year}{2004})
  .

\bibitem{Lindstrom_2015}
\bibinfo{author}{Lindstr\"om, A.}, \bibinfo{author}{Mirbt, S.},
  \bibinfo{author}{Sanyal, B.} \& \bibinfo{author}{Klintenberg, M.}
\newblock \bibinfo{title}{High resistivity in undoped {CdTe}: carrier
  compensation of {Te} antisites and {Cd} vacancies}.
\newblock \emph{\bibinfo{journal}{J. Phys. D}}
  \textbf{\bibinfo{volume}{49}}~(3), \bibinfo{pages}{035101}
  (\bibinfo{year}{2015}) .

\bibitem{Sokol_2010}
\bibinfo{author}{Sokol, A.~A.}, \bibinfo{author}{Walsh, A.} \&
  \bibinfo{author}{Catlow, C. R.~A.}
\newblock \bibinfo{title}{{Oxygen interstitial structures in close-packed metal
  oxides}}.
\newblock \emph{\bibinfo{journal}{Chemical Physics Letters}}
  \textbf{\bibinfo{volume}{492}}~(1), \bibinfo{pages}{44--48}
  (\bibinfo{year}{2010}) .

\bibitem{Osterbacka_2022}
\bibinfo{author}{\"Osterbacka, N.}, \bibinfo{author}{Ambrosio, F.} \&
  \bibinfo{author}{Wiktor, J.}
\newblock \bibinfo{title}{{Charge Localization in Defective $\rm BiVO_4$}}.
\newblock \emph{\bibinfo{journal}{J. Phys. Chem. C}}
  \textbf{\bibinfo{volume}{126}}~(6), \bibinfo{pages}{2960--2970}
  (\bibinfo{year}{2022}) .

\bibitem{Krajewska_2021}
\bibinfo{author}{Krajewska, C.~J.} \emph{et~al.}
\newblock \bibinfo{title}{{Enhanced visible light absorption in layered $\rm
  Cs_3Bi_2Br_9$ through mixed-valence Sn(II)/Sn(IV) doping}}.
\newblock \emph{\bibinfo{journal}{Chem. Sci.}} \textbf{\bibinfo{volume}{12}},
  \bibinfo{pages}{14686--14699} (\bibinfo{year}{2021}) .

\bibitem{Mosquera_2021}
\bibinfo{author}{Mosquera-Lois, I.} \& \bibinfo{author}{Kavanagh, S.~R.}
\newblock \bibinfo{title}{In search of hidden defects}.
\newblock \emph{\bibinfo{journal}{Matter}} \textbf{\bibinfo{volume}{4}}~(8),
  \bibinfo{pages}{2602--2605} (\bibinfo{year}{2021}) .

\bibitem{Kavanagh_2021}
\bibinfo{author}{Kavanagh, S.~R.}, \bibinfo{author}{Walsh, A.} \&
  \bibinfo{author}{Scanlon, D.~O.}
\newblock \bibinfo{title}{{Rapid Recombination by Cadmium Vacancies in
  {CdTe}}}.
\newblock \emph{\bibinfo{journal}{ACS Energy Lett.}}
  \textbf{\bibinfo{volume}{6}}~(4), \bibinfo{pages}{1392--1398}
  (\bibinfo{year}{2021}) .

\bibitem{Kehoe_2011}
\bibinfo{author}{Kehoe, A.~B.}, \bibinfo{author}{Scanlon, D.~O.} \&
  \bibinfo{author}{Watson, G.~W.}
\newblock \bibinfo{title}{{Role of Lattice Distortions in the Oxygen Storage
  Capacity of Divalently Doped {$\rm CeO_2$}}}.
\newblock \emph{\bibinfo{journal}{Chem. Mater.}}
  \textbf{\bibinfo{volume}{23}}~(20), \bibinfo{pages}{4464--4468}
  (\bibinfo{year}{2011}) .

\bibitem{Lany_2004}
\bibinfo{author}{Lany, S.} \& \bibinfo{author}{Zunger, A.}
\newblock \bibinfo{title}{{Metal-Dimer Atomic Reconstruction Leading to Deep
  Donor States of the Anion Vacancy in {II-VI} and Chalcopyrite
  Semiconductors}}.
\newblock \emph{\bibinfo{journal}{Phys. Rev. Lett.}}
  \textbf{\bibinfo{volume}{93}}, \bibinfo{pages}{156404} (\bibinfo{year}{2004})
  .

\bibitem{Goyal_2020}
\bibinfo{author}{Goyal, A.} \emph{et~al.}
\newblock \bibinfo{title}{{On the Dopability of Semiconductors and Governing
  Material Properties}}.
\newblock \emph{\bibinfo{journal}{Chem. Mater.}}
  \textbf{\bibinfo{volume}{32}}~(11), \bibinfo{pages}{4467--4480}
  (\bibinfo{year}{2020}) .

\bibitem{Pickard_2006}
\bibinfo{author}{Pickard, C.~J.} \& \bibinfo{author}{Needs, R.~J.}
\newblock \bibinfo{title}{{High-Pressure Phases of Silane}}.
\newblock \emph{\bibinfo{journal}{Phys. Rev. Lett.}}
  \textbf{\bibinfo{volume}{97}}, \bibinfo{pages}{045504} (\bibinfo{year}{2006})
  .

\bibitem{Morris_2008}
\bibinfo{author}{Morris, A.~J.}, \bibinfo{author}{Pickard, C.~J.} \&
  \bibinfo{author}{Needs, R.~J.}
\newblock \bibinfo{title}{Hydrogen/silicon complexes in silicon from
  computational searches}.
\newblock \emph{\bibinfo{journal}{Phys. Rev. B}} \textbf{\bibinfo{volume}{78}},
  \bibinfo{pages}{184102} (\bibinfo{year}{2008}) .

\bibitem{Morris_2009}
\bibinfo{author}{Morris, A.~J.}, \bibinfo{author}{Pickard, C.~J.} \&
  \bibinfo{author}{Needs, R.~J.}
\newblock \bibinfo{title}{Hydrogen/nitrogen/oxygen defect complexes in silicon
  from computational searches}.
\newblock \emph{\bibinfo{journal}{Phys. Rev. B}} \textbf{\bibinfo{volume}{80}},
  \bibinfo{pages}{144112} (\bibinfo{year}{2009}) .

\bibitem{Morris_2011}
\bibinfo{author}{Morris, A.~J.}, \bibinfo{author}{Grey, C.~P.},
  \bibinfo{author}{Needs, R.~J.} \& \bibinfo{author}{Pickard, C.~J.}
\newblock \bibinfo{title}{{Energetics of hydrogen/lithium complexes in silicon
  analyzed using the Maxwell construction}}.
\newblock \emph{\bibinfo{journal}{Phys. Rev. B}} \textbf{\bibinfo{volume}{84}},
  \bibinfo{pages}{224106} (\bibinfo{year}{2011}) .

\bibitem{Mulroue_2011}
\bibinfo{author}{Mulroue, J.}, \bibinfo{author}{Morris, A.~J.} \&
  \bibinfo{author}{Duffy, D.~M.}
\newblock \bibinfo{title}{Ab initio study of intrinsic defects in zirconolite}.
\newblock \emph{\bibinfo{journal}{Phys. Rev. B}} \textbf{\bibinfo{volume}{84}},
  \bibinfo{pages}{094118} (\bibinfo{year}{2011}) .

\bibitem{Coutinho_2020}
\bibinfo{author}{Coutinho, J.}, \bibinfo{author}{Markevich, V.~P.} \&
  \bibinfo{author}{Peaker, A.~R.}
\newblock \bibinfo{title}{Characterisation of negative-{U} defects in
  semiconductors}.
\newblock \emph{\bibinfo{journal}{J. Condens. Matter Phys.}}
  \textbf{\bibinfo{volume}{32}}~(32), \bibinfo{pages}{323001}
  (\bibinfo{year}{2020}) .

\bibitem{kavanagh2022impact}
\bibinfo{author}{Kavanagh, S.~R.}, \bibinfo{author}{Scanlon, D.~O.},
  \bibinfo{author}{Walsh, A.} \& \bibinfo{author}{Freysoldt, C.}
\newblock \bibinfo{title}{Impact of metastable defect structures on carrier
  recombination in solar cells}.
\newblock \emph{\bibinfo{journal}{Faraday Discuss.}} \bibinfo{pages}{--}
  (\bibinfo{year}{2022}) .

\bibitem{Watkins_1983}
\bibinfo{author}{Watkins, G.~D.}
\newblock \bibinfo{title}{Deep levels in semiconductors}.
\newblock \emph{\bibinfo{journal}{Physica B+C}} \textbf{\bibinfo{volume}{117}},
  \bibinfo{pages}{9--15} (\bibinfo{year}{1983}) .

\bibitem{Watkins_1993}
\bibinfo{author}{Watkins, G.~D.}
\newblock \bibinfo{title}{{35 Years of Defects in Semiconductors: What Next?}}
\newblock \emph{\bibinfo{journal}{Mater. Sci. Forum}}
  \textbf{\bibinfo{volume}{143}}, \bibinfo{pages}{9--20} (\bibinfo{year}{1993})
  .

\bibitem{Watkins_1996}
\bibinfo{author}{Watkins, G.~D.}
\newblock \bibinfo{title}{{Intrinsic defects in II-VI semiconductors}}.
\newblock \emph{\bibinfo{journal}{J. Cryst. Growth}}
  \textbf{\bibinfo{volume}{159}}, \bibinfo{pages}{338--344}
  (\bibinfo{year}{1996}) .

\bibitem{Watkins_1997}
\bibinfo{author}{Watkins, G.}
\newblock \bibinfo{title}{{Native Defects and their Interactions with
  Impurities in Silicon.}}
\newblock \emph{\bibinfo{journal}{OPL}} \textbf{\bibinfo{volume}{469}},
  \bibinfo{pages}{139–150} (\bibinfo{year}{1997}) .

\bibitem{Watkins_2000}
\bibinfo{author}{Watkins, G.~D.}
\newblock \bibinfo{title}{Intrinsic defects in silicon}.
\newblock \emph{\bibinfo{journal}{Mater. Sci. Semicond. Process.}}
  \textbf{\bibinfo{volume}{3}}~(4), \bibinfo{pages}{227--235}
  (\bibinfo{year}{2000}) .

\bibitem{Watkins_2001}
\bibinfo{author}{Watkins, G.}
\newblock \bibinfo{title}{{What We Have Learned about Intrinsic Defects in
  Silicon: A Help in Understanding Diamond?}}
\newblock \emph{\bibinfo{journal}{PSS (a)}} \textbf{\bibinfo{volume}{186}}~(2),
  \bibinfo{pages}{176} (\bibinfo{year}{2001}) .

\bibitem{Coulson_1957}
\bibinfo{author}{Coulson, C.~A.} \& \bibinfo{author}{Kearsley, M.~J.}
\newblock \bibinfo{title}{{Colour centres in irradiated diamonds. I}}.
\newblock \emph{\bibinfo{journal}{Proc. R. Soc. A}}
  \textbf{\bibinfo{volume}{241}}~(1227), \bibinfo{pages}{433--454}
  (\bibinfo{year}{1957}) .

\bibitem{El-Maghraby_1998}
\bibinfo{author}{El-Maghraby, M.} \& \bibinfo{author}{Shinozuka, Y.}
\newblock \bibinfo{title}{{Structural Change of a Tetrahedral Four-Site System
  with Arbitrary Electron Occupancy}}.
\newblock \emph{\bibinfo{journal}{J. Phys. Soc. Japan}}
  \textbf{\bibinfo{volume}{67}}~(10), \bibinfo{pages}{3524--3535}
  (\bibinfo{year}{1998}) .

\bibitem{Stoneham_2001_vacancies}
\bibinfo{author}{Stoneham, A.}
\newblock \emph{\bibinfo{title}{Vacancies in valence crystals}},
  Ch.~\bibinfo{chapter}{27}, \bibinfo{pages}{851--883}
  (\bibinfo{publisher}{Oxford University Press}, \bibinfo{address}{Oxford},
  \bibinfo{year}{2007}).

\bibitem{Carvalho_2010}
\bibinfo{author}{Carvalho, A.}, \bibinfo{author}{Tagantsev, A.~K.},
  \bibinfo{author}{\"Oberg, S.}, \bibinfo{author}{Briddon, P.~R.} \&
  \bibinfo{author}{Setter, N.}
\newblock \bibinfo{title}{Cation-site intrinsic defects in {Zn}-doped {CdTe}}.
\newblock \emph{\bibinfo{journal}{Phys. Rev. B}} \textbf{\bibinfo{volume}{81}},
  \bibinfo{pages}{075215} (\bibinfo{year}{2010}) .

\bibitem{Lanoo_1981}
\bibinfo{author}{Lannoo, M.} \& \bibinfo{author}{Bourgeon, J.}
\newblock \emph{\bibinfo{title}{Point Defects in Semiconductors {I}:
  Experimental Aspects}} Vol.~\bibinfo{volume}{22}
  (\bibinfo{publisher}{Springer}, \bibinfo{address}{Berlin},
  \bibinfo{year}{1981}).

\bibitem{Lany_2005}
\bibinfo{author}{Lany, S.} \& \bibinfo{author}{Zunger, A.}
\newblock \bibinfo{title}{Anion vacancies as a source of persistent
  photoconductivity in {II-VI} and chalcopyrite semiconductors}.
\newblock \emph{\bibinfo{journal}{Phys. Rev. B}} \textbf{\bibinfo{volume}{72}},
  \bibinfo{pages}{035215} (\bibinfo{year}{2005}) .

\bibitem{Chanier_2008}
\bibinfo{author}{Chanier, T.}, \bibinfo{author}{Opahle, I.},
  \bibinfo{author}{Sargolzaei, M.}, \bibinfo{author}{Hayn, R.} \&
  \bibinfo{author}{Lannoo, M.}
\newblock \bibinfo{title}{{Magnetic State around Cation Vacancies in II--VI
  Semiconductors}}.
\newblock \emph{\bibinfo{journal}{Phys. Rev. Lett.}}
  \textbf{\bibinfo{volume}{100}}, \bibinfo{pages}{026405}
  (\bibinfo{year}{2008}) .

\bibitem{Schultz_2009}
\bibinfo{author}{Schultz, P.~A.} \& \bibinfo{author}{von Lilienfeld, O.~A.}
\newblock \bibinfo{title}{Simple intrinsic defects in gallium arsenide}.
\newblock \emph{\bibinfo{journal}{Model Simul Mat Sci Eng}}
  \textbf{\bibinfo{volume}{17}}~(8), \bibinfo{pages}{084007}
  (\bibinfo{year}{2009}) .

\bibitem{Feichtinger_1986}
\bibinfo{author}{Feichtinger, H.}
\newblock \emph{\bibinfo{title}{Deep Centers in Semiconductors}},
  Ch.~\bibinfo{chapter}{4}, \bibinfo{pages}{168--223}
  (\bibinfo{publisher}{Wiley}, \bibinfo{address}{Weinheim},
  \bibinfo{year}{2000}).

\bibitem{Pham_2020}
\bibinfo{author}{Pham, T.~D.} \& \bibinfo{author}{Deskins, N.~A.}
\newblock \bibinfo{title}{{Efficient Method for Modeling Polarons Using
  Electronic Structure Methods}}.
\newblock \emph{\bibinfo{journal}{J. Chem. Theory Comput.}}
  \textbf{\bibinfo{volume}{16}}~(8), \bibinfo{pages}{5264--5278}
  (\bibinfo{year}{2020}) .

\bibitem{Wang_2021}
\bibinfo{author}{Wang, Z.}, \bibinfo{author}{Malyi, O.~I.},
  \bibinfo{author}{Zhao, X.} \& \bibinfo{author}{Zunger, A.}
\newblock \bibinfo{title}{Mass enhancement in $3d$ and
  $s\text{\ensuremath{-}}p$ perovskites from symmetry breaking}.
\newblock \emph{\bibinfo{journal}{Phys. Rev. B}}
  \textbf{\bibinfo{volume}{103}}, \bibinfo{pages}{165110}
  (\bibinfo{year}{2021}) .

\bibitem{Huang_2022}
\bibinfo{author}{Huang, M.} \emph{et~al.}
\newblock \bibinfo{title}{{DASP}: Defect and dopant ab-initio simulation
  package}.
\newblock \emph{\bibinfo{journal}{J. Semicond.}}
  \textbf{\bibinfo{volume}{43}}~(4), \bibinfo{pages}{042101}
  (\bibinfo{year}{2022}) .

\bibitem{Gake_2021}
\bibinfo{author}{Gake, T.}, \bibinfo{author}{Kumagai, Y.},
  \bibinfo{author}{Takahashi, A.} \& \bibinfo{author}{Oba, F.}
\newblock \bibinfo{title}{{Point defects in $p$-type transparent conductive
  $\mathrm{Cu}M{\mathrm{O}}_{2}$ ($M=\mathrm{Al}$, {Ga}, {In}) from first
  principles}}.
\newblock \emph{\bibinfo{journal}{Phys. Rev. Materials}}
  \textbf{\bibinfo{volume}{5}}, \bibinfo{pages}{104602} (\bibinfo{year}{2021})
  .

\bibitem{Kumagai_2021}
\bibinfo{author}{Kumagai, Y.}, \bibinfo{author}{Tsunoda, N.},
  \bibinfo{author}{Takahashi, A.} \& \bibinfo{author}{Oba, F.}
\newblock \bibinfo{title}{Insights into oxygen vacancies from high-throughput
  first-principles calculations}.
\newblock \emph{\bibinfo{journal}{Phys. Rev. Materials}}
  \textbf{\bibinfo{volume}{5}}, \bibinfo{pages}{123803} (\bibinfo{year}{2021})
  .

\bibitem{Ong_pymatgen_2013}
\bibinfo{author}{Ong, S.~P.} \emph{et~al.}
\newblock \bibinfo{title}{{Python Materials Genomics (pymatgen): A robust,
  open-source python library for materials analysis}}.
\newblock \emph{\bibinfo{journal}{Comput. Mater. Sci.}}
  \textbf{\bibinfo{volume}{68}}, \bibinfo{pages}{314--319}
  (\bibinfo{year}{2013}) .

\bibitem{Bahn_ase_2002}
\bibinfo{author}{Bahn, S.} \& \bibinfo{author}{Jacobsen, K.}
\newblock \bibinfo{title}{An object-oriented scripting interface to a legacy
  electronic structure code}.
\newblock \emph{\bibinfo{journal}{Comput Sci Eng}}
  \textbf{\bibinfo{volume}{4}}~(3), \bibinfo{pages}{56--66}
  (\bibinfo{year}{2002}) .

\bibitem{Kresse_1993}
\bibinfo{author}{Kresse, G.} \& \bibinfo{author}{Hafner, J.}
\newblock \bibinfo{title}{Ab initio molecular dynamics for liquid metals}.
\newblock \emph{\bibinfo{journal}{Phys. Rev. B}} \textbf{\bibinfo{volume}{47}},
  \bibinfo{pages}{558--561} (\bibinfo{year}{1993}) .

\bibitem{Kresse_1994}
\bibinfo{author}{Kresse, G.} \& \bibinfo{author}{Hafner, J.}
\newblock \bibinfo{title}{Ab initio molecular-dynamics simulation of the
  liquid-metal--amorphous-semiconductor transition in germanium}.
\newblock \emph{\bibinfo{journal}{Phys. Rev. B}} \textbf{\bibinfo{volume}{49}},
  \bibinfo{pages}{14251--14269} (\bibinfo{year}{1994}) .

\bibitem{Kresse_1996}
\bibinfo{author}{Kresse, G.} \& \bibinfo{author}{Furthmüller, J.}
\newblock \bibinfo{title}{Efficiency of ab-initio total energy calculations for
  metals and semiconductors using a plane-wave basis set}.
\newblock \emph{\bibinfo{journal}{Comput. Mater. Sci.}}
  \textbf{\bibinfo{volume}{6}}~(1), \bibinfo{pages}{15--50}
  (\bibinfo{year}{1996}) .

\bibitem{Heyd_2003}
\bibinfo{author}{Heyd, J.}, \bibinfo{author}{Scuseria, G.~E.} \&
  \bibinfo{author}{Ernzerhof, M.}
\newblock \bibinfo{title}{Hybrid functionals based on a screened coulomb
  potential}.
\newblock \emph{\bibinfo{journal}{J. Chem. Phys.}}
  \textbf{\bibinfo{volume}{118}}~(18), \bibinfo{pages}{8207--8215}
  (\bibinfo{year}{2003}) .

\bibitem{Alkauskas_2011}
\bibinfo{author}{Alkauskas, A.}, \bibinfo{author}{Broqvist, P.} \&
  \bibinfo{author}{Pasquarello, A.}
\newblock \bibinfo{title}{Defect levels through hybrid density functionals:
  Insights and applications}.
\newblock \emph{\bibinfo{journal}{Phys. Status Solidi B}}
  \textbf{\bibinfo{volume}{248}}~(4), \bibinfo{pages}{775--789}
  (\bibinfo{year}{2011}) .

\bibitem{Alkauskas_2014}
\bibinfo{author}{Alkauskas, A.}, \bibinfo{author}{Yan, Q.} \&
  \bibinfo{author}{Van~de Walle, C.~G.}
\newblock \bibinfo{title}{First-principles theory of nonradiative carrier
  capture via multiphonon emission}.
\newblock \emph{\bibinfo{journal}{Phys. Rev. B}} \textbf{\bibinfo{volume}{90}},
  \bibinfo{pages}{075202} (\bibinfo{year}{2014}) .

\bibitem{Lyons_2010}
\bibinfo{author}{Lyons, J.~L.}, \bibinfo{author}{Janotti, A.} \&
  \bibinfo{author}{Van~de Walle, C.~G.}
\newblock \bibinfo{title}{{Carbon impurities and the yellow luminescence in
  GaN}}.
\newblock \emph{\bibinfo{journal}{Appl. Phys. Lett.}}
  \textbf{\bibinfo{volume}{97}}~(15), \bibinfo{pages}{152108}
  (\bibinfo{year}{2010}) .

\bibitem{Deak_2005}
\bibinfo{author}{De{\'{a}}k, P.}, \bibinfo{author}{Gali, A.},
  \bibinfo{author}{S{\'{o}}lyom, A.}, \bibinfo{author}{Buruzs, A.} \&
  \bibinfo{author}{Frauenheim, T.}
\newblock \bibinfo{title}{Electronic structure of boron-interstitial clusters
  in silicon}.
\newblock \emph{\bibinfo{journal}{J. Condens. Matter Phys.}}
  \textbf{\bibinfo{volume}{17}}~(22), \bibinfo{pages}{S2141--S2153}
  (\bibinfo{year}{2005}) .

\bibitem{Huang_2021}
\bibinfo{author}{Huang, Y.-T.}, \bibinfo{author}{Kavanagh, S.~R.},
  \bibinfo{author}{Scanlon, D.~O.}, \bibinfo{author}{Walsh, A.} \&
  \bibinfo{author}{Hoye, R. L.~Z.}
\newblock \bibinfo{title}{Perovskite-inspired materials for photovoltaics and
  beyond{\textemdash}from design to devices}.
\newblock \emph{\bibinfo{journal}{Nanotechnology}}
  \textbf{\bibinfo{volume}{32}}~(13), \bibinfo{pages}{132004}
  (\bibinfo{year}{2021}) .

\bibitem{Zunger_2008_assessment}
\bibinfo{author}{Lany, S.} \& \bibinfo{author}{Zunger, A.}
\newblock \bibinfo{title}{Assessment of correction methods for the band-gap
  problem and for finite-size effects in supercell defect calculations: Case
  studies for {ZnO} and {GaAs}}.
\newblock \emph{\bibinfo{journal}{Phys. Rev. B}} \textbf{\bibinfo{volume}{78}},
  \bibinfo{pages}{235104} (\bibinfo{year}{2008}) .

\bibitem{Freysoldt_2014}
\bibinfo{author}{Freysoldt, C.} \emph{et~al.}
\newblock \bibinfo{title}{First-principles calculations for point defects in
  solids}.
\newblock \emph{\bibinfo{journal}{Rev. Mod. Phys.}}
  \textbf{\bibinfo{volume}{86}}, \bibinfo{pages}{253--305}
  (\bibinfo{year}{2014}) .

\bibitem{Kumagai_2014}
\bibinfo{author}{Kumagai, Y.} \& \bibinfo{author}{Oba, F.}
\newblock \bibinfo{title}{Electrostatics-based finite-size corrections for
  first-principles point defect calculations}.
\newblock \emph{\bibinfo{journal}{Phys. Rev. B}} \textbf{\bibinfo{volume}{89}},
  \bibinfo{pages}{195205} (\bibinfo{year}{2014}) .

\bibitem{ertural_development_2019}
\bibinfo{author}{Ertural, C.}, \bibinfo{author}{Steinberg, S.} \&
  \bibinfo{author}{Dronskowski, R.}
\newblock \bibinfo{title}{Development of a robust tool to extract {Mulliken}
  and {Löwdin} charges from plane waves and its application to solid-state
  materials}.
\newblock \emph{\bibinfo{journal}{RSC Advances}}
  \textbf{\bibinfo{volume}{9}}~(51), \bibinfo{pages}{29821--29830}
  (\bibinfo{year}{2019}) .

\bibitem{dronskowski_crystal_1993}
\bibinfo{author}{Dronskowski, R.} \& \bibinfo{author}{Bloechl, P.~E.}
\newblock \bibinfo{title}{Crystal orbital {Hamilton} populations ({COHP}):
  energy-resolved visualization of chemical bonding in solids based on
  density-functional calculations}.
\newblock \emph{\bibinfo{journal}{J. Phys. Chem.}}
  \textbf{\bibinfo{volume}{97}}~(33), \bibinfo{pages}{8617--8624}
  (\bibinfo{year}{1993}) .

\bibitem{deringer_crystal_2011}
\bibinfo{author}{Deringer, V.~L.}, \bibinfo{author}{Tchougréeff, A.~L.} \&
  \bibinfo{author}{Dronskowski, R.}
\newblock \bibinfo{title}{Crystal {Orbital} {Hamilton} {Population} ({COHP})
  {Analysis} {As} {Projected} from {Plane}-{Wave} {Basis} {Sets}}.
\newblock \emph{\bibinfo{journal}{J. Phys. Chem. A}}
  \textbf{\bibinfo{volume}{115}}~(21), \bibinfo{pages}{5461--5466}
  (\bibinfo{year}{2011}) .

\bibitem{maintz_analytic_2013}
\bibinfo{author}{Maintz, S.}, \bibinfo{author}{Deringer, V.~L.},
  \bibinfo{author}{Tchougréeff, A.~L.} \& \bibinfo{author}{Dronskowski, R.}
\newblock \bibinfo{title}{Analytic projection from plane-wave and {PAW}
  wavefunctions and application to chemical-bonding analysis in solids}.
\newblock \emph{\bibinfo{journal}{J. Comput. Chem.}}
  \textbf{\bibinfo{volume}{34}}~(29), \bibinfo{pages}{2557--2567}
  (\bibinfo{year}{2013}) .

\bibitem{nelson_lobster_2020}
\bibinfo{author}{Nelson, R.} \emph{et~al.}
\newblock \bibinfo{title}{{LOBSTER}: {Local} orbital projections, atomic
  charges, and chemical-bonding analysis from projector-augmented-wave-based
  density-functional theory}.
\newblock \emph{\bibinfo{journal}{J. Comput. Chem.}}
  \textbf{\bibinfo{volume}{41}}~(21), \bibinfo{pages}{1931--1940}
  (\bibinfo{year}{2020}) .

\bibitem{Guo_2018}
\bibinfo{author}{Guo, L.} \emph{et~al.}
\newblock \bibinfo{title}{{Tunable Quasi-One-Dimensional Ribbon Enhanced Light
  Absorption in $\rm Sb_2Se_3$ Thin-film Solar Cells Grown by Close-Space
  Sublimation}}.
\newblock \emph{\bibinfo{journal}{Solar RRL}}
  \textbf{\bibinfo{volume}{2}}~(10), \bibinfo{pages}{1800128}
  (\bibinfo{year}{2018}) .

\bibitem{Wang_2022}
\bibinfo{author}{Wang, X.}, \bibinfo{author}{Li, Z.},
  \bibinfo{author}{Kavanagh, S.~R.}, \bibinfo{author}{Ganose, A.~M.} \&
  \bibinfo{author}{Walsh, A.}
\newblock \bibinfo{title}{Lone pair driven anisotropy in antimony chalcogenide
  semiconductors}.
\newblock \emph{\bibinfo{journal}{Phys. Chem. Chem. Phys.}}
  \textbf{\bibinfo{volume}{24}}, \bibinfo{pages}{7195--7202}
  (\bibinfo{year}{2022}) .

\bibitem{Caruso_2015}
\bibinfo{author}{Caruso, F.}, \bibinfo{author}{Filip, M.~R.} \&
  \bibinfo{author}{Giustino, F.}
\newblock \bibinfo{title}{{Excitons in one-dimensional van der Waals materials:
  ${\mathrm{Sb}}_{2}{\mathrm{S}}_{3}$ nanoribbons}}.
\newblock \emph{\bibinfo{journal}{Phys. Rev. B}} \textbf{\bibinfo{volume}{92}},
  \bibinfo{pages}{125134} (\bibinfo{year}{2015}) .

\bibitem{Song_2017}
\bibinfo{author}{Song, H.} \emph{et~al.}
\newblock \bibinfo{title}{{Highly Anisotropic $\rm Sb_2Se_3$ Nanosheets: Gentle
  Exfoliation from the Bulk Precursors Possessing 1D Crystal Structure}}.
\newblock \emph{\bibinfo{journal}{J. Adv. Mater.}}
  \textbf{\bibinfo{volume}{29}}~(29), \bibinfo{pages}{1700441}
  (\bibinfo{year}{2017}) .

\bibitem{Yang_2018}
\bibinfo{author}{Yang, W.} \emph{et~al.}
\newblock \bibinfo{title}{{Adjusting the Anisotropy of 1D $\rm Sb_2Se_3$
  Nanostructures for Highly Efficient Photoelectrochemical Water Splitting}}.
\newblock \emph{\bibinfo{journal}{J. Adv. Energy Mater.}}
  \textbf{\bibinfo{volume}{8}}~(14), \bibinfo{pages}{1702888}
  (\bibinfo{year}{2018}) .

\bibitem{Gusmao_2019}
\bibinfo{author}{Gusm\~ao, R.}, \bibinfo{author}{Sofer, Z.},
  \bibinfo{author}{Luxa, J.} \& \bibinfo{author}{Pumera, M.}
\newblock \bibinfo{title}{{Antimony Chalcogenide van der Waals Nanostructures
  for Energy Conversion and Storage}}.
\newblock \emph{\bibinfo{journal}{ACS Sustain. Chem. Eng.}}
  \textbf{\bibinfo{volume}{7}}~(18), \bibinfo{pages}{15790--15798}
  (\bibinfo{year}{2019}) .

\bibitem{Deringer_2011}
\bibinfo{author}{Deringer, V.~L.}, \bibinfo{author}{Tchougréeff, A.~L.} \&
  \bibinfo{author}{Dronskowski, R.}
\newblock \bibinfo{title}{{Crystal Orbital Hamilton Population (COHP) Analysis
  As Projected from Plane-Wave Basis Sets}}.
\newblock \emph{\bibinfo{journal}{J. Phys. Chem. A}}
  \textbf{\bibinfo{volume}{115}}~(21), \bibinfo{pages}{5461--5466}
  (\bibinfo{year}{2011}) .

\bibitem{Dronskowski_1993}
\bibinfo{author}{Dronskowski, R.} \& \bibinfo{author}{Bloechl, P.~E.}
\newblock \bibinfo{title}{{Crystal orbital Hamilton populations (COHP):
  energy-resolved visualization of chemical bonding in solids based on
  density-functional calculations}}.
\newblock \emph{\bibinfo{journal}{J. Phys. Chem.}}
  \textbf{\bibinfo{volume}{97}}~(33), \bibinfo{pages}{8617--8624}
  (\bibinfo{year}{1993}) .

\bibitem{Maintz_2013}
\bibinfo{author}{Maintz, S.}, \bibinfo{author}{Deringer, V.~L.},
  \bibinfo{author}{Tchougréeff, A.~L.} \& \bibinfo{author}{Dronskowski, R.}
\newblock \bibinfo{title}{{Analytic projection from plane-wave and PAW
  wavefunctions and application to chemical-bonding analysis in solids}}.
\newblock \emph{\bibinfo{journal}{J. Comput. Chem.}}
  \textbf{\bibinfo{volume}{34}}~(29), \bibinfo{pages}{2557--2567}
  (\bibinfo{year}{2013}) .

\bibitem{Zhang_2022}
\bibinfo{author}{Zhang, B.} \& \bibinfo{author}{Qian, X.}
\newblock \bibinfo{title}{Competing superior electronic structure and complex
  defect chemistry in quasi-one-dimensional antimony chalcogenide photovoltaic
  absorbers}.
\newblock \emph{\bibinfo{journal}{ACS Appl. Energy Mater.}}
  \textbf{\bibinfo{volume}{5}}~(1), \bibinfo{pages}{492--502}
  (\bibinfo{year}{2022}) .

\bibitem{Cai_2020}
\bibinfo{author}{Cai, Z.}, \bibinfo{author}{Dai, C.-M.} \&
  \bibinfo{author}{Chen, S.}
\newblock \bibinfo{title}{{Intrinsic Defect Limit to the Electrical
  Conductivity and a Two-Step p-Type Doping Strategy for Overcoming the
  Efficiency Bottleneck of $\rm Sb_2S_3$-Based Solar Cells}}.
\newblock \emph{\bibinfo{journal}{Solar RRL}} \textbf{\bibinfo{volume}{4}}~(4),
  \bibinfo{pages}{1900503} (\bibinfo{year}{2020}) .

\bibitem{Savory_2019}
\bibinfo{author}{Savory, C.} \& \bibinfo{author}{Scanlon, D.~O.}
\newblock \bibinfo{title}{The complex defect chemistry of antimony selenide}.
\newblock \emph{\bibinfo{journal}{J. Mater. Chem. A}}
  \textbf{\bibinfo{volume}{7}}, \bibinfo{pages}{10739--10744}
  (\bibinfo{year}{2019}) .

\bibitem{Huang_2019}
\bibinfo{author}{Huang, M.}, \bibinfo{author}{Xu, P.}, \bibinfo{author}{Han,
  D.}, \bibinfo{author}{Tang, J.} \& \bibinfo{author}{Chen, S.}
\newblock \bibinfo{title}{{Complicated and Unconventional Defect Properties of
  the Quasi-One-Dimensional Photovoltaic Semiconductor $\rm Sb_2Se_3$}}.
\newblock \emph{\bibinfo{journal}{ACS Appl. Mater. Interfaces}}
  \textbf{\bibinfo{volume}{11}}~(17), \bibinfo{pages}{15564--15572}
  (\bibinfo{year}{2019}) .

\bibitem{Liu_2017}
\bibinfo{author}{Liu, X.} \emph{et~al.}
\newblock \bibinfo{title}{{Enhanced $\rm Sb_2Se_3$ solar cell performance
  through theory-guided defect control}}.
\newblock \emph{\bibinfo{journal}{Prog Photovolt.}}
  \textbf{\bibinfo{volume}{25}}~(10), \bibinfo{pages}{861--870}
  (\bibinfo{year}{2017}) .

\bibitem{Zhao_2021}
\bibinfo{author}{Zhao, R.}, \bibinfo{author}{Yang, X.}, \bibinfo{author}{Shi,
  H.} \& \bibinfo{author}{Du, M.-H.}
\newblock \bibinfo{title}{{Intrinsic and complex defect engineering of
  quasi-one-dimensional ribbons ${\mathrm{Sb}}_{2}{\mathrm{S}}_{3}$ for
  photovoltaics performance}}.
\newblock \emph{\bibinfo{journal}{Phys. Rev. Materials}}
  \textbf{\bibinfo{volume}{5}}, \bibinfo{pages}{054605} (\bibinfo{year}{2021})
  .

\bibitem{Tumelero_2016}
\bibinfo{author}{Tumelero, M.~A.}, \bibinfo{author}{Faccio, R.} \&
  \bibinfo{author}{Pasa, A.~A.}
\newblock \bibinfo{title}{{Unraveling the Native Conduction of Trichalcogenides
  and Its Ideal Band Alignment for New Photovoltaic Interfaces}}.
\newblock \emph{\bibinfo{journal}{J. Phys. Chem. C}}
  \textbf{\bibinfo{volume}{120}}~(3), \bibinfo{pages}{1390--1399}
  (\bibinfo{year}{2016}) .

\bibitem{Stoliaroff_2020}
\bibinfo{author}{Stoliaroff, A.} \emph{et~al.}
\newblock \bibinfo{title}{{Deciphering the Role of Key Defects in $\rm
  Sb_2Se_3$, a Promising Candidate for Chalcogenide-Based Solar Cells}}.
\newblock \emph{\bibinfo{journal}{ACS Appl. Energy Mater.}}
  \textbf{\bibinfo{volume}{3}}~(3), \bibinfo{pages}{2496--2509}
  (\bibinfo{year}{2020}) .

\bibitem{Han_2017}
\bibinfo{author}{Han, D.}, \bibinfo{author}{Du, M.-H.}, \bibinfo{author}{Dai,
  C.-M.}, \bibinfo{author}{Sun, D.} \& \bibinfo{author}{Chen, S.}
\newblock \bibinfo{title}{Influence of defects and dopants on the photovoltaic
  performance of {$\rm Bi_2S_3$}: first-principles insights}.
\newblock \emph{\bibinfo{journal}{J. Mater. Chem. A}}
  \textbf{\bibinfo{volume}{5}}, \bibinfo{pages}{6200--6210}
  (\bibinfo{year}{2017}) .

\bibitem{Agoston_2009}
\bibinfo{author}{{\'{A}}goston, P.}, \bibinfo{author}{Erhart, P.},
  \bibinfo{author}{Klein, A.} \& \bibinfo{author}{Albe, K.}
\newblock \bibinfo{title}{Geometry, electronic structure and thermodynamic
  stability of intrinsic point defects in indium oxide}.
\newblock \emph{\bibinfo{journal}{J. Condens. Matter Phys.}}
  \textbf{\bibinfo{volume}{21}}~(45), \bibinfo{pages}{455801}
  (\bibinfo{year}{2009}) .

\bibitem{Erhart_2005}
\bibinfo{author}{Erhart, P.}, \bibinfo{author}{Klein, A.} \&
  \bibinfo{author}{Albe, K.}
\newblock \bibinfo{title}{{First-principles study of the structure and
  stability of oxygen defects in zinc oxide}}.
\newblock \emph{\bibinfo{journal}{Phys. Rev. B}} \textbf{\bibinfo{volume}{72}},
  \bibinfo{pages}{085213} (\bibinfo{year}{2005}) .

\bibitem{Evarestov_1996}
\bibinfo{author}{R.A.~Evarestov, P.} \& \bibinfo{author}{Jacobs, A.~L.}
\newblock \bibinfo{title}{{Oxygen interstitials in magnesium oxide: A
  band-model study}}.
\newblock \emph{\bibinfo{journal}{Phys. Rev. B}} \textbf{\bibinfo{volume}{54}},
  \bibinfo{pages}{8969} (\bibinfo{year}{1996}) .

\bibitem{Kotomin_1998}
\bibinfo{author}{Kotomin, E.} \& \bibinfo{author}{Popov, A.}
\newblock \bibinfo{title}{Radiation-induced point defects in simple oxides.}
\newblock \emph{\bibinfo{journal}{Nucl. Instrum. Methods Phys. Res., Sect. B}}
  \textbf{\bibinfo{volume}{141}}, \bibinfo{pages}{1--15} (\bibinfo{year}{1998})
  .

\bibitem{Burbano_2011}
\bibinfo{author}{Burbano, M.}, \bibinfo{author}{Scanlon, D.~O.} \&
  \bibinfo{author}{Watson, G.~W.}
\newblock \bibinfo{title}{{Sources of Conductivity and Doping Limits in CdO
  from Hybrid Density Functional Theory}}.
\newblock \emph{\bibinfo{journal}{J. Am. Chem. Soc.}}
  \textbf{\bibinfo{volume}{133}}~(38), \bibinfo{pages}{15065--15072}
  (\bibinfo{year}{2011}) .

\bibitem{Scanlon_2012_Onthe}
\bibinfo{author}{Scanlon, D.~O.} \& \bibinfo{author}{Watson, G.~W.}
\newblock \bibinfo{title}{{On the possibility of p-type $\rm SnO_2$}}.
\newblock \emph{\bibinfo{journal}{J. Mater. Chem.}}
  \textbf{\bibinfo{volume}{22}}, \bibinfo{pages}{25236--25245}
  (\bibinfo{year}{2012}) .

\bibitem{Godinho_2009}
\bibinfo{author}{Godinho, K.~G.}, \bibinfo{author}{Walsh, A.} \&
  \bibinfo{author}{Watson, G.~W.}
\newblock \bibinfo{title}{{Energetic and Electronic Structure Analysis of
  Intrinsic Defects in SnO2}}.
\newblock \emph{\bibinfo{journal}{J. Phys. Chem. C}}
  \textbf{\bibinfo{volume}{113}}~(1), \bibinfo{pages}{439--448}
  (\bibinfo{year}{2009}) .

\bibitem{Scanlon_2011_Nature}
\bibinfo{author}{Scanlon, D.~O.} \emph{et~al.}
\newblock \bibinfo{title}{Nature of the band gap and origin of the conductivity
  of ${\mathrm{pbo}}_{2}$ revealed by theory and experiment}.
\newblock \emph{\bibinfo{journal}{Phys. Rev. Lett.}}
  \textbf{\bibinfo{volume}{107}}, \bibinfo{pages}{246402}
  (\bibinfo{year}{2011}) .

\bibitem{Keating_2012}
\bibinfo{author}{Keating, P. R.~L.}, \bibinfo{author}{Scanlon, D.~O.},
  \bibinfo{author}{Morgan, B.~J.}, \bibinfo{author}{Galea, N.~M.} \&
  \bibinfo{author}{Watson, G.~W.}
\newblock \bibinfo{title}{{Analysis of Intrinsic Defects in $\rm CeO_2$ Using a
  Koopmans-Like GGA+U Approach}}.
\newblock \emph{\bibinfo{journal}{J. Phys. Chem. C}}
  \textbf{\bibinfo{volume}{116}}~(3), \bibinfo{pages}{2443--2452}
  (\bibinfo{year}{2012}) .

\bibitem{Scanlon_2013_Defect}
\bibinfo{author}{Scanlon, D.~O.}
\newblock \bibinfo{title}{Defect engineering of basno${}_{3}$ for
  high-performance transparent conducting oxide applications}.
\newblock \emph{\bibinfo{journal}{Phys. Rev. B}} \textbf{\bibinfo{volume}{87}},
  \bibinfo{pages}{161201} (\bibinfo{year}{2013}) .

\bibitem{Walsh_2009}
\bibinfo{author}{Walsh, A.}, \bibinfo{author}{Da~Silva, J. L.~F.} \&
  \bibinfo{author}{Wei, S.-H.}
\newblock \bibinfo{title}{{Interplay between Order and Disorder in the High
  Performance of Amorphous Transparent Conducting Oxides}}.
\newblock \emph{\bibinfo{journal}{Chem. Mater.}}
  \textbf{\bibinfo{volume}{21}}~(21), \bibinfo{pages}{5119--5124}
  (\bibinfo{year}{2009}) .

\bibitem{Wilson_2008}
\bibinfo{author}{Wilson, D.~J.}, \bibinfo{author}{Sokol, A.~A.},
  \bibinfo{author}{French, S.~A.} \& \bibinfo{author}{Catlow, C. R.~A.}
\newblock \bibinfo{title}{Defect structures in the silver halides}.
\newblock \emph{\bibinfo{journal}{Phys. Rev. B}} \textbf{\bibinfo{volume}{77}},
  \bibinfo{pages}{064115} (\bibinfo{year}{2008}) .

\bibitem{Agiorgousis_2014}
\bibinfo{author}{Agiorgousis, M.~L.}, \bibinfo{author}{Sun, Y.-Y.},
  \bibinfo{author}{Zeng, H.} \& \bibinfo{author}{Zhang, S.}
\newblock \bibinfo{title}{{Strong Covalency-Induced Recombination Centers in
  Perovskite Solar Cell Material $\rm CH_3NH_3PbI_3$}}.
\newblock \emph{\bibinfo{journal}{J. Am. Chem. Soc.}}
  \textbf{\bibinfo{volume}{136}}~(41), \bibinfo{pages}{14570--14575}
  (\bibinfo{year}{2014}) .

\bibitem{Whalley_2017}
\bibinfo{author}{Whalley, L.~D.}, \bibinfo{author}{Crespo-Otero, R.} \&
  \bibinfo{author}{Walsh, A.}
\newblock \bibinfo{title}{{H-Center and V-Center Defects in Hybrid Halide
  Perovskites}}.
\newblock \emph{\bibinfo{journal}{ACS Energy Lett.}}
  \textbf{\bibinfo{volume}{2}}~(12), \bibinfo{pages}{2713--2714}
  (\bibinfo{year}{2017}) .

\bibitem{Whalley_2021}
\bibinfo{author}{Whalley, L.~D.} \emph{et~al.}
\newblock \bibinfo{title}{{Giant Huang–Rhys Factor for Electron Capture by
  the Iodine Intersitial in Perovskite Solar Cells}}.
\newblock \emph{\bibinfo{journal}{J. Am. Chem. Soc.}}
  \textbf{\bibinfo{volume}{143}}~(24), \bibinfo{pages}{9123--9128}
  (\bibinfo{year}{2021}) .

\bibitem{Motti_2019}
\bibinfo{author}{Motti, S.~G.} \emph{et~al.}
\newblock \bibinfo{title}{{Defect Activity in Lead Halide Perovskites}}.
\newblock \emph{\bibinfo{journal}{Adv. Mater.}}
  \textbf{\bibinfo{volume}{31}}~(47), \bibinfo{pages}{1901183}
  (\bibinfo{year}{2019}) .

\bibitem{Kang_2017}
\bibinfo{author}{Kang, J.} \& \bibinfo{author}{Wang, L.-W.}
\newblock \bibinfo{title}{{High Defect Tolerance in Lead Halide Perovskite
  CsPbBr3}}.
\newblock \emph{\bibinfo{journal}{J. Phys. Chem}}
  \textbf{\bibinfo{volume}{8}}~(2), \bibinfo{pages}{489--493}
  (\bibinfo{year}{2017}) .

\bibitem{Zhao_2016}
\bibinfo{author}{Zhao, Y.} \emph{et~al.}
\newblock \bibinfo{title}{{Correlations between Immobilizing Ions and
  Suppressing Hysteresis in Perovskite Solar Cells}}.
\newblock \emph{\bibinfo{journal}{ACS Energy Lett.}}
  \textbf{\bibinfo{volume}{1}}~(1), \bibinfo{pages}{266--272}
  (\bibinfo{year}{2016}) .

\bibitem{Xiao_2016}
\bibinfo{author}{Xiao, Z.}, \bibinfo{author}{Meng, W.}, \bibinfo{author}{Wang,
  J.} \& \bibinfo{author}{Yan, Y.}
\newblock \bibinfo{title}{{Defect properties of the two-dimensional $\rm
  (CH3NH3)_2Pb(SCN)_2I_2$ perovskite: a density-functional theory study}}.
\newblock \emph{\bibinfo{journal}{Phys. Chem. Chem. Phys.}}
  \textbf{\bibinfo{volume}{18}}, \bibinfo{pages}{25786--25790}
  (\bibinfo{year}{2016}) .

\bibitem{Meggiolaro_2020}
\bibinfo{author}{Meggiolaro, D.}, \bibinfo{author}{Ricciarelli, D.},
  \bibinfo{author}{Alasmari, A.~A.}, \bibinfo{author}{Alasmary, F. A.~S.} \&
  \bibinfo{author}{De~Angelis, F.}
\newblock \bibinfo{title}{{Tin versus Lead Redox Chemistry Modulates Charge
  Trapping and Self-Doping in Tin/Lead Iodide Perovskites}}.
\newblock \emph{\bibinfo{journal}{J. Phys. Chem}}
  \textbf{\bibinfo{volume}{11}}~(9), \bibinfo{pages}{3546--3556}
  (\bibinfo{year}{2020}) .

\bibitem{liao_2006}
\bibinfo{author}{Liao, Y.}
\newblock \bibinfo{title}{Practical electron microscopy and database}.
\newblock \emph{\bibinfo{journal}{An Online Book}}  (\bibinfo{year}{2006}) .

\bibitem{Hiley_2015}
\bibinfo{author}{Hiley, C.~I.} \emph{et~al.}
\newblock \bibinfo{title}{{Incorporation of square-planar $\rm Pd^{2+}$ in
  fluorite $\rm CeO_2$: hydrothermal preparation{,} local structure{,} redox
  properties and stability}}.
\newblock \emph{\bibinfo{journal}{J. Mater. Chem. A}}
  \textbf{\bibinfo{volume}{3}}, \bibinfo{pages}{13072--13079}
  (\bibinfo{year}{2015}) .

\bibitem{Hegde_2015}
\bibinfo{author}{Hegde, M.} \& \bibinfo{author}{Bera, P.}
\newblock \bibinfo{title}{{Noble metal ion substituted $\rm CeO_2$ catalysts:
  Electronic interaction between noble metal ions and $\rm CeO_2$ lattice}}.
\newblock \emph{\bibinfo{journal}{Catal. Today}}
  \textbf{\bibinfo{volume}{253}}, \bibinfo{pages}{40--50}
  (\bibinfo{year}{2015}) .

\bibitem{Huang_more_2021}
\bibinfo{author}{Huang, M.} \emph{et~al.}
\newblock \bibinfo{title}{{More Se Vacancies in $\rm Sb_2Se_3$ under Se-Rich
  Conditions: An Abnormal Behavior Induced by Defect-Correlation in Compensated
  Compound Semiconductors}}.
\newblock \emph{\bibinfo{journal}{Small}} \textbf{\bibinfo{volume}{17}}~(36),
  \bibinfo{pages}{2102429} (\bibinfo{year}{2021}) .

\bibitem{lian_revealing_2021}
\bibinfo{author}{Lian, W.} \emph{et~al.}
\newblock \bibinfo{title}{Revealing composition and structure dependent
  deep-level defect in antimony trisulfide photovoltaics}.
\newblock \emph{\bibinfo{journal}{Nat Commun}}
  \textbf{\bibinfo{volume}{12}}~(1), \bibinfo{pages}{3260}
  (\bibinfo{year}{2021}) .

\bibitem{guo_scalable_2019}
\bibinfo{author}{Guo, L.} \emph{et~al.}
\newblock \bibinfo{title}{Scalable and efficient {$\rm Sb_2S_3$} thin-film
  solar cells fabricated by close space sublimation}.
\newblock \emph{\bibinfo{journal}{APL Materials}}
  \textbf{\bibinfo{volume}{7}}~(4), \bibinfo{pages}{041105}
  (\bibinfo{year}{2019}) .

\bibitem{Zhang_2020}
\bibinfo{author}{Zhang, Z.}, \bibinfo{author}{Qiao, L.},
  \bibinfo{author}{Mora-Perez, C.}, \bibinfo{author}{Long, R.} \&
  \bibinfo{author}{Prezhdo, O.~V.}
\newblock \bibinfo{title}{{Pb dimerization greatly accelerates charge losses in
  $\rm MAPbI(CH3NH3)_2Pb(SCN)_2I2_3$: Time-domain ab initio analysis}}.
\newblock \emph{\bibinfo{journal}{J. Chem. Phys.}}
  \textbf{\bibinfo{volume}{152}}~(6), \bibinfo{pages}{064707}
  (\bibinfo{year}{2020}) .

\bibitem{Cai_2021}
\bibinfo{author}{Cai, L.}, \bibinfo{author}{Wang, S.}, \bibinfo{author}{Huang,
  M.}, \bibinfo{author}{Wu, Y.-N.} \& \bibinfo{author}{Chen, S.}
\newblock \bibinfo{title}{First-principles identification of deep energy levels
  of sulfur impurities in silicon and their carrier capture cross sections}.
\newblock \emph{\bibinfo{journal}{J. Phys. D: Appl. Phys}}
  \textbf{\bibinfo{volume}{54}}~(33), \bibinfo{pages}{335103}
  (\bibinfo{year}{2021}) .

\bibitem{Krasikov_2018}
\bibinfo{author}{Krasikov, D.} \& \bibinfo{author}{Sankin, I.}
\newblock \bibinfo{title}{{Beyond thermodynamic defect models: A kinetic
  simulation of arsenic activation in CdTe}}.
\newblock \emph{\bibinfo{journal}{Phys. Rev. Materials}}
  \textbf{\bibinfo{volume}{2}}, \bibinfo{pages}{103803} (\bibinfo{year}{2018})
  .

\bibitem{Yang_2016}
\bibinfo{author}{Yang, J.-H.}, \bibinfo{author}{Shi, L.},
  \bibinfo{author}{Wang, L.-W.} \& \bibinfo{author}{Wei, S.-H.}
\newblock \bibinfo{title}{{Non-Radiative Carrier Recombination Enhanced by
  Two-Level Process: A First-Principles Study}}.
\newblock \emph{\bibinfo{journal}{Sci. Rep.}} \textbf{\bibinfo{volume}{6}}~(1)
  (\bibinfo{year}{2016}) .

\bibitem{MaoHua_2005}
\bibinfo{author}{Du, M.-H.} \& \bibinfo{author}{Zhang, S.~B.}
\newblock \bibinfo{title}{{$DX$ centers in GaAs and GaSb}}.
\newblock \emph{\bibinfo{journal}{Phys. Rev. B}} \textbf{\bibinfo{volume}{72}},
  \bibinfo{pages}{075210} (\bibinfo{year}{2005}) .

\bibitem{Kim_2019}
\bibinfo{author}{Kim, S.}, \bibinfo{author}{Hood, S.~N.} \&
  \bibinfo{author}{Walsh, A.}
\newblock \bibinfo{title}{Anharmonic lattice relaxation during nonradiative
  carrier capture}.
\newblock \emph{\bibinfo{journal}{Phys. Rev. B}}
  \textbf{\bibinfo{volume}{100}}, \bibinfo{pages}{041202}
  (\bibinfo{year}{2019}) .

\bibitem{Dobaczewski_1995}
\bibinfo{author}{Dobaczewski, L.}, \bibinfo{author}{Kaczor, P.},
  \bibinfo{author}{Missous, M.}, \bibinfo{author}{Peaker, A.~R.} \&
  \bibinfo{author}{Zytkiewicz, Z.~R.}
\newblock \bibinfo{title}{{Structure of the DX state formed by donors in
  (Al,Ga)As and Ga(As,P)}}.
\newblock \emph{\bibinfo{journal}{Int. J. Appl. Phys.}}
  \textbf{\bibinfo{volume}{78}}~(4), \bibinfo{pages}{2468--2477}
  (\bibinfo{year}{1995}) .

\bibitem{Yamaguchi_1991}
\bibinfo{author}{Yamaguchi, E.}, \bibinfo{author}{Shiraishi, K.} \&
  \bibinfo{author}{Ohno, T.}
\newblock \bibinfo{title}{{First Principle Calculation of the DX-Center
  Ground-States in GaAs, $\rm Al_xGa_{1-x}As$ Alloys and AlAs/GaAs
  Superlattices}}.
\newblock \emph{\bibinfo{journal}{JPSJ}} \textbf{\bibinfo{volume}{60}}~(9),
  \bibinfo{pages}{3093--3107} (\bibinfo{year}{1991}) .

\bibitem{Li_2005}
\bibinfo{author}{Li, J.}, \bibinfo{author}{Wei, S.-H.} \&
  \bibinfo{author}{Wang, L.-W.}
\newblock \bibinfo{title}{{Stability of the $D{X}^{\ensuremath{-}}$ Center in
  GaAs Quantum Dots}}.
\newblock \emph{\bibinfo{journal}{Phys. Rev. Lett.}}
  \textbf{\bibinfo{volume}{94}}, \bibinfo{pages}{185501} (\bibinfo{year}{2005})
  .

\bibitem{Saito_1992}
\bibinfo{author}{Saito, M.}, \bibinfo{author}{Oshiyama, A.} \&
  \bibinfo{author}{Sugino, O.}
\newblock \bibinfo{title}{{Validity of the broken-bond model for the DX center
  in GaAs}}.
\newblock \emph{\bibinfo{journal}{Phys. Rev. B}} \textbf{\bibinfo{volume}{45}},
  \bibinfo{pages}{13745--13748} (\bibinfo{year}{1992}) .

\bibitem{Kundu_2019}
\bibinfo{author}{Kundu, A.} \emph{et~al.}
\newblock \bibinfo{title}{{Effect of local chemistry and structure on thermal
  transport in doped GaAs}}.
\newblock \emph{\bibinfo{journal}{Phys. Rev. Materials}}
  \textbf{\bibinfo{volume}{3}}, \bibinfo{pages}{094602} (\bibinfo{year}{2019})
  .

\bibitem{Du_dx_2005}
\bibinfo{author}{Du, M.~H.} \& \bibinfo{author}{Zhang, S.~B.}
\newblock \bibinfo{title}{{DX} centers in {GaAs} and {GaSb}}.
\newblock \emph{\bibinfo{journal}{Phys. Rev. B Condens. Matter}}
  \textbf{\bibinfo{volume}{72}}~(7) (\bibinfo{year}{2005}) .

\end{thebibliography}
\includepdf[pages=-]{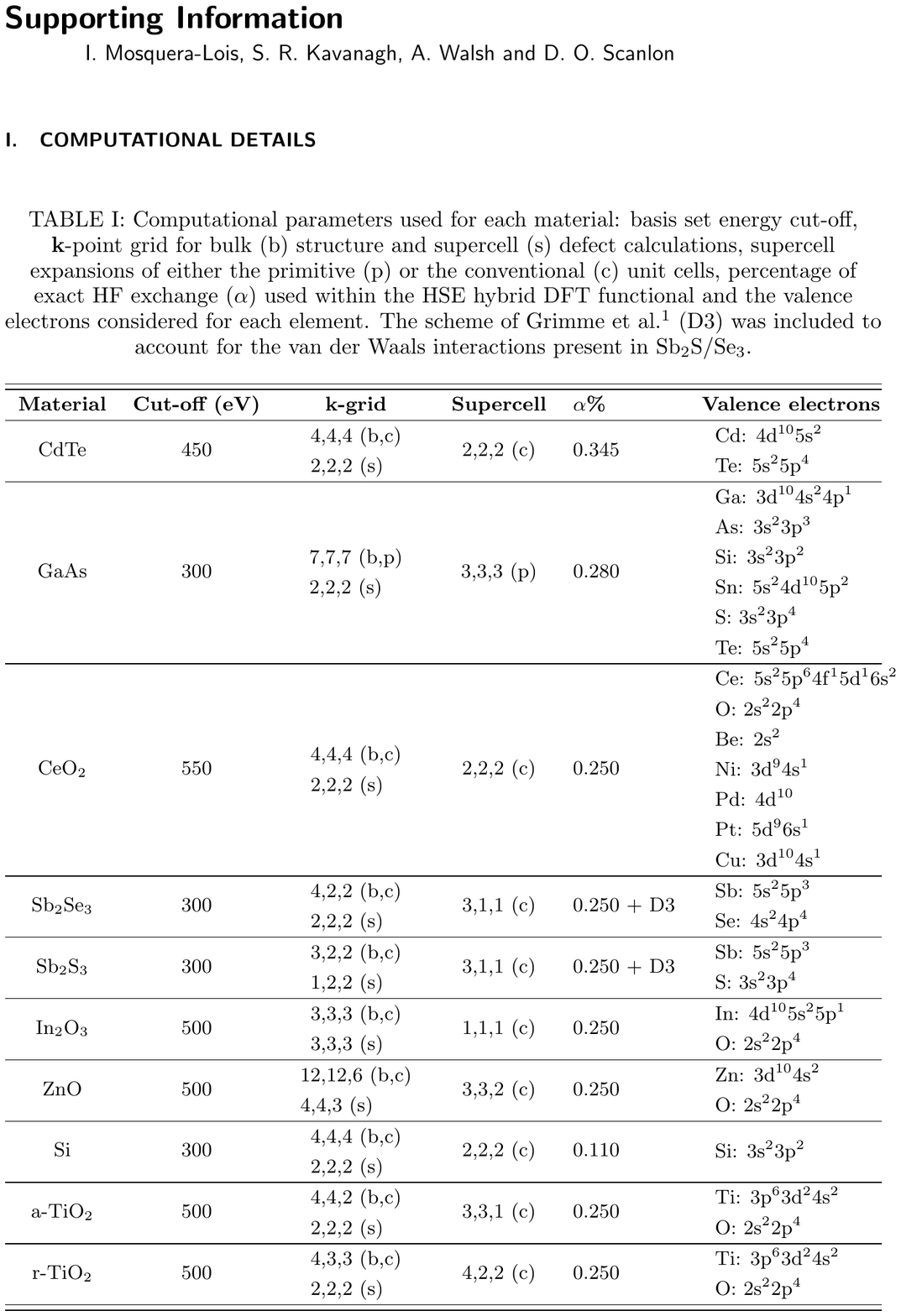}
\end{document}


\title{Supporting Information} 
\author{I. Mosquera-Lois, S. R. Kavanagh, A. Walsh and D. O. Scanlon}
\maketitle

\section{Computational details}\label{sec:comp_details}
\renewcommand{\arraystretch}{0.8}
\begin{table}[ht] 
	\begin{center}
		\caption{Computational parameters used for each material: basis set energy cut-off, {\bf k}-point grid for bulk (b) structure and supercell (s) defect calculations, supercell expansions of either the primitive (p) or the conventional (c) unit cells, percentage of exact HF exchange ($\alpha$) used within the HSE hybrid DFT functional and the valence electrons considered for each element. The scheme of Grimme et al.\cite{Grimme_2010} (D3) was included to account for the van der Waals interactions present in \ce{Sb2S/Se3}.}
		\label{tab:bulk_struct_computational_details}
		\setlength{\tabcolsep}{6pt} 
		\renewcommand{\arraystretch}{1.30} 
		\begin{tabular*}{\textwidth}{ccccll}
			\hline \hline
			{\bf Material} & {\bf Cut-off (eV)} & {\bf k-grid} &{\bf Supercell} 
			&  {\bf $\bf \alpha$\%} & 
			{\bf Valence electrons} \\
			\hline
			CdTe & 450 & 
			\begin{tabular}{l}
				4,4,4 (b,c) \\
				2,2,2 (s)
			\end{tabular}
			& 2,2,2 (c)
			& 0.345 &
			\begin{tabular}{l}
				Cd: $\rm 4d^{10} 5s^{2}$ \\
				Te: $\rm 5s^2 5p^4$
			\end{tabular}
			\\
			\hline
			GaAs &300& 
			\begin{tabular}{l}
				7,7,7 (b,p) \\
				2,2,2 (s)
			\end{tabular}
    		& 3,3,3 (p)
    		& 0.280 &
			\begin{tabular}{l}
    			Ga: $\rm 3d^{10} 4s^2 4p^1$ \\
    			As: $\rm 3s^2 3p^3$ \\
            	Si: $\rm 3s^2 3p^2$ \\
            	Sn: $\rm 5s^{2} 4d^{10} 5p^2$\\
            	S: $\rm 3s^2 3p^4$ \\
            	Te: $\rm 5s^2 5p^4$ 
			\end{tabular}
			\\ 
			\hline
			\ce{CeO2}  & 550 & 
			\begin{tabular}{l}
				4,4,4 (b,c) \\ 
				2,2,2 (s)
			\end{tabular}  
			& 2,2,2 (c) 
		    & 0.250 	
		    &\begin{tabular}{l}
				Ce: $\rm 5s^2 5p^6 4f^1 5d^1 6s^2$ \\
				O: $\rm 2s^2 2p^4$ \\
				Be: $\rm 2s^2$\\
				Ni: $\rm 3d^9 4s^1$\\
				Pd: $\rm 4d^{10} $\\
				Pt: $\rm 5d^9 6s^1$\\
				Cu: $\rm 3d^{10} 4s^1$
    		\end{tabular} 
    		\\
    		\hline
    		\ce{Sb2Se3} & 300 & 
    		\begin{tabular}{l}
				4,2,2 (b,c)\\
				2,2,2 (s)
			\end{tabular} 
			& 3,1,1 (c) & 0.250 + D3\tnote{a}& 	\begin{tabular}{l}
				Sb: $\rm 5s^2 5p^3$ \\
				Se: $\rm 4s^2 4p^4$
			\end{tabular} \\
			\hline 
			\ce{Sb2S3} &  300 &
			\begin{tabular}{l}
				3,2,2 (b,c)\\
                1,2,2 (s)
			\end{tabular} 
			& 3,1,1 (c) & 0.250 + D3\tnote{a}&
			\begin{tabular}{l}
				Sb: $\rm 5s^2 5p^3$ \\
				S: $\rm 3s^2 3p^4$ \\
			\end{tabular} \\ 
			\hline
			\ce{In2O3} &  500 &
			\begin{tabular}{l}
				 3,3,3 (b,c)\\
                 3,3,3 (s)
			\end{tabular}  
			& 1,1,1 (c) & 0.250&
			\begin{tabular}{l}
				In: $\rm 4d^{10} 5s^2 5p^1$ \\
				O: $\rm 2s^2 2p^4$ \\
			\end{tabular} \\ 
			\hline
			\ce{ZnO} &  500 &
			\begin{tabular}{l}
				 12,12,6 (b,c)\\
                 4,4,3 (s)
			\end{tabular}  
			& 3,3,2 (c) & 0.250 &
			\begin{tabular}{l}
				Zn: $\rm 3d^{10} 4s^2$ \\
				O: $\rm 2s^2 2p^4$ 
			\end{tabular} \\  
			\hline
			\ce{Si} &  300 &
			\begin{tabular}{l}
			4,4,4 (b,c)\\ 
			2,2,2 (s)
			\end{tabular}  
			& 2,2,2 (c) & 0.110 &
			\begin{tabular}{l}
				Si: $\rm 3s^{2} 3p^2$ 
			\end{tabular} \\ 
			\hline
			\ce{a-\ce{TiO2}} &  500 &
			\begin{tabular}{l}
			4,4,2 (b,c)\\ 
			2,2,2 (s)
			\end{tabular}  
			& 3,3,1 (c) & 0.250 &
			\begin{tabular}{l}
				Ti: $\rm 3p^{6} 3d^2 4s^2$\\
				O: $\rm 2s^2 2p^4$ 
			\end{tabular} \\ 
			\hline
			\ce{r-\ce{TiO2}} &  500 &
			\begin{tabular}{l}
			4,3,3 (b,c)\\ 
			2,2,2 (s)
			\end{tabular}  
			& 4,2,2 (c) & 0.250 &
			\begin{tabular}{l}
				Ti: $\rm 3p^{6} 3d^2 4s^2$\\
				O: $\rm 2s^2 2p^4$ 
			\end{tabular} \\ 
	\hline\hline
\end{tabular*}
\end{center}
\end{table}

\newpage
\section{Bulk crystals}
\renewcommand{\arraystretch}{1.2}
\begin{table}[ht]
\begin{center}
		\caption{Calculated lattice parameters (in \AA) of the conventional cells for the studied materials and 
		percentage differences from the experimental lattice parameters.}
		\label{tab:bulk_struct_parametes}
		\setlength{\tabcolsep}{8pt} 
		\renewcommand{\arraystretch}{1.30} 
		\begin{tabular}{cccccccc}
			\hline \hline
			{\bf Material} & {\bf a} & {\bf $\Delta$a\%} &{\bf b} & {\bf $\Delta$b\%} & {\bf c} & {\bf $\Delta$c\%} & {\bf Ref.} \\
			\hline
			CdTe & 6.543 & 1 & 6.543& 1 & 6.543 & 1 & Strauuss\cite{Strauss_1977}
			\\
			\hline
			GaAs &5.662& -0.1 & 5.662& -0.1& 5.662 &-0.1  & Madelung\cite{Madelung2004}
			\\
            \hline  
			\ce{CeO2} \tnote{1} & 5.394 & -0.3 &5.394 &-0.3 & 5.394 & -0.3  & Rossignol\cite{Rossignol_2003}
			\\
			\hline
			\ce{Sb2Se3} & 3.961 & -0.6 & 11.494 & -1.3 & 11.928 & 1.1 & Voutsa\cite{Voutsas_1985}	\\
			\hline 
			\ce{Sb2S3} & 3.841 & 0.1 &  11.063 & -1.5 & 11.362 & 0.4 & Bayliss\cite{Bayliss}	\\ 
			\hline 
			\ce{In2O3} & 10.205 & 0.8 &  10.205 & 0.8 & 10.205 & 0.8 & González\cite{Gonzalez_2001}	\\
			\hline 
			\ce{ZnO} & 3.25 & -0.04 & 3.25  & -0.04 & 5.20 & -0.18 & Kisi\cite{Kisi_1989}	\\
			\hline
			Si & 5.449 & 0.3 & 5.449 & 0.3 & 5.449 & 0.3 & Bond\cite{Bond_1960}\\
			\hline
			a-\ce{TiO2} & 3.774 & -0.3 & 3.774 & -0.3 & 9.579 & 0.7& Burdett\cite{Burdett_1987}\\
			\hline
			r-\ce{TiO2} & 2.950 & -0.3 & 4.597 & 0.05 & 4.597 & 0.05 & Shiiba\cite{Shiiba_2010}\\
			\hline \hline
		\end{tabular}
\end{center}
\end{table}

\newpage
\section{Defect Structure Searching (\texttt{ShakeNBreak}) approach: Parameter optimisation}
\begin{figure}[ht]
    \centering
    \includegraphics[width=1.0\textwidth]{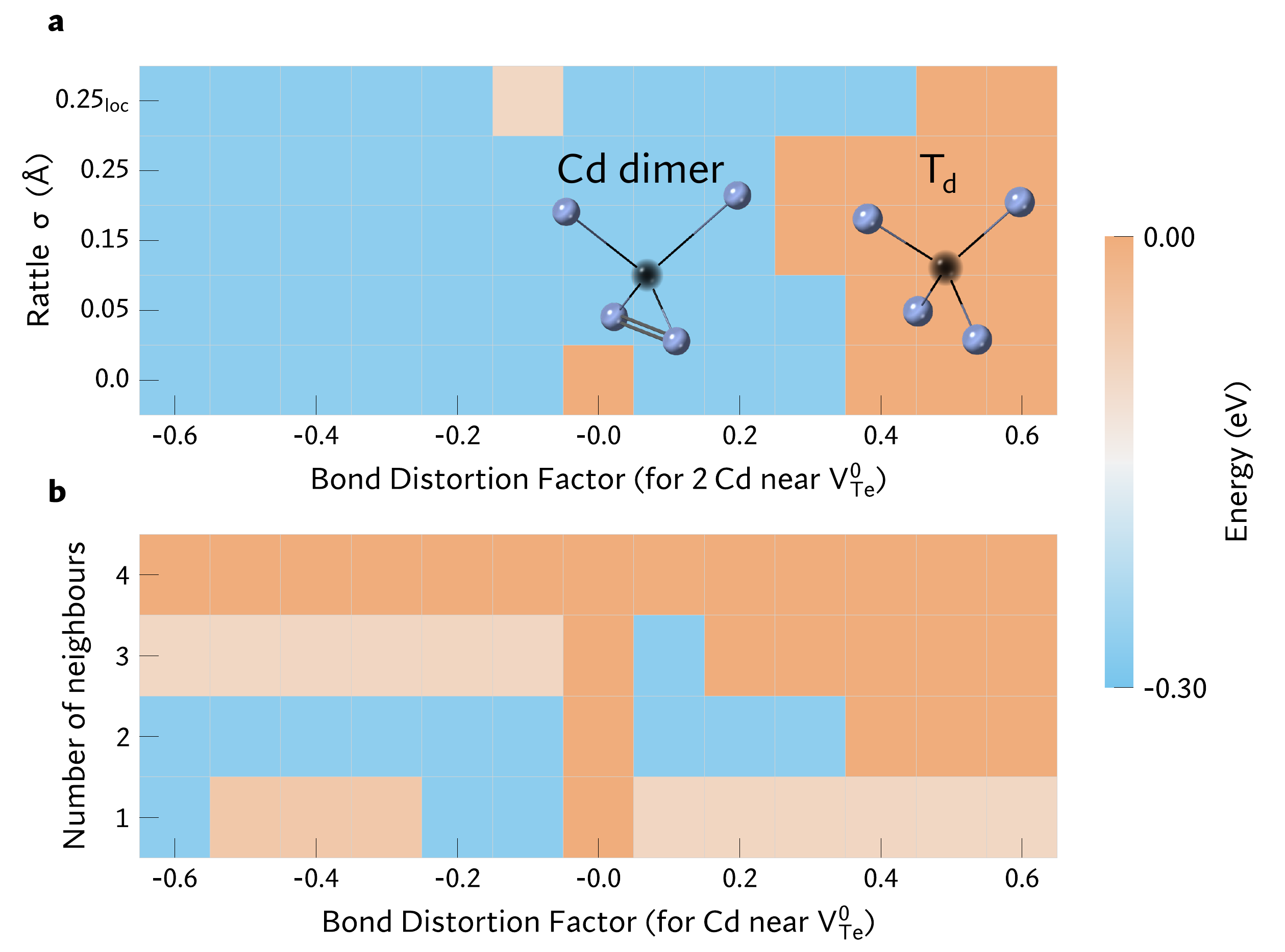}
    \caption{Relative energy of final structures for \kvc{V}{Te}{0} in CdTe (either `Cd dimer' (-0.3 eV), `Tetrahedral' (0 eV) or a $\rm C_{3\textrm{v}}$ reconstruction (-0.1 eV)), for different rattle standard deviations ({\bf a}) and number of distorted defect neighbours ({\bf b}). The wider interval of bond distortions leading to the defect ground state is obtained with  a standard deviation of 0.25 \AA ~localised to the atoms within 5 \AA ~from the defect ($0.25_{\rm loc}$). Regarding the number of neighbours, best performance is obtained when distorting two cadmium atoms, as expected considering that the defect has two extra electrons in the Cd dangling bonds.}
    \label{fig:BDM_param_opt_V_Te}
\end{figure}

\begin{figure}[ht]
    \centering
    \includegraphics[width=1.0\textwidth]{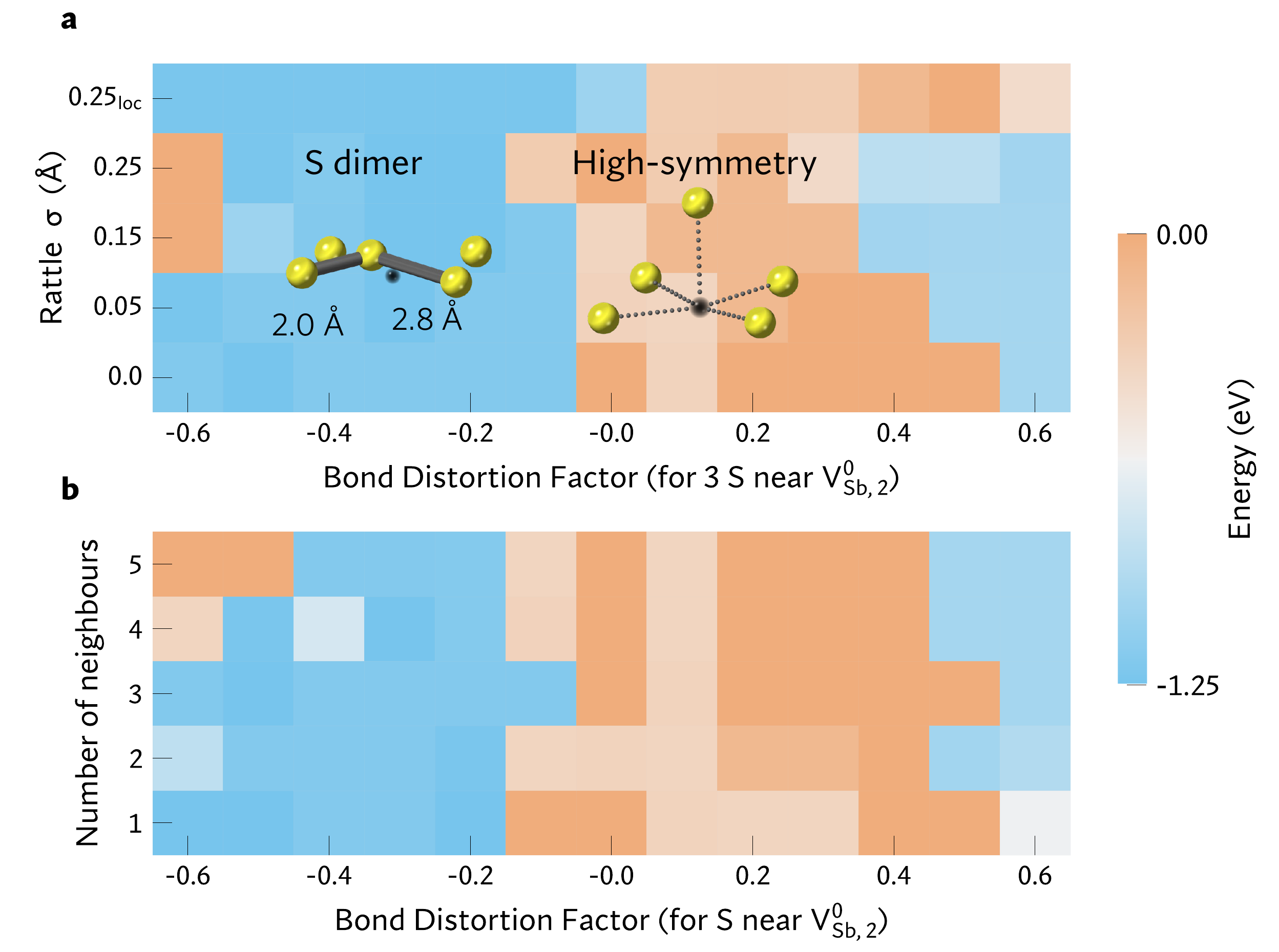}
    \caption{Relative energy of final structures for \kvc{V}{Sb,2}{0} in \ce{Sb2S3} (either `S dimer' (-1.25 eV) or `High-symmetry' configuration (0 eV)), for different rattle standard deviations ({\bf a}) and number of distorted defect neighbours ({\bf b}). Here, the bond distortion parameters have a minor influence (with two and three distorted neighbours performing slightly better), which suggests that any symmetry breaking is enough to escape the local minimum and find the favourable `S dimer' configuration.}
    \label{fig:BDM_param_opt_V_Sb}
\end{figure}

\begin{figure}[ht]
    \centering
    \includegraphics[width=1.0\textwidth]{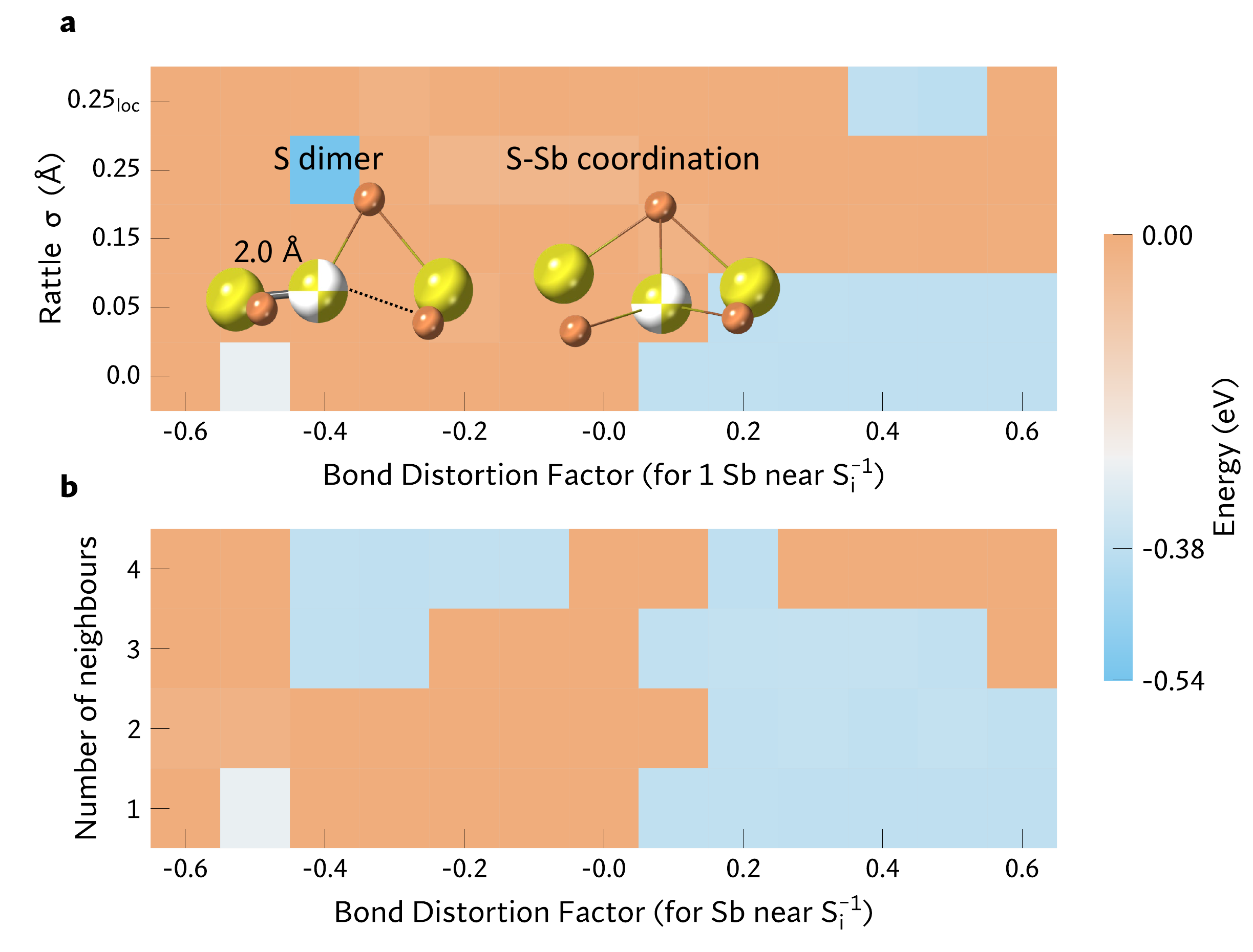}
    \caption{Relative energy of final structures for $\rm S_{i}^{-1}$ in \ce{Sb2S3} (either two different arrangements of `S dimer' (\SI{-0.55}{eV}, \SI{-0.38}{eV}) or weak S-Sb coordination (\SI{0}{eV})), for different rattle standard deviations ({\bf a}) and number of distorted defect neighbours ({\bf b}). Notably, the lowest energy structure is only obtained when applying a total rattle standard deviation of \SI{0.25}{\AA}. In this case, the number of distorted neighbours seems to have a minor role -- with three and one performing slightly better.}
    \label{fig:BDM_param_opt_S_i}
\end{figure}

\subsection{Comparison of local and total rattle}
\begin{figure}[ht]
\centering
    \begin{subfigure}{0.45\linewidth}
    \centering
    \includegraphics[width=\textwidth]{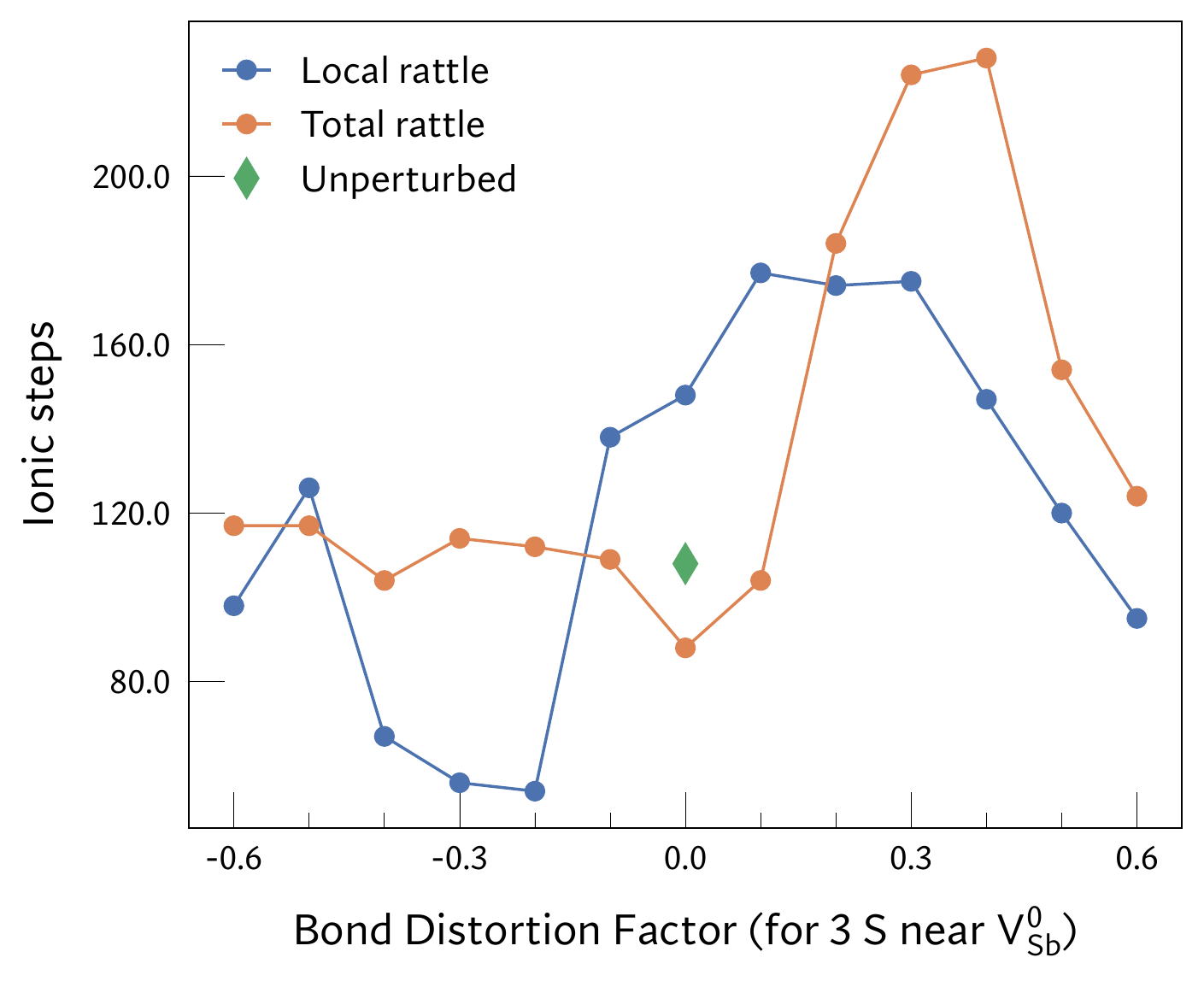}
    \caption{\small $V_{\rm Sb}^0$}
    \end{subfigure}
    \begin{subfigure}{0.45\linewidth}
    \centering
    \includegraphics[width=\textwidth]{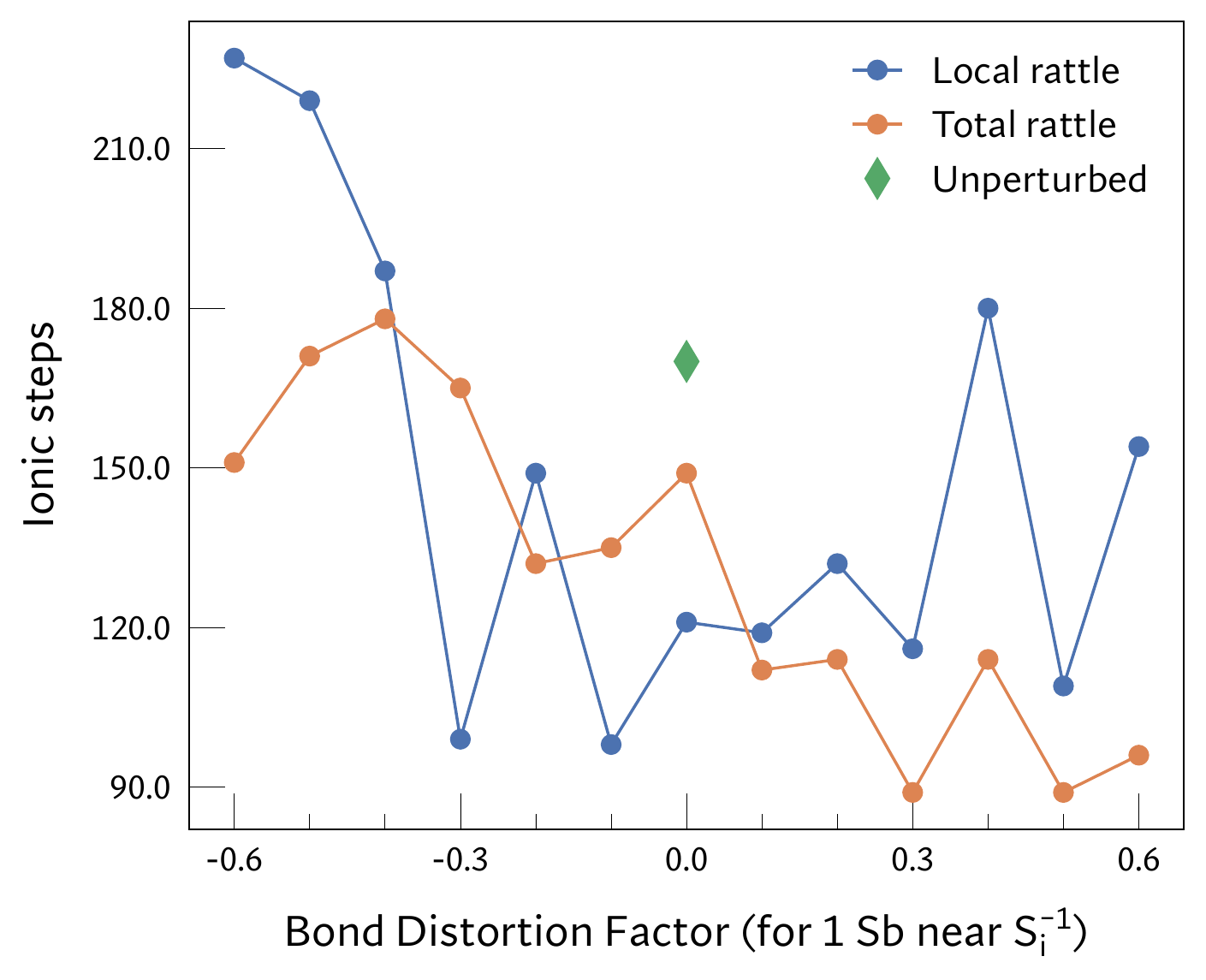}
    \caption{$\rm S_{\rm i}^{-1}$}
    \end{subfigure}
    \caption{\small Total number of ionic steps for each bond distortion with a local (restricted to 5 \AA~ sphere about the defect, $\sigma = 0.25$) and a total rattle ($\sigma = 0.25$) in the relaxation of $V_{\rm Sb}^0$ (a) and $\rm S_{\rm i}^{-1}$ (b) in \ce{Sb2S3}. Surprisingly, the local rattle does not lead to a significant decrease in the number of ionic steps compared to its total counterpart.} \label{fig:ionic_steps_local_total_rattle}
\end{figure}

\section{Comparison with other structure searching approaches}
To compare the performance of our method with existing approaches in the literature, it was applied to the neutral silicon self-interstitial and the neutral oxygen vacancy in anatase. These systems were selected as they were previously investigated using the evolutionary approach of Arrigoni and Madsen.\cite{Arrigoni_evolutionary_2021}
\subsection{Silicon self-interstitial}
\begin{figure}[ht]
	\hspace*{\fill}
	\begin{subfigure}[b]{0.30\textwidth}
		\includegraphics[width=\textwidth]{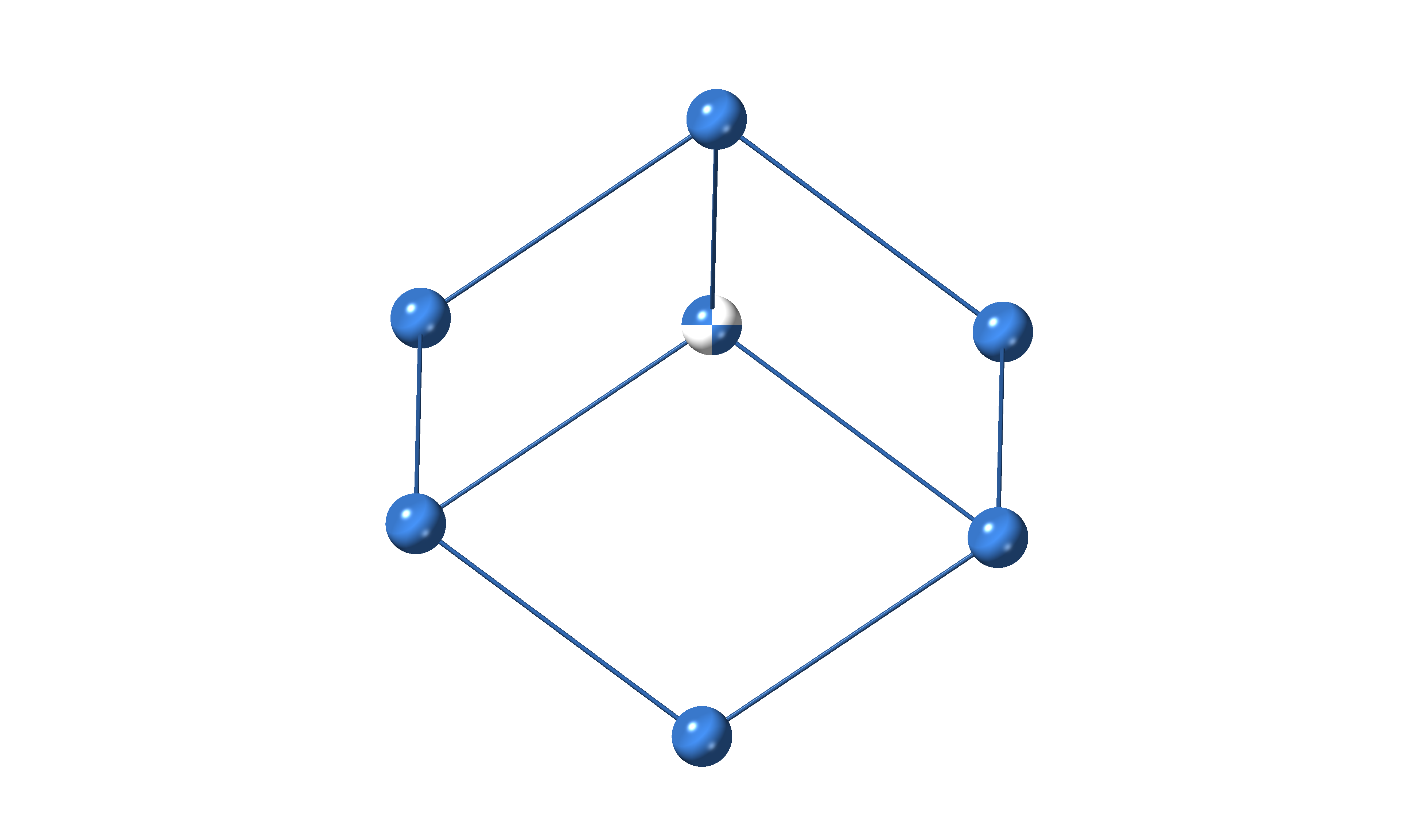}
		\caption{$\rm Si_T$}
	\end{subfigure}
	\hspace*{\fill}
	\begin{subfigure}[b]{0.30\textwidth}
		\includegraphics[width=\textwidth]{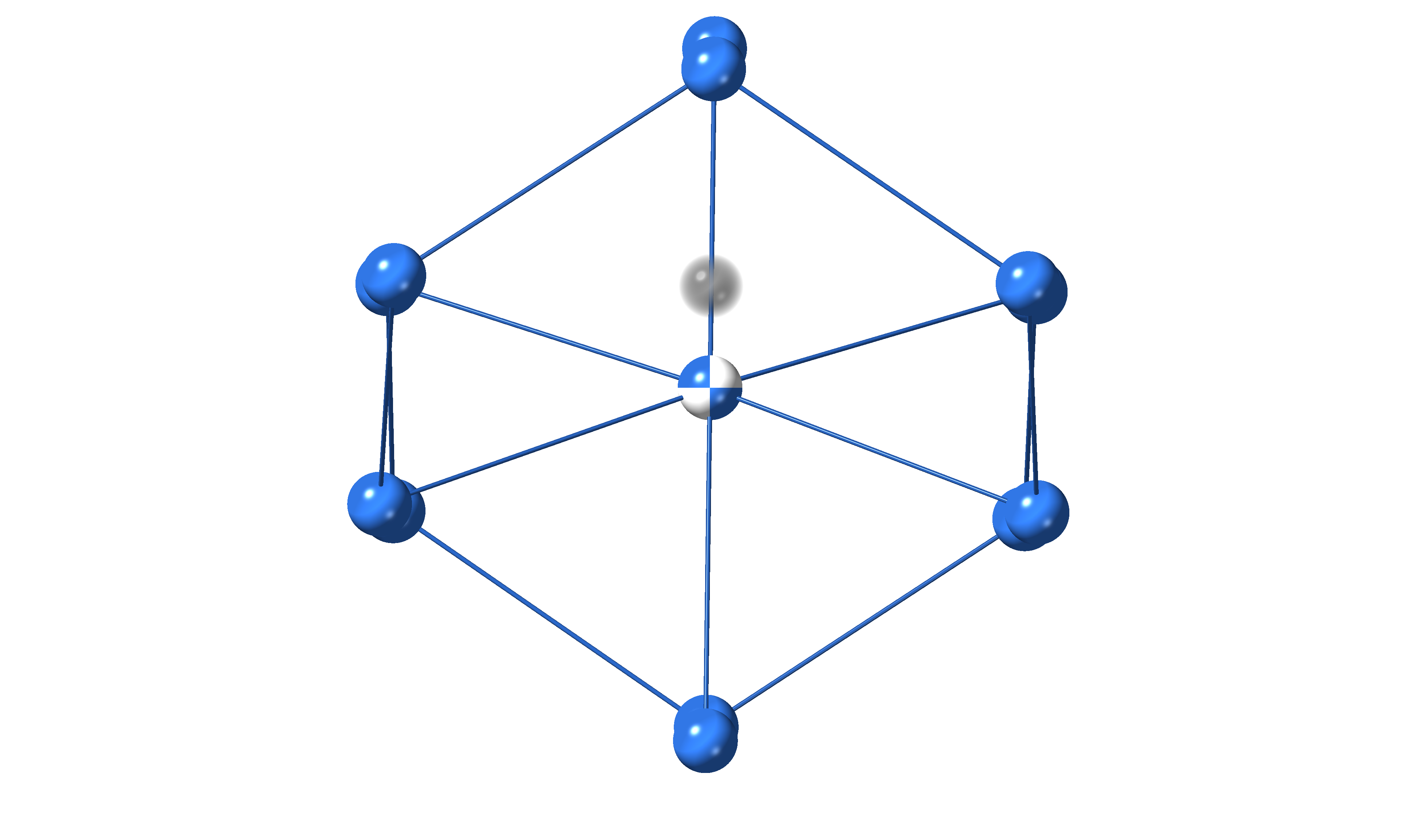}
		\caption{$\rm Si_H$}
	\end{subfigure}
	\hspace*{\fill}
	\begin{subfigure}[b]{0.3\textwidth}
		\includegraphics[width=\textwidth]{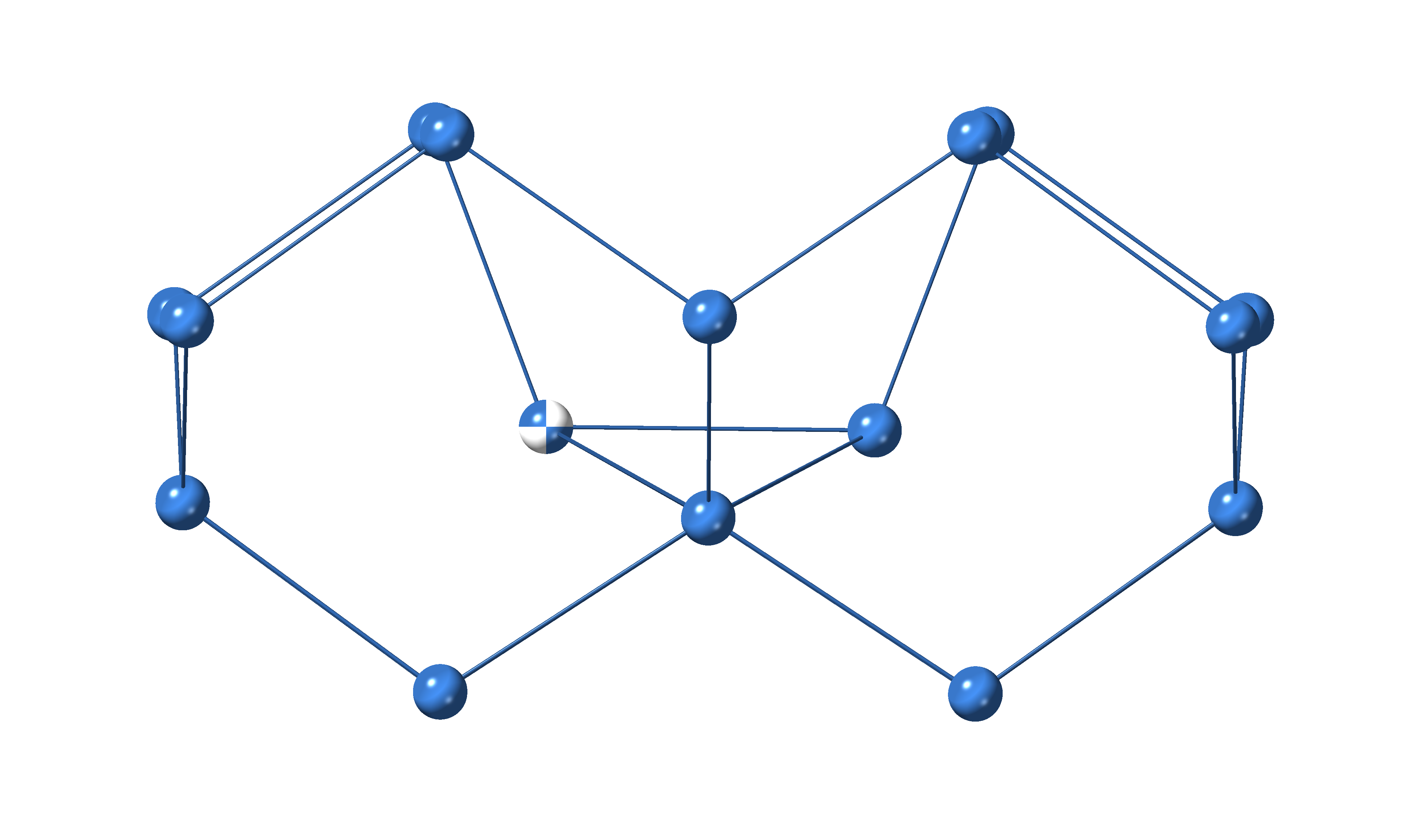}
		\caption{$\rm Si_{<110>}$}
	\end{subfigure}	
	\hspace*{\fill}
	\caption{\small Lowest energy structures for the silicon self-interstitial: a) $\rm Si_T$ b) $\rm Si_H$ and c) $\rm Si_{<110>}$. The silicon interstitial is shown with a different pattern and the original tetrahedral position is shown in grey.} \label{fig:Si_i}
\end{figure}

The high technological relevance of the silicon self-interstitial has motivated a large number of studies on its structural and migration properties. These have identified several low energy configurations: Si in an tetrahedral interstitial ($\rm Si_T$), Si in a hexagonal interstitial ($\rm Si_H$) and the split $<110>$ configuration ($\rm Si_{<110>}$), where two Si move from their original site and form a dumbbell configuration along the $<110>$ direction\cite{Centoni_2005,Ganchenkova_2015,Lee_1998,Needs_1999,Stewart_2020,Rinke_2009,Bruneval_2012,Gao_2013,Budde_1998,Al-Mushadani_2003,Al-Mushadani_2003,Mattsson_2008,Leung_1999}. These are shown in \cref{fig:Si_i}. Previous studies agree that the hexagonal and split configurations are the most stable, differing by a small energy difference (order of meV), with the latter likely corresponding with the ground state\cite{Budde_1998,Al-Mushadani_2003,Morris_2008,Centoni_2005,Bloch_1993,Bar-Yam_1984}. Significantly higher in energy lies the tetrahedral structure ($\approx0.4$ eV) \cite{Morris_2008,Rinke_2009,Bruneval_2012,Gao_2013}. 

By applying our method, we identified the two lowest energy structures, thereby exhibiting similar results to the evolutionary approach of Arrigoni and Madsen, yet with a simpler implementation. In agreement with their result, we identified the ground state as the hexagonal configuration. We find it to lie 7 meV lower in energy than the split arrangement, opposed to their result of 23 meV using local DFT (LDA) and $\boldsymbol{\Gamma}$-point only {\bf k}-point sampling. As expected, the latter agrees better with our initial result (31 meV), obtained prior to optimising the final structures with a denser {\bf k}-mesh. On this point, it is noteworthy how the approximate exploration of the PES (i.e. the $\boldsymbol{\Gamma}$-point approximation) qualitatively described the landscape despite the minute energy differences between local minima. These minor energy differences between structures explain the debate across different studies (and energy functionals) regarding the actual ground state.\cite{Leung_1999,Rinke_2009,Arrigoni_2021,Morris_2008,Mattsson_2008} Semi-local functionals often favour the split configuration\cite{Morris_2008,Mattsson_2008,Centoni_2005,Al-Mushadani_2003,Bloch_1993,Bar-Yam_1984}, while higher levels of theory (QMC\cite{Leung_1999} or \ce{G_0W_0} corrections\cite{Rinke_2009}) agree that the hexagonal structure lies lower in energy, by 160 and 60 meV, respectively.\\
Finally, we note that a tetrahedral configuration ($\rm Si_T$) was found to have no local stability with our hybrid DFT functional, with unperturbed relaxation of this initial geometry without symmetry constraints yielding the hexagonal arrangement ($\rm Si_H$), as also reported in other studies\cite{Leung_1999}.

\subsection{Neutral oxygen vacancy in anatase}
\begin{figure}[ht]
	\hspace*{\fill}
	\begin{subfigure}[b]{0.21\textwidth}
		\includegraphics[width=\textwidth]{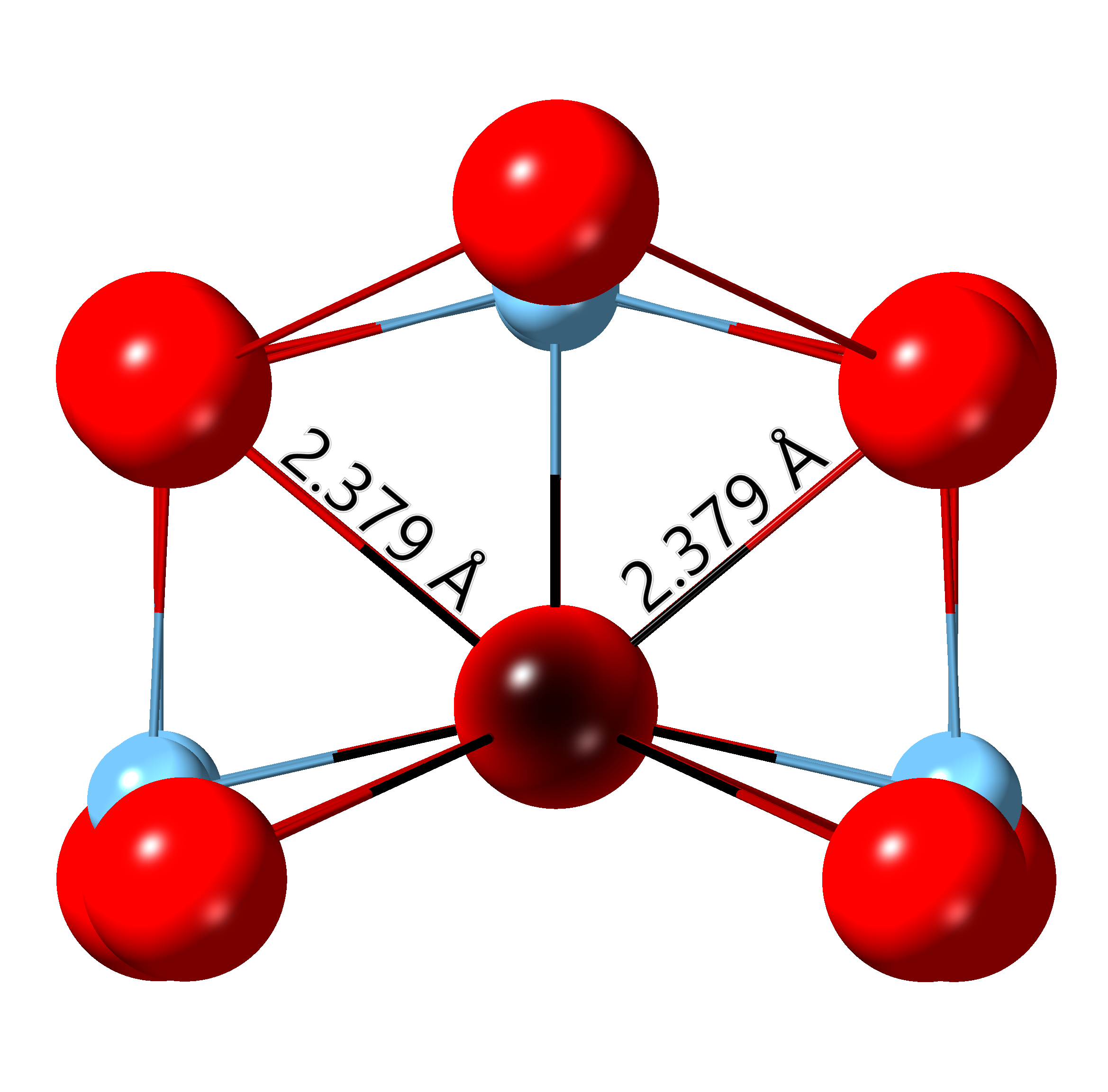}
		\caption{Simple \ce{V_{O}^0}} 
	\end{subfigure}
	\hspace*{\fill}
	\begin{subfigure}[b]{0.22\textwidth}
		\includegraphics[width=\textwidth]{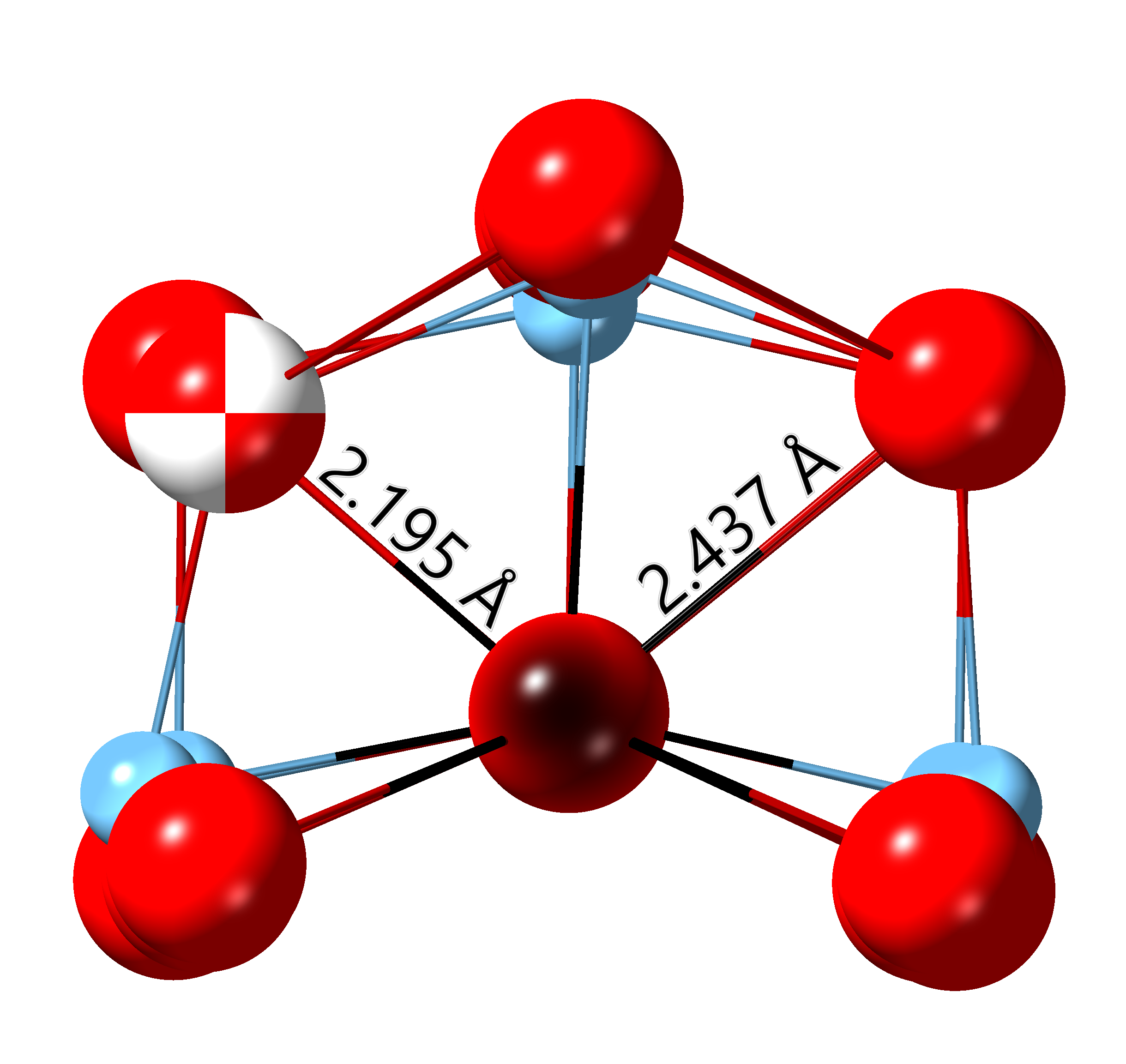}
		\caption{Split \ce{V_{O}^0}}
	\end{subfigure}
	\hspace*{\fill}
	\begin{subfigure}[b]{0.22\textwidth}
		\includegraphics[width=\textwidth]{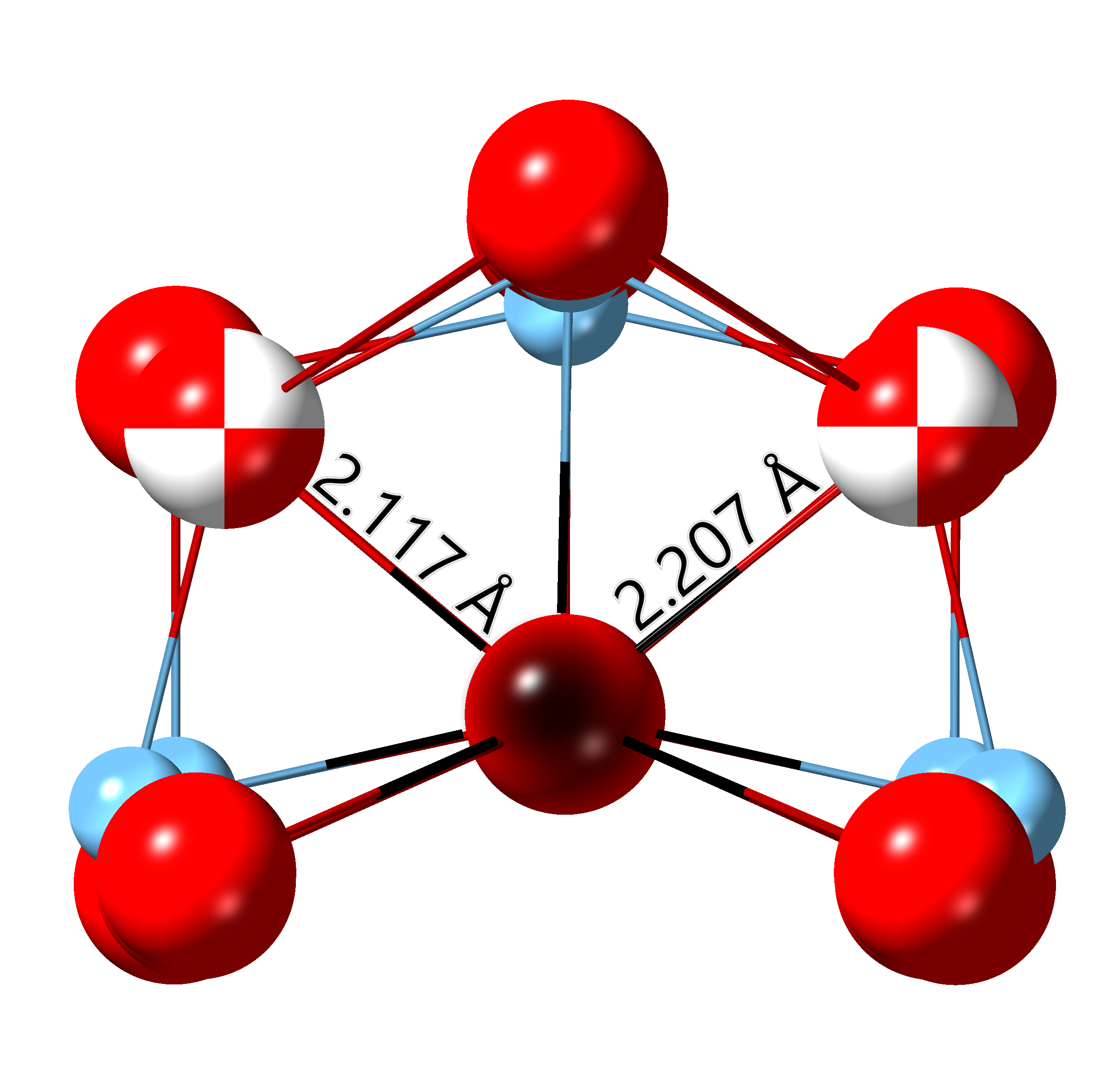}
		\caption{Delocalised \ce{V_{O}^0}}
	\end{subfigure}	
	\hspace*{\fill}
	\caption{\small Lowest energy structures for the neutral oxygen vacancy in anatase \ce{TiO2}: a) Simple \ce{V_{O}^0} b) Split \ce{V_{O}^0} and c) Delocalised \ce{V_{O}^0}. Ti atoms in blue, O in red, \ce{V_{O}} in black and displaced O neighbours represented with a different pattern.} \label{fig:V_O^0}
\end{figure}

The key role of anatase in photo-catalysis\cite{Schneider_2014,Linsebigler_1995}, solar cells\cite{Gratzel_2011,Bach_1998}, spintronic devices\cite{Janisch_2005} and as a promising transparent conductive oxide\cite{Hitosugi_2005,Furubayashi_2005} have motivated intense study of its intrinsic defects\cite{Na-Phattalung_2006}. In particular, significant efforts have been dedicated to the oxygen vacancy\cite{Bouzoubaa_2005,Shao_2014,Setvin_2014,Morgan_2010,Morgan_2009,Finazzi_2008,Mattioli_2008,Deak_2011,Deak_2012,Arrigoni_2020,Yang_2010}, as it renders intrinsic anatase n-type\cite{Deak_2015}.
Several studies\cite{Morgan_2009,Arrigoni_2020,Finazzi_2008,Mattioli_2008} have reported two possible configurations: the simple vacancy (analogous to the ideal defect structure, Fig. \ref{fig:V_O^0} (a)) and the split configuration, where one of the neighbouring oxygen atoms moves towards the vacancy (Fig. \ref{fig:V_O^0} (b)); 
with the latter lying 0.38 eV lower in energy (using HSE(15\%))\cite{Arrigoni_2020}.
In terms of the electronic structure, the former is characterised by two localised electrons in the vacancy site while for the latter, one electron is localised on a neighbouring Ti while the other one is excited to a delocalised conduction band state.   
Notably, this lower energy configuration has been missed by previous hybrid studies\cite{Deak_2012,Boonchun_2016}, as a standard relaxation from the ideal structure gets trapped on the simple vacancy basin. 

By applying their evolutionary algorithm, Arrigoni and Madsen identified, besides these structures, 
a new metastable configuration, termed the delocalised vacancy since the two electrons are delocalised in the conduction band (Fig. \ref{fig:V_O^0} (c)). 
Lying 59 meV higher in energy than the split structure, it is characterised by two neighbouring oxygen atoms moving towards the vacancy. 
These reduce their bond lengths from an initial value of 2.45 \AA~to 2.12 and 2.21 \AA. In comparison, our method also identified these structures, finding the simple vacancy and delocalised vacancy to lie 0.37 and 0.03 eV higher in energy than the split structure, respectively -- in good agreement with the result of Arrigoni (0.38, 0.06 eV). 
Hence, both approaches rendered similar configurations and relative stabilities.

\section{Symmetry breaking reconstructions}
\begin{figure}[ht]
    \centering
    \includegraphics[width=0.5\textwidth]{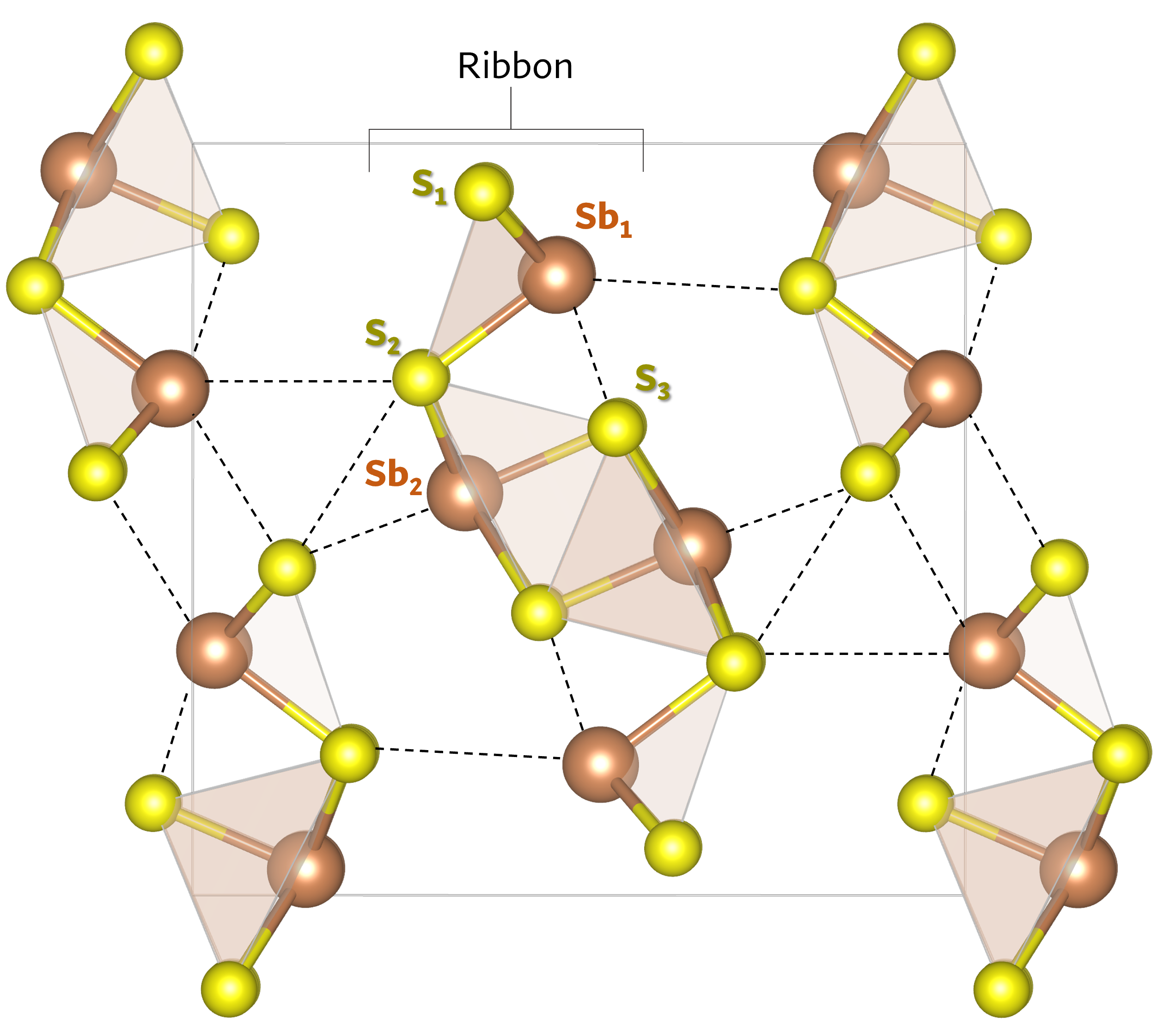}
    \caption{\small The crystal structure of \ce{Sb2S3} (and \ce{Sb2Se3}) comprises chains of [\ce{Sb4S6}]$\rm _{n}$ ([\ce{Sb4Se6}]$\rm _{n}$) (termed ribbons), which are bonded by weak van der Waals forces (dashed lines). Each ribbon has two and three symmetry inequivalent sites for antimony and sulfur, respectively, with the Sb cations occupying the centre of a pyramid formed by the closest S anions.}
    \label{fig:diagram_Sb2S3}
\end{figure}

\subsection{Rebonding: dimerisation}

\begin{figure}
    \includegraphics[width=0.8\linewidth]{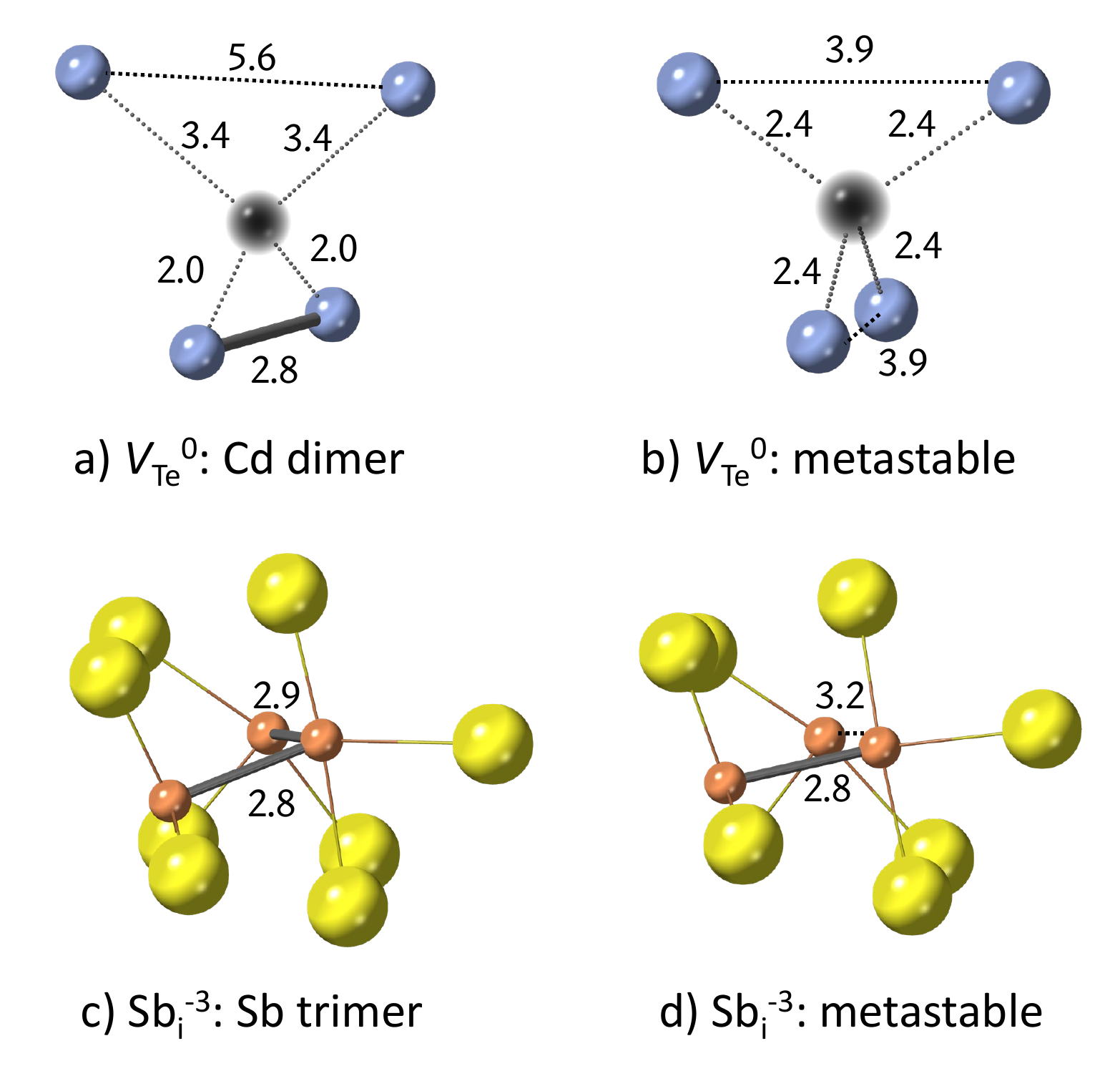}
    \caption{Ground state (left: a,c) and metastable (right: b,d) structures identified by our method and by relaxing the high-symmetry configuration, respectively, for cation-cation dimerisation. (Energies and displacements reported in Table 1). Vacancy in black, Cd in blue, Sb in orange and S in yellow.}
    \label{fig:structs_cation_cation}
\end{figure}

\begin{figure}[ht]
	\includegraphics[width=0.75\linewidth]{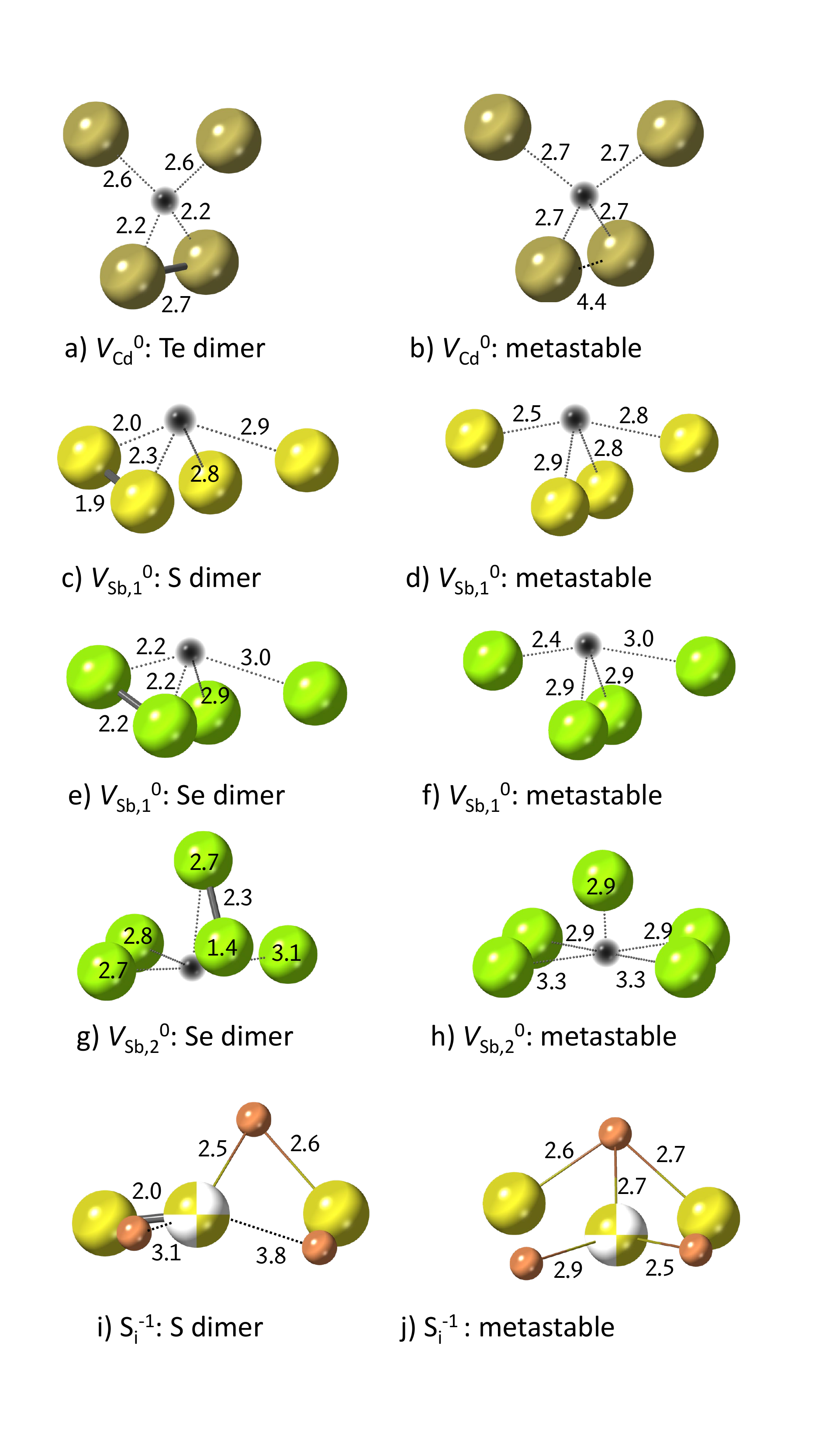}
    \caption{Ground state (left: a,c,e,g,i) and metastable (right: b,d,f,h,j) structures identified by our method and by relaxing the high-symmetry configuration, respectively, for anion-anion dimerisation. (Energies and displacements reported in Table 1). Te in gold, Sb in orange, S in yellow and Se in green. Vacancy shown in shaded black and $\rm S_i$ displayed in a different pattern. Pseudo-bonds between a vacancy and its neighbours shown with dotted lines, and distances in \AA.}
    \label{fig:structs_anion_anion}
\end{figure}

\begin{figure}[ht]
    \includegraphics[width=0.7\linewidth]{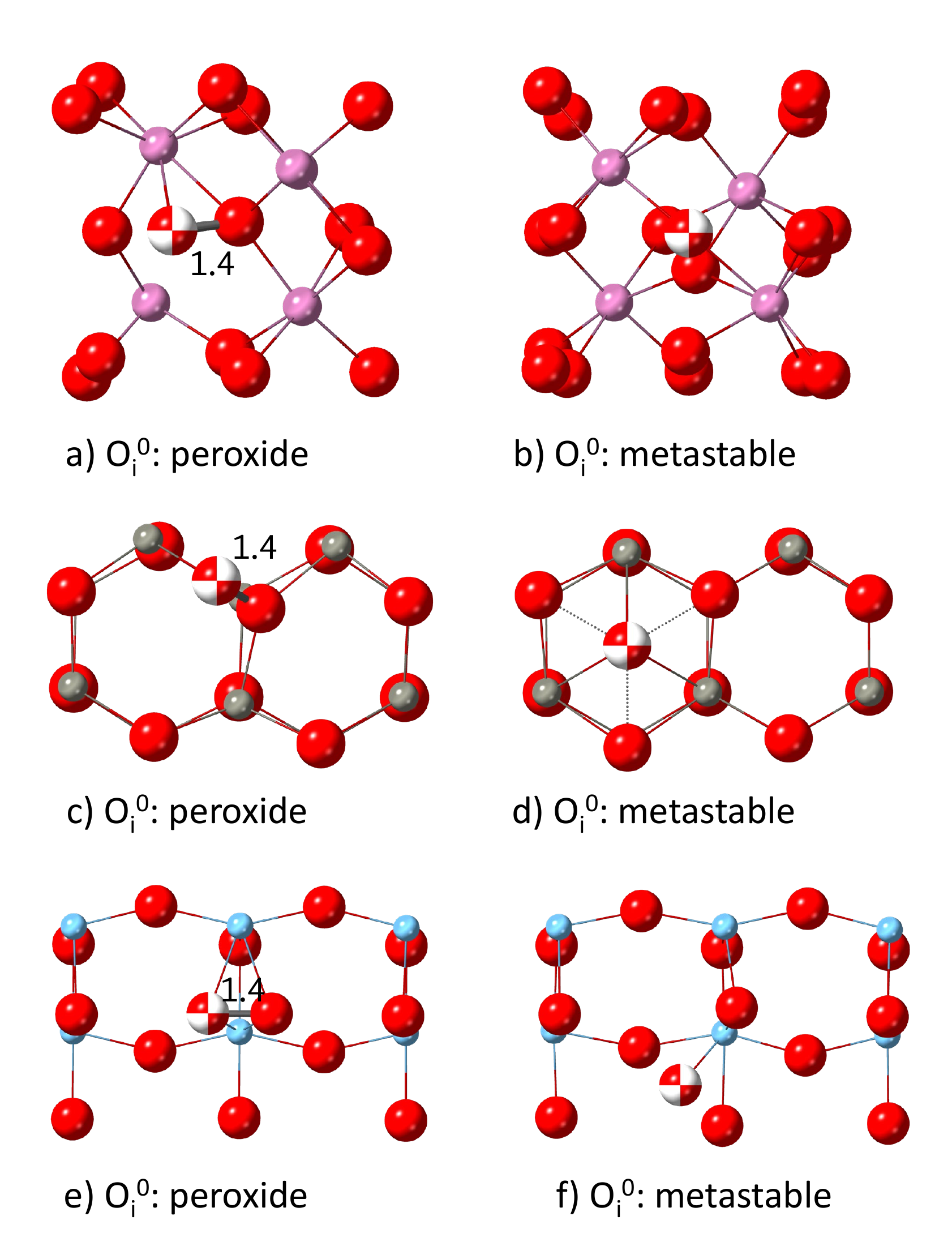}
    \caption{Ground state (left: a,c,e) and metastable (right: b,d,f) structures identified by our method and by relaxing the high-symmetry configuration, respectively, for the anion-anion dimerisation in \ce{In2O3}, \ce{ZnO} and anatase \ce{TiO2}. (Energies and displacements reported in Table 1). In in purple, Zn in grey, Ti in blue and O in red with $\rm O_i$ shown in a different pattern for clarity.}
    \label{fig:structs_anion_anion_peroxides}
\end{figure}

\begin{table}[ht]
    \caption{Integrated Crystal Orbital Hamilton Population (ICOHP) for the S-S/Se-Se bonds formed by the under-coordinated neighbours of \kv{V}{Sb} in \ce{Sb2S/Se3}.}
    \centering
    \setlength{\tabcolsep}{8pt} 
    \label{tab:V_Sb_ICOHP}
    \begin{tabular}{ccc|cc}
    \hline \hline
    \multirow{3}{*}{Charge} & \multicolumn{4}{c}{\ce{ICOHP (eV)}}\\
    \hline
    & \multicolumn{2}{c}{\ce{Sb2S3}} & \multicolumn{2}{c}{\ce{Sb2Se3}}\\
    & Site 1 & Site 2 & Site 1 & Site 2\\
    \hline
         -1 & -6.87 & -6.84 & -4.97 & -4.96\\
         0 & -9.15 & -7.74 & -6.60 & -5.54\\
         +1 & -13.36 & -13.41 & -6.69& -10.12\\
         +2 & -14.50 & -14.84 & -11.09 & -10.83\\
        \hline \hline
    \end{tabular}
\end{table}

\begin{figure}[ht]
    \includegraphics[width=0.9\linewidth]{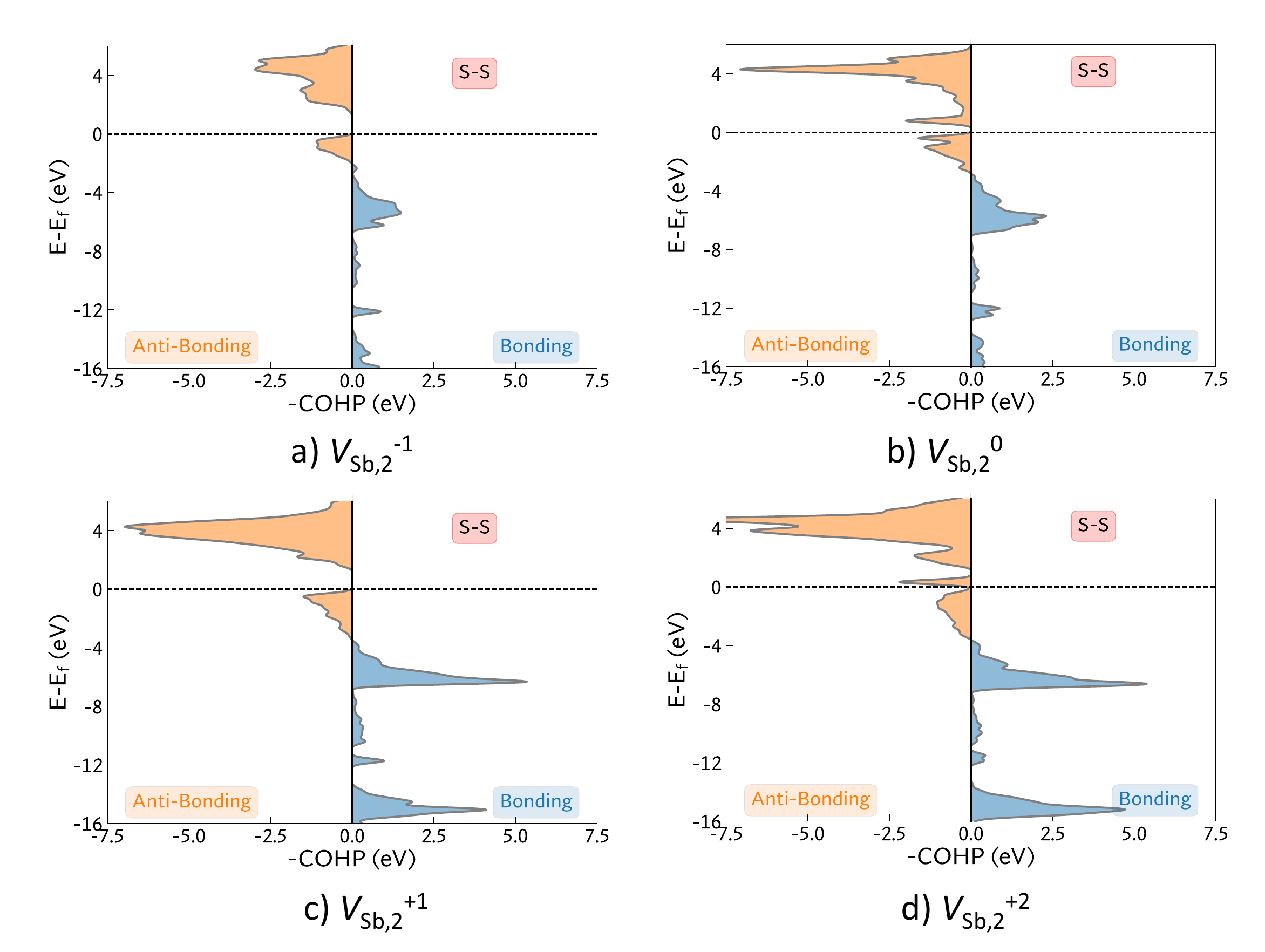}
    \caption{Crystal Orbital Hamilton Population (COHP) analysis for the sulfur-sulfur bonds formed by the vacancy neighbours in the ground state structures of \kv{V}{Sb,2} in \ce{Sb2S3}, illustrating the stronger bonding character of the S-S dimers for the positive charge states.}
    \label{fig:COHP_V_Sb,1_Sb2S3}
\end{figure}

\begin{figure}[ht]
    \includegraphics[width=0.9\linewidth]{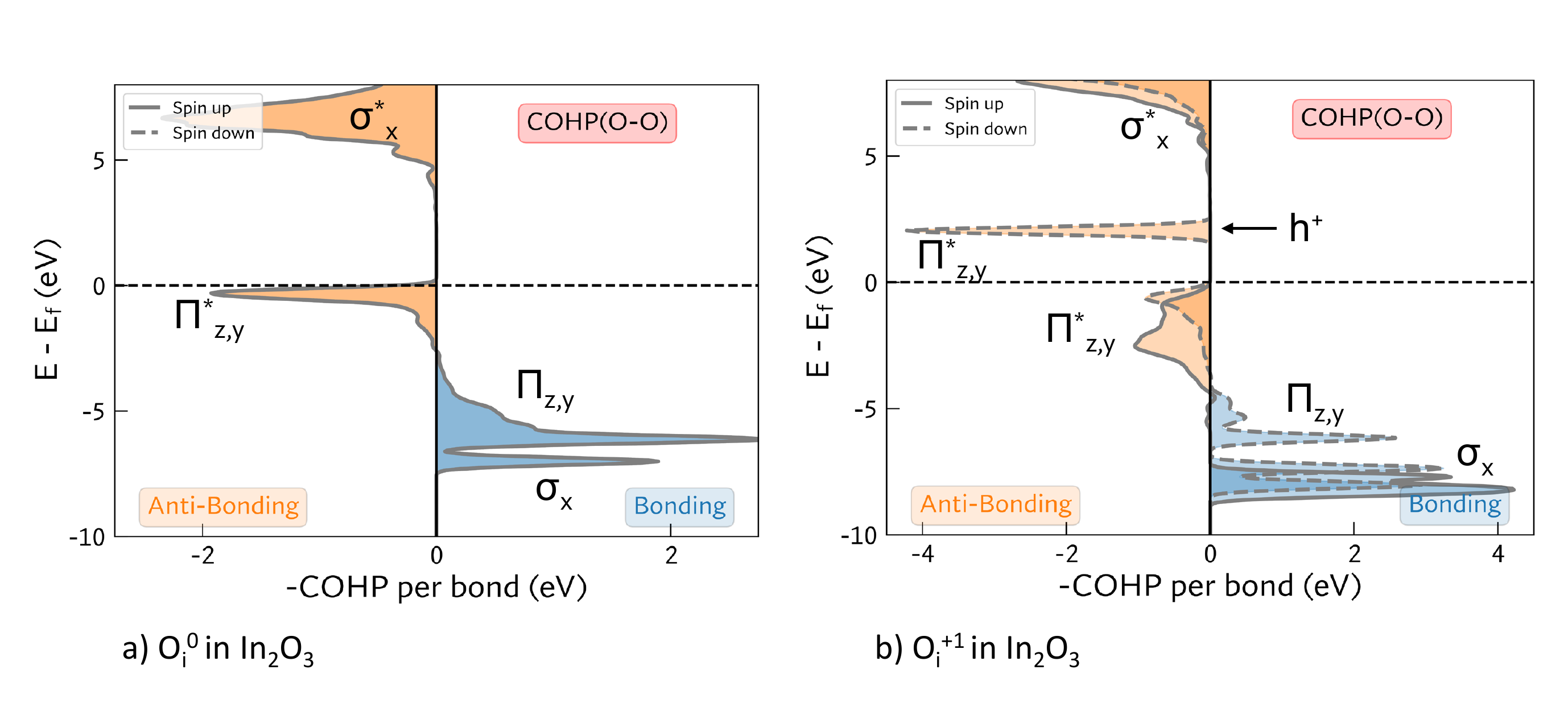}
    \caption{Crystal Orbital Hamilton Population (COHP) analysis for the oxygen dimer bonds in the ground state structures of \kv{O}{i} in \ce{In2O3}. The orbital character of the different states is assigned by comparison with the molecular orbital diagram of a peroxide anion. For positive charge state, the hole is localised on the antibonding $\pi$* state, which appears within the band gap.}
    \label{fig:COHP_peroxides_O_i_In2O3}
\end{figure}

\begin{figure}[ht]
	\includegraphics[width=0.8\textwidth]{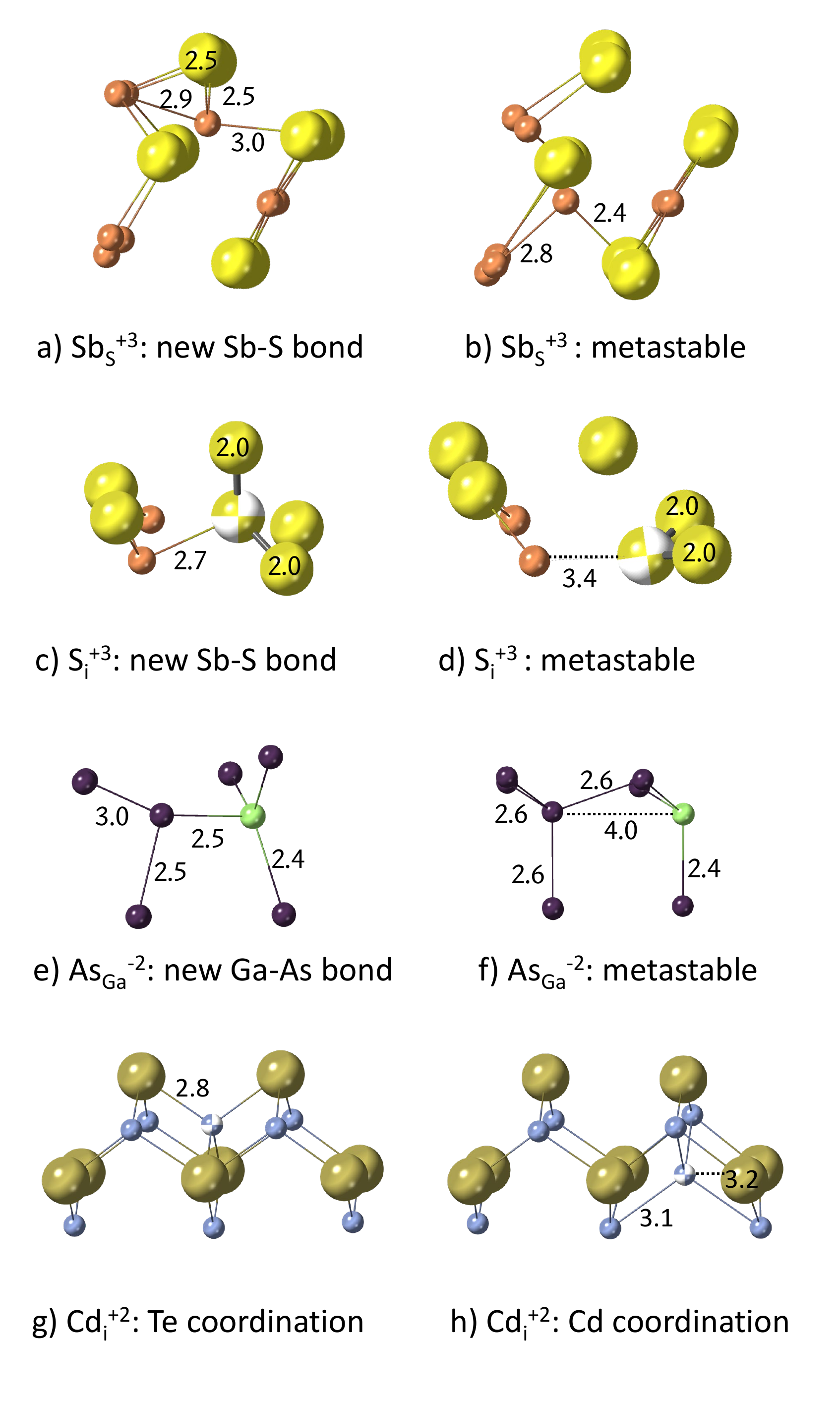}
    \caption{Ground state (left: a,c,e,g) and metastable (right: b,d,f,h) structures identified by our method and by relaxing the high-symmetry configuration, respectively, for the cation-anion rebonding reconstruction. (Energies and displacements reported in Table 2). Vacancy in black, Te in gold, Cd in blue, Sb in orange, S in yellow, Ga in green and As in purple. Interstitials shown with a different pattern and distances in \AA.}
    \label{fig:structs_cation_anion}
\end{figure}

\subsection{Crystal field and Jahn-Teller effects}

\begin{figure}[ht]
	\includegraphics[width=0.7\textwidth]{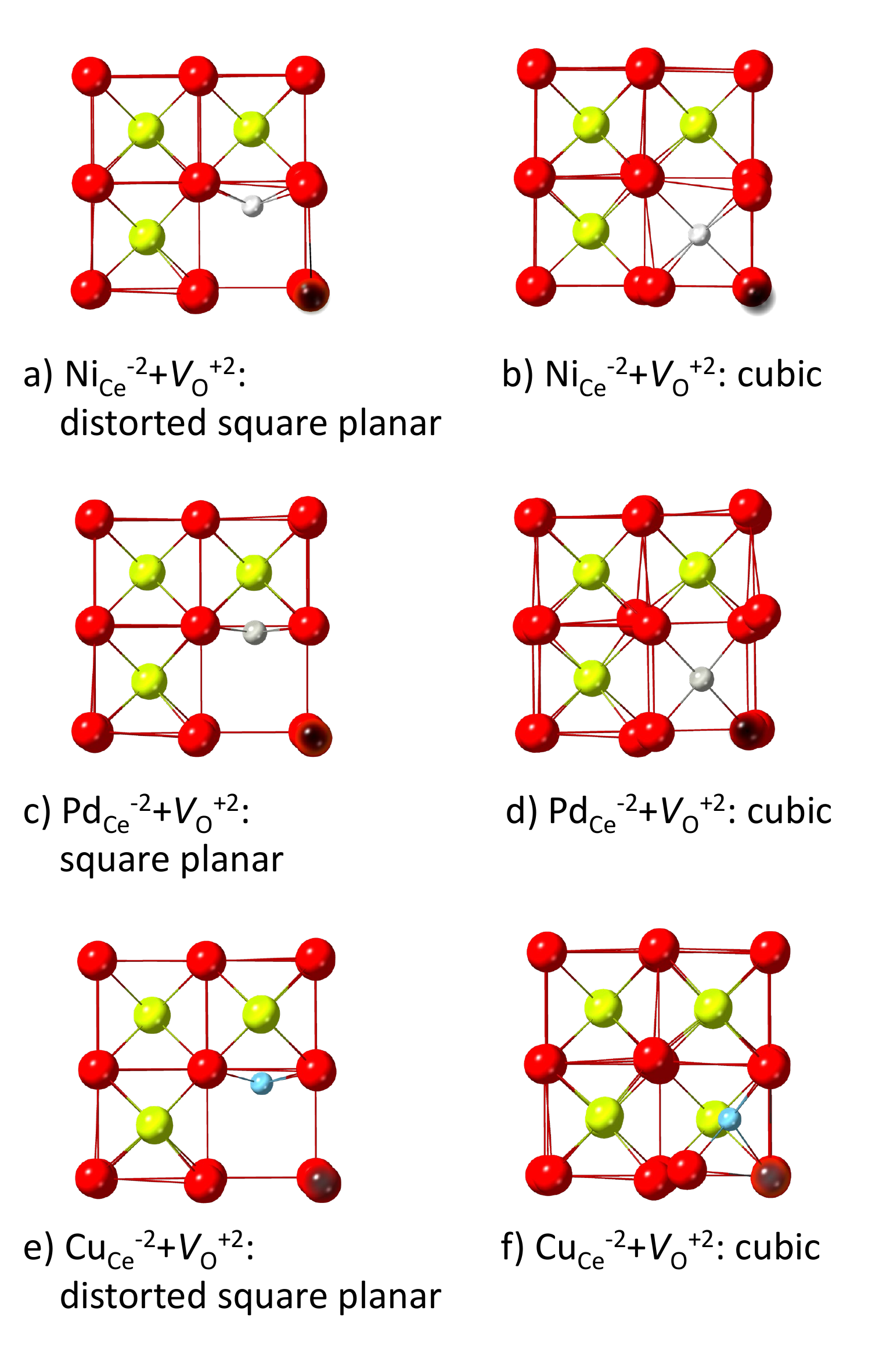}
    \caption{Ground state (left: a,c,e) and metastable (right: b,d,f) structures identified by our method and by relaxing the high-symmetry configuration, respectively, for crystal field and Jahn-Teller reconstructions in ceria. (Energies and displacements reported in Table 3). \kv{V}{O} in black, Ce in green, O in red, Ni in light grey, Pd in dark grey and Cu in blue.}
    \label{fig:structs_crystal_field}
\end{figure}

\begin{figure}[ht]
    \begin{center}
	\includegraphics[width=0.6\textwidth]{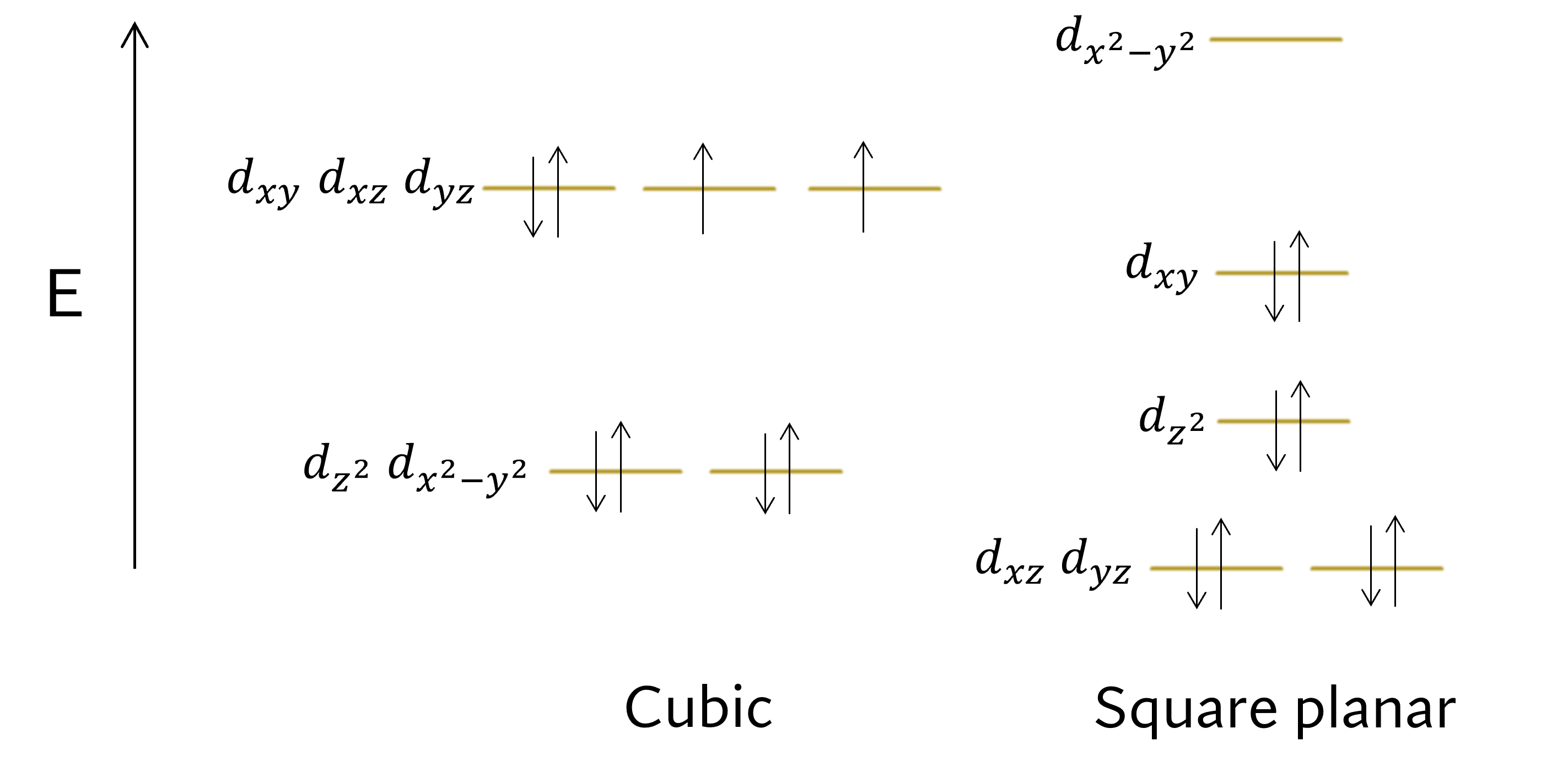}	
	\caption{\small Electron energy level diagram for $d^8$ dopants on \ce{CeO2} showing the crystal field splitting for the initial cubic coordination and the more favourable square-planar arrangement, which leads to a gain in electronic energy.}\label{fig:d8_electron_diagram}
	\end{center}
\end{figure}

\begin{figure}[ht]
\includegraphics[width=0.85\textwidth]{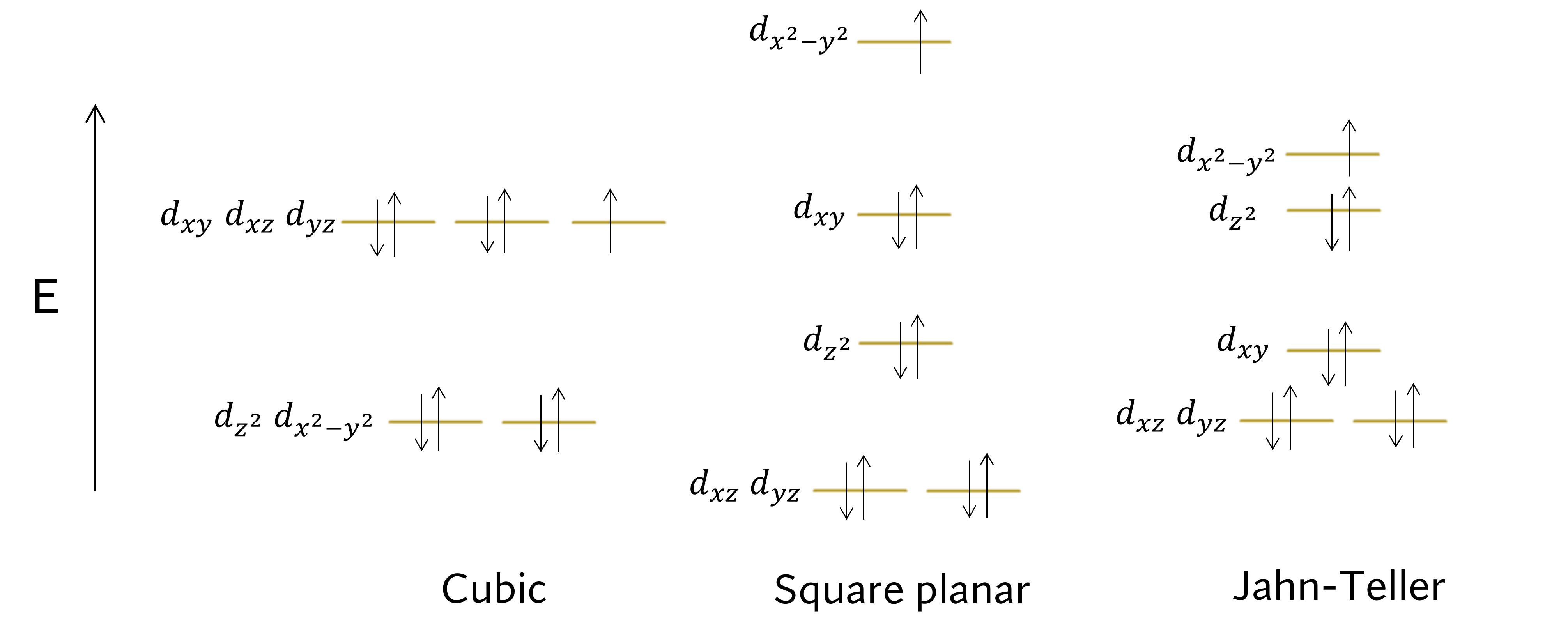}
\caption{Electron energy level diagram for a $d^9$ dopant, showing the advantage of the Jahn-Teller distortion as it lowers the energy of the partially-occupied $d_{x^2-y^2}$ orbital, in comparison with the square-planar arrangement.}\label{fig:d9_electron_diagram}
\end{figure}

\clearpage
\newpage
\subsection{Electrostatically driven}
\begin{figure}[ht]
	\includegraphics[width=0.8\textwidth]{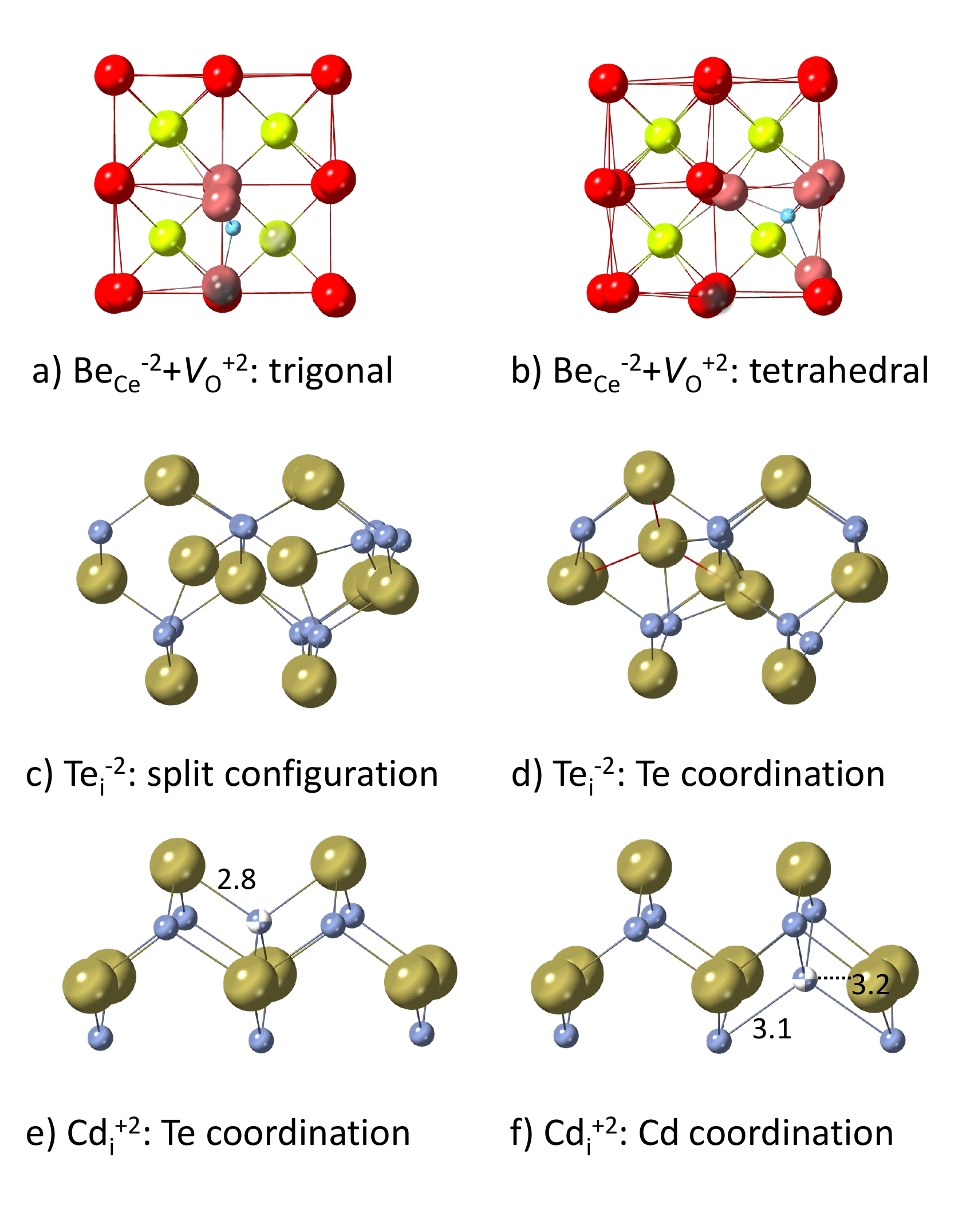}
	\caption{Ground state (left: a,c,e) and metastable (right: b,d,f) structures identified by our method and by relaxing the high-symmetry configuration, respectively, for electrostatically driven reconstructions. (Energies and displacements reported in Table 4). Be in blue, Ce in green, O in red, Te in gold and Cd in blue. Vacancy shown in shaded black.}
	\label{fig:electrostatically_driven}
\end{figure}

\section{Defect bound polarons}
\subsection{Neutral oxygen vacancy in \ce{CeO2}}
\begin{figure}[ht]
	\includegraphics[width=0.8\textwidth]{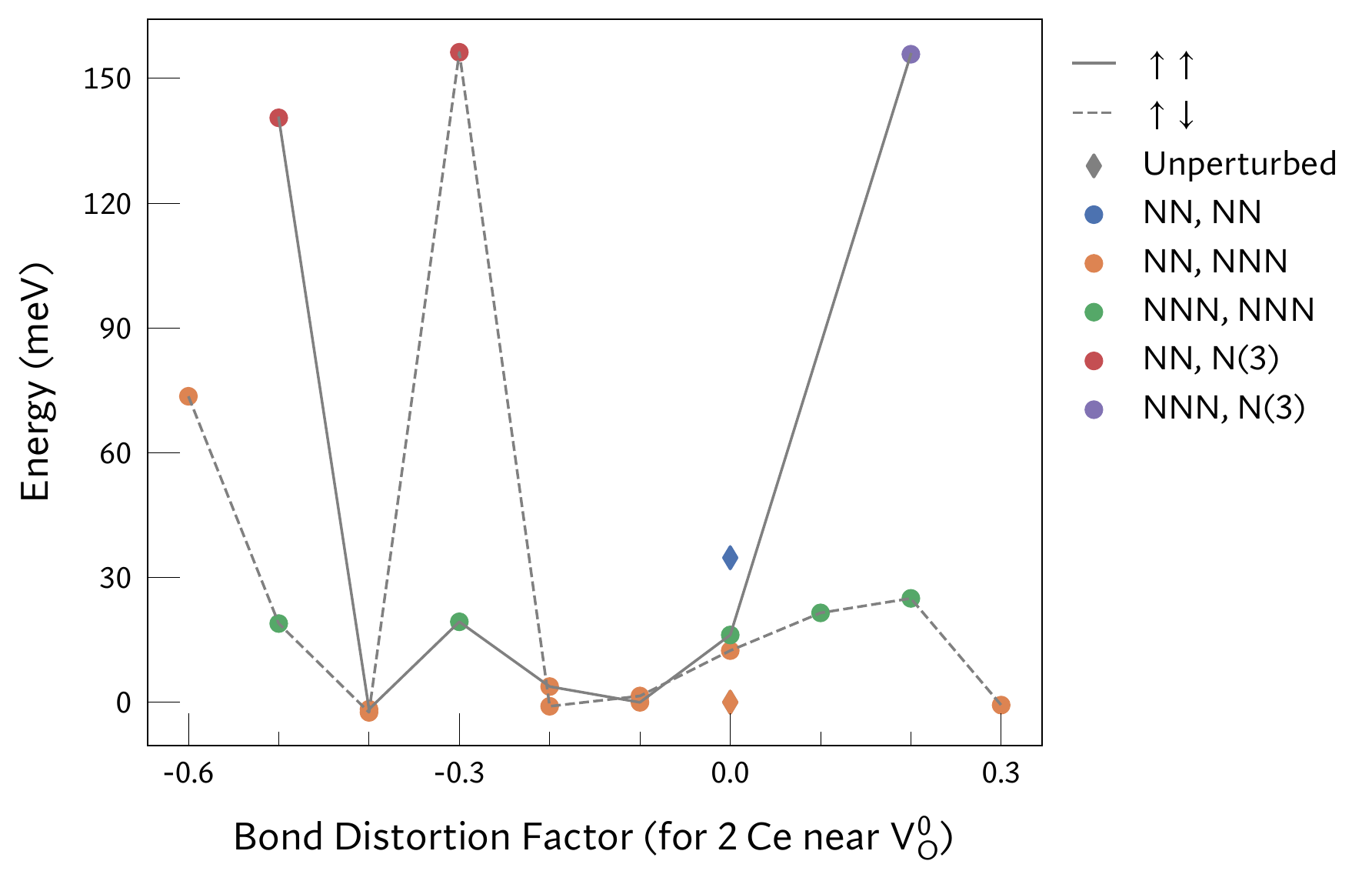}
	\caption{Relative energy (in meV) of the different configurations identified for the neutral oxygen vacancy in \ce{CeO2}. The method was applied both constraining the difference of the number of electrons in the up and down spin component to 2 (solid line) and also to 0 (dashed line). Standard relaxation from the high symmetry structure shown with a diamond (Unperturbed). The different colours indicate where the two Ce(III) ions are located relative to the vacancy: nearest neighbour (NN), next nearest neighbour (NNN) or in the third coordination shell (N(3)).}
	\label{fig:vac_O_ceo2}
\end{figure}

\begin{figure}[ht]
    \begin{subfigure}[b]{0.3\textwidth}
		\includegraphics[width=\textwidth]{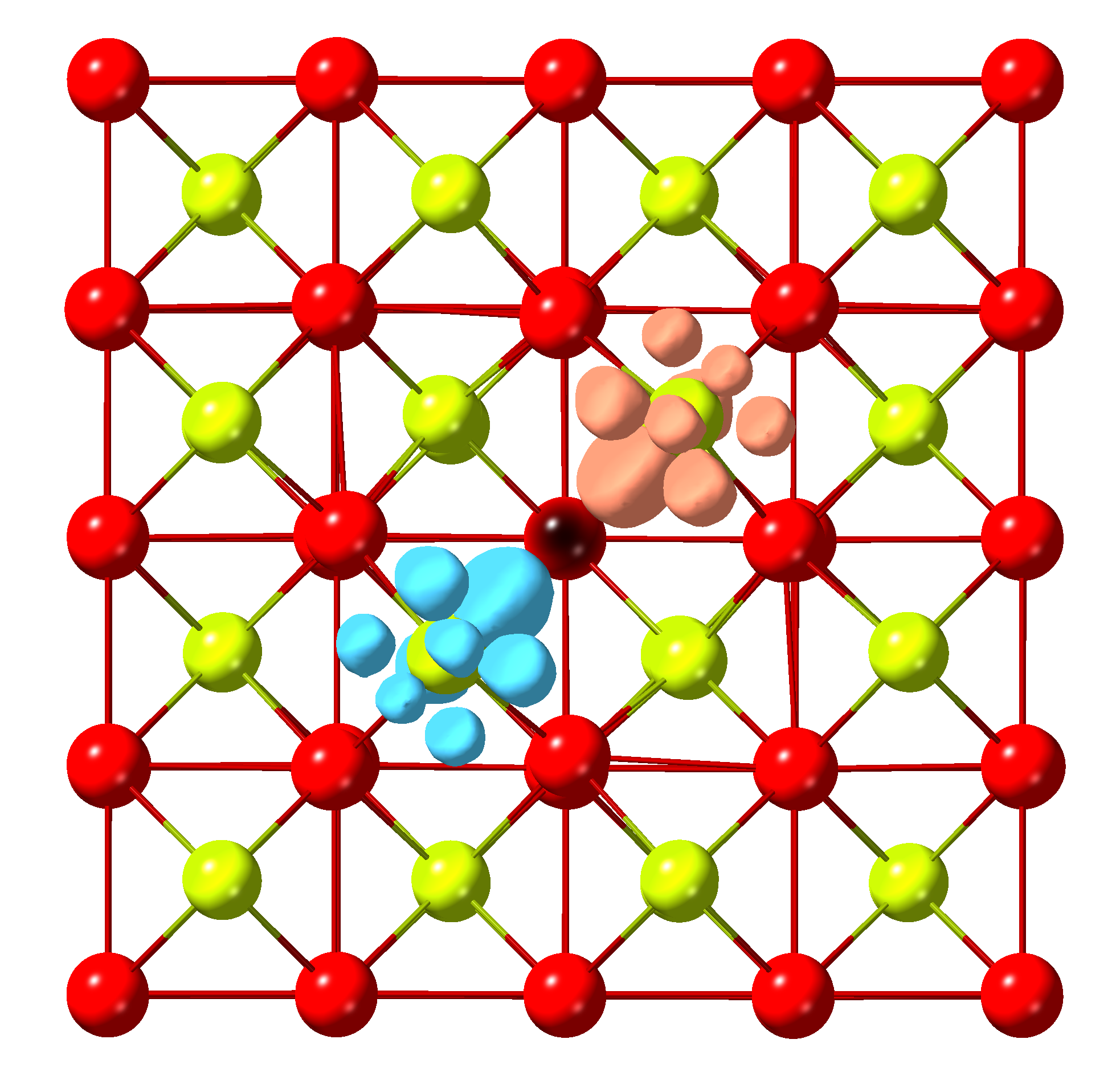}
		\caption{NN, NN ($\uparrow\downarrow$) }
	\end{subfigure}
    \begin{subfigure}[b]{0.3\textwidth}
		\includegraphics[width=\textwidth]{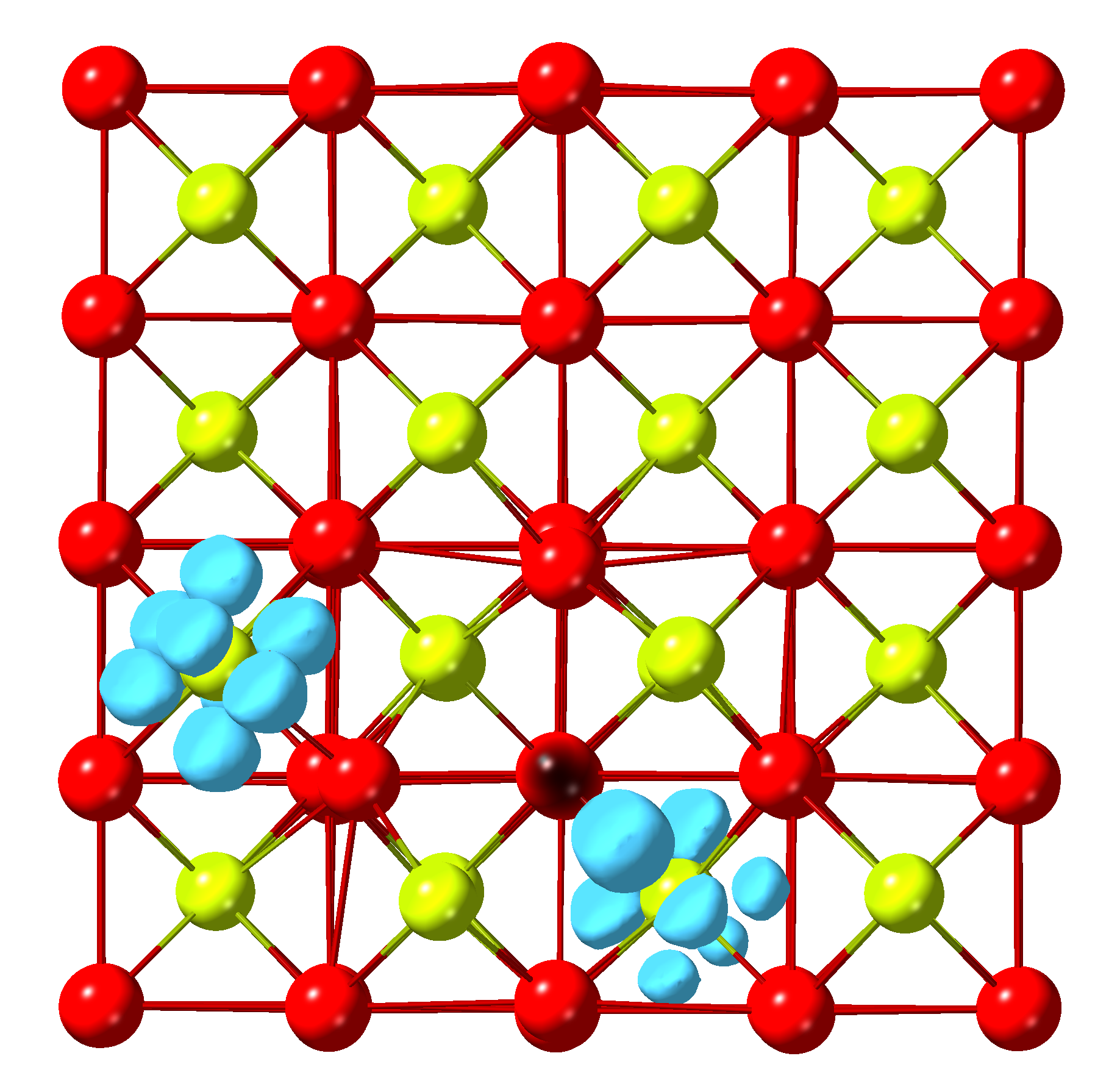}		
		\caption{NN, NNN ($\uparrow\uparrow$) }
	\end{subfigure}
	\begin{subfigure}[b]{0.3\textwidth}
		\includegraphics[width=\textwidth]{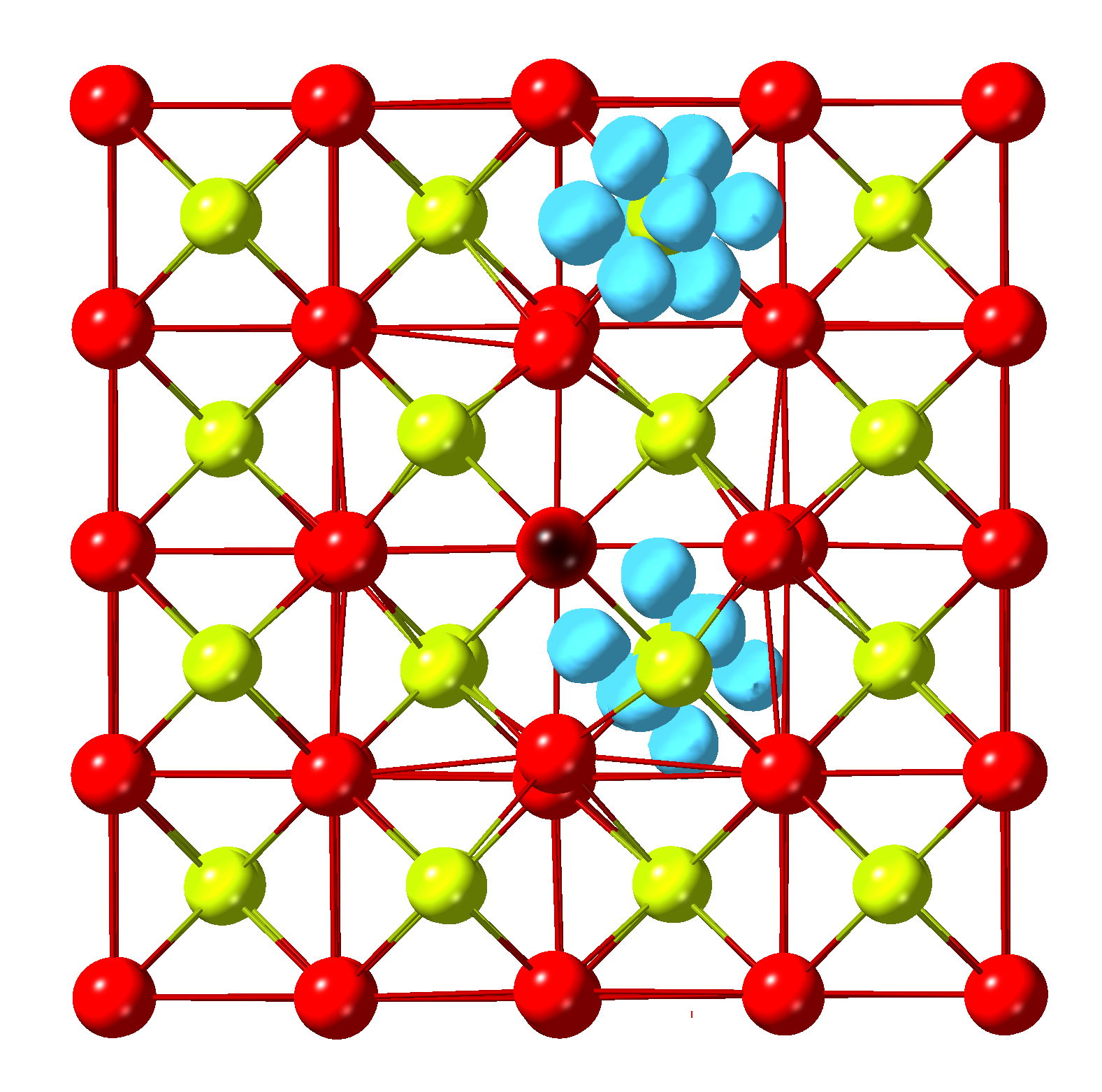}		
		\caption{NNN, NNN ($\uparrow\uparrow$) }
	\end{subfigure}
	\begin{subfigure}[b]{0.3\textwidth}
		\includegraphics[width=\textwidth]{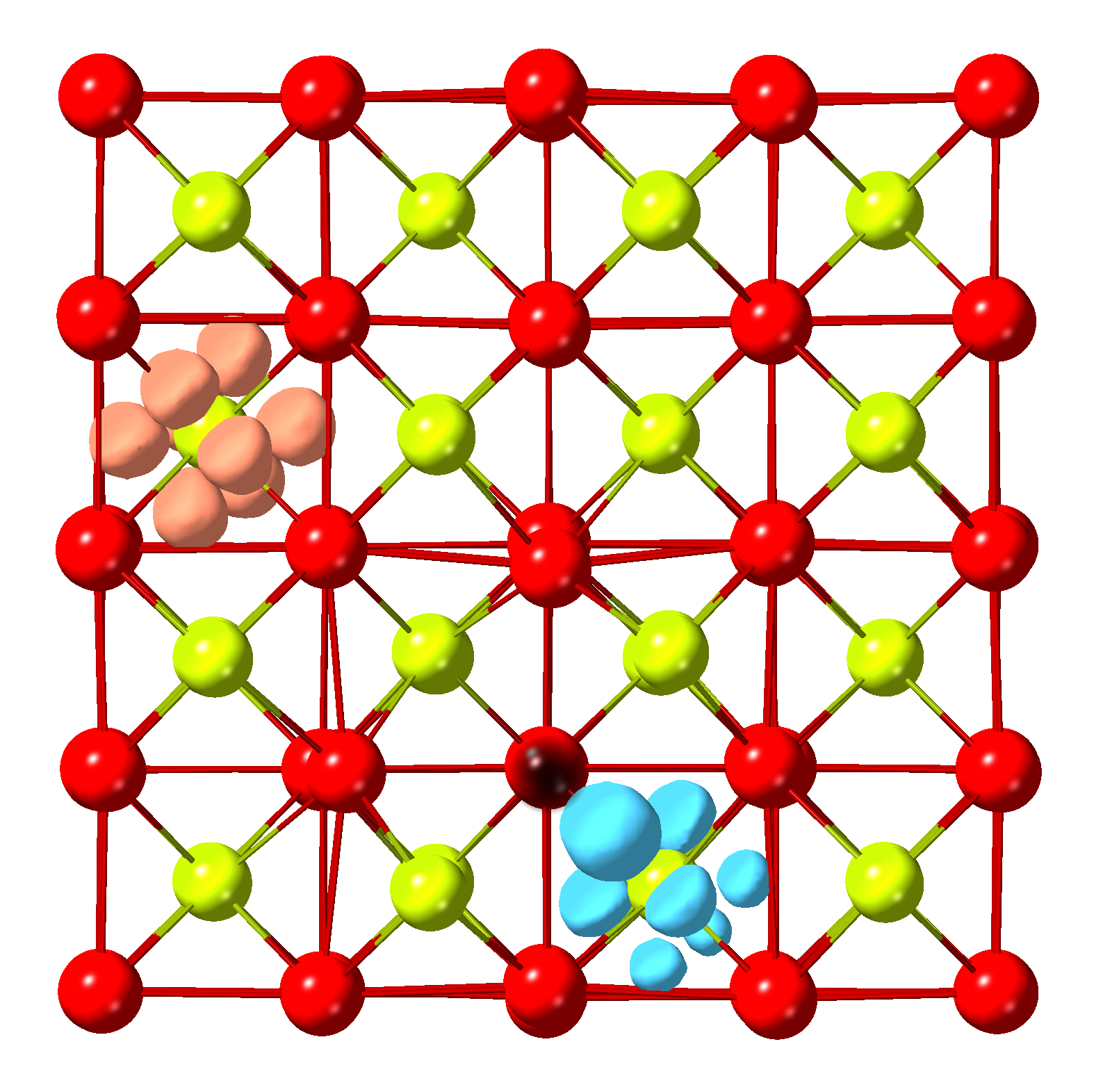}
		\caption{NN, N(3) ($\uparrow\downarrow$)}
	\end{subfigure}
	\begin{subfigure}[b]{0.3\textwidth}
		\includegraphics[width=\textwidth]{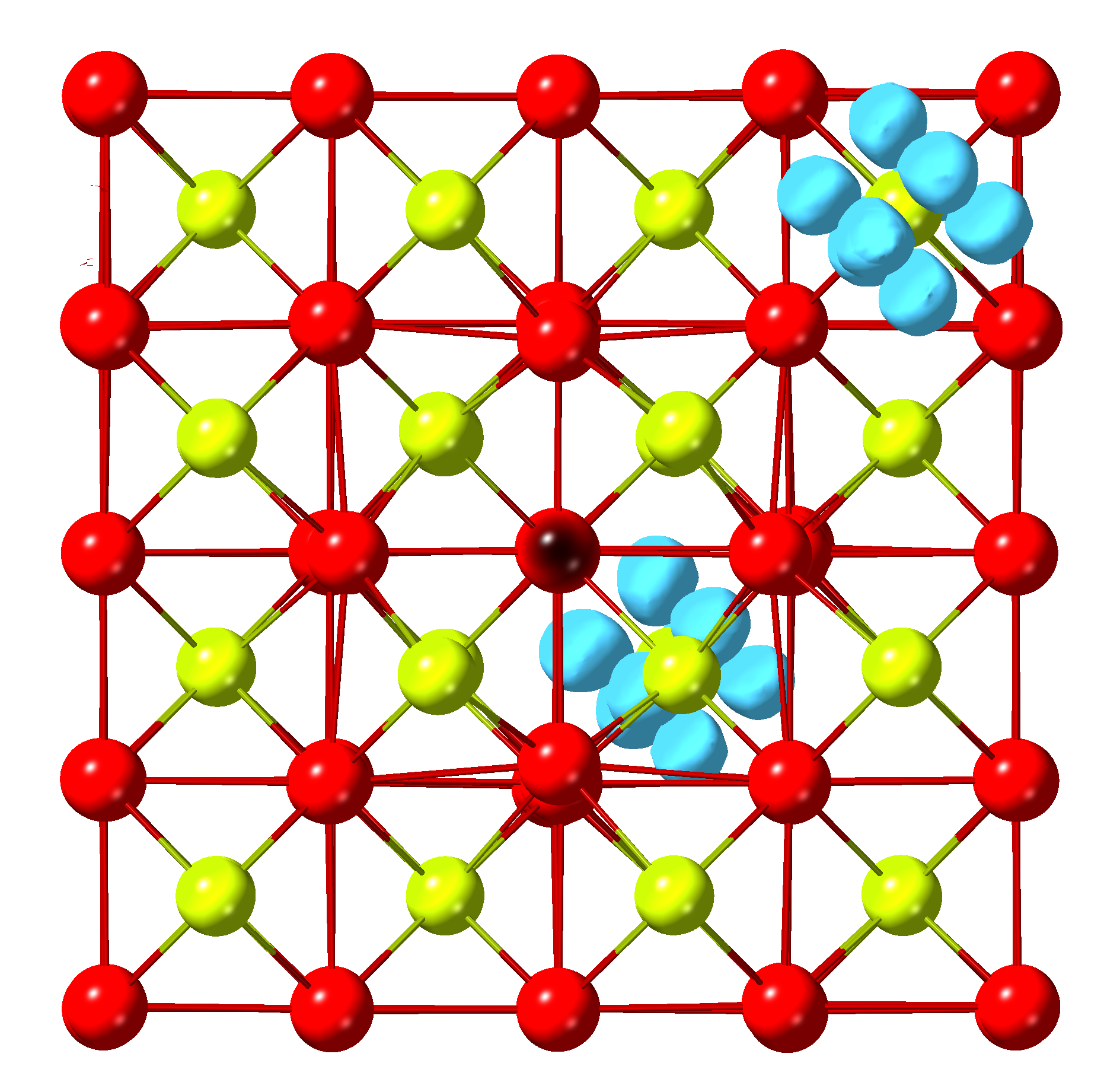}
		\caption{NNN, N(3) ($\uparrow\uparrow$)}
	\end{subfigure}
    \caption{Structure with spin density plot for the different configurations identified with our method for \kvc{V}{O}{0} in \ce{CeO2}. \kv{V}{O} shown in black, Ce in green and O in red. The spin density of different channels is shown with different colours. In all cases, the isosurfaces correspond to 10.5\% of the maximum charge density.}
    \label{fig:vac_O_0_ceo2_density_plots}
\end{figure}

In order to test the ability of the method to identify defect bound polarons, we first considered the neutral oxygen vacancy in \ce{CeO2}. Upon vacancy formation, two electrons localise on two Ce ions, reducing them to Ce(III). The location of the Ce(III) ions is however not clear, with some studies reporting both Ce(III) as nearest neighbours of the vacancy (NN, NN)\cite{keating_analysis_2012,Fabris_2005,Castleton_2007,Nakayama_2012}, next nearest neighbours (NNN, NNN)\cite{Wang_2009,Shoko_2010}, or even a combination of the two (NN, NNN)\cite{allen_2014}. We therefore applied the method to this defect to test whether it was able to identify these configurations. In order to thoroughly sample both spin alignments, we applied it twice: setting the number of unpaired electrons to 2 (spin-aligned, $\uparrow\uparrow$) and also to 0 (anti-aligned, $\uparrow\downarrow$) (Fig. \ref{fig:vac_O_ceo2}, \ref{fig:vac_O_0_ceo2_density_plots}). This results in all previously reported configurations being identified: both Ce(III) in the first coordination shell (NN, NN), both in the second coordination shell (NNN, NNN), one in the first and the other in the second (NN, NN) and also combinations of first/second and third coordination shells (NN, N(3) and NNN, N(3)). The energy differences between these states are of the order of meV, explaining the debate in the literature regarding the most stable configuration. These results demonstrate the ability of the method to identify different polaronic states, even when their energy differences are small.

\subsection{Neutral oxygen vacancy in \ce{TiO2}}
\begin{figure}[ht]
	\includegraphics[width=1.0\textwidth]{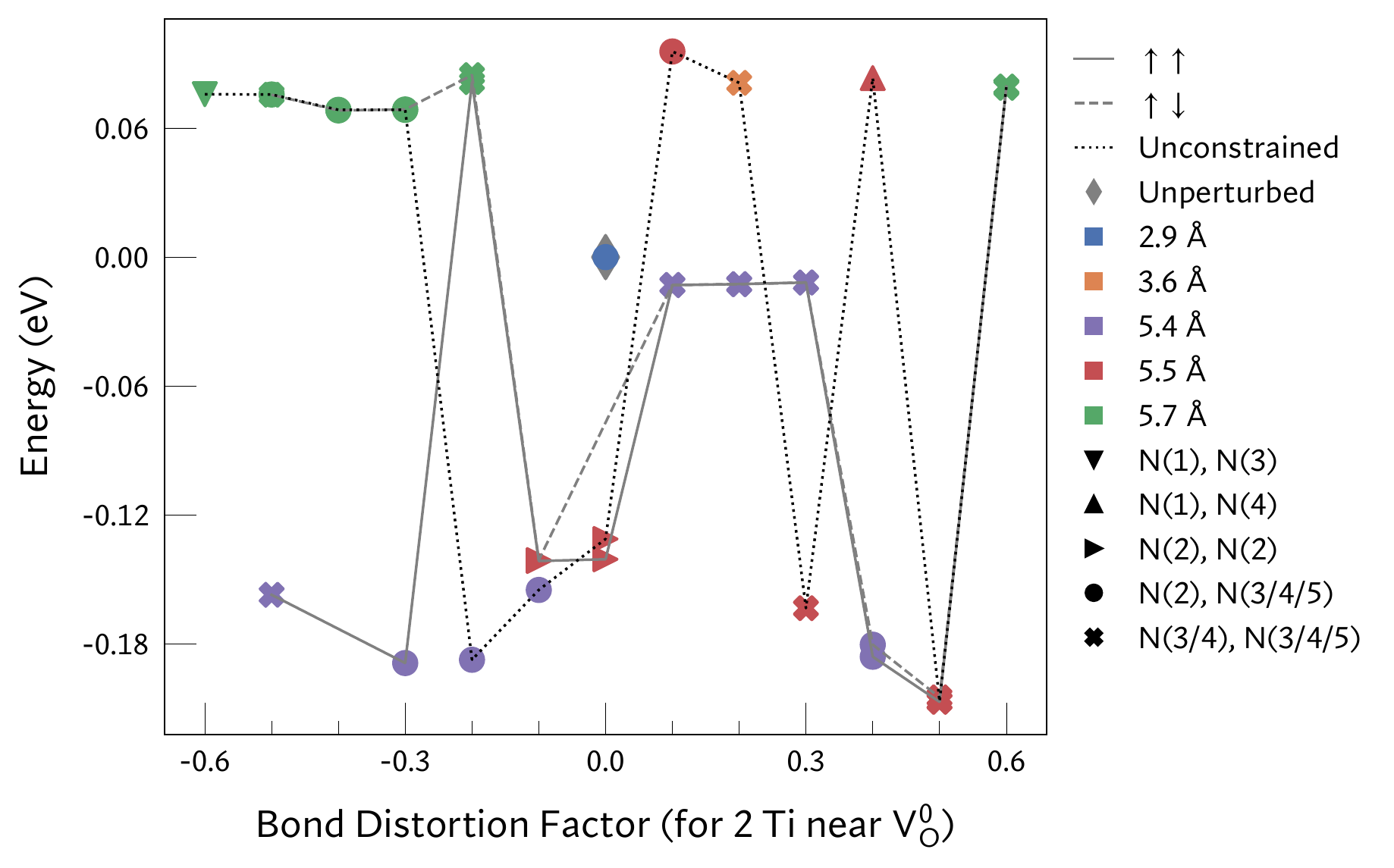}
	\caption{Relative energy (in eV) of the different configurations identified for the neutral oxygen vacancy in rutile \ce{TiO2}. The method was applied by either constraining the difference of the number of electrons in the up and down spin component to 2 (solid line), to 0 (dashed line) and without constraints (dotted line). The different colours indicate the distance between the two Ti(III) ions while the marker shapes correspond to where the Ti(III) are located relative to the vacancy: first coordination shell (N(1), i.e. nearest neighbour), second coordination shell (N(2), i.e. next nearest neighbour) and so on. As shown, spin-aligned and anti-aligned states have negligible energy differences.}
	\label{fig:vac_O_tio2}
\end{figure}

A similar situation occurs for the neutral oxygen vacancy in rutile \ce{TiO2}. Upon vacancy formation, the two electrons donated by the vacancy can localise on different Ti ions, reducing them to Ti(III). To test whether our approach could find the different polaronic configurations and investigate the influence of enforcing magnetisation constraints, we applied the method with different setups: i) without any magnetisation constraints, ii) setting the number of unpaired electrons to 2 (spin-aligned, $\uparrow\uparrow$) and also to 0 (anti-aligned, $\uparrow\downarrow$). As shown in Figure \ref{fig:vac_O_tio2}, $\uparrow\uparrow$ and $\uparrow\downarrow$ show negligible energy differences, in agreement with previous studies\cite{Deak_2012,Morgan_2010}. 

\begin{figure}[ht]
	\includegraphics[width=0.75\textwidth]{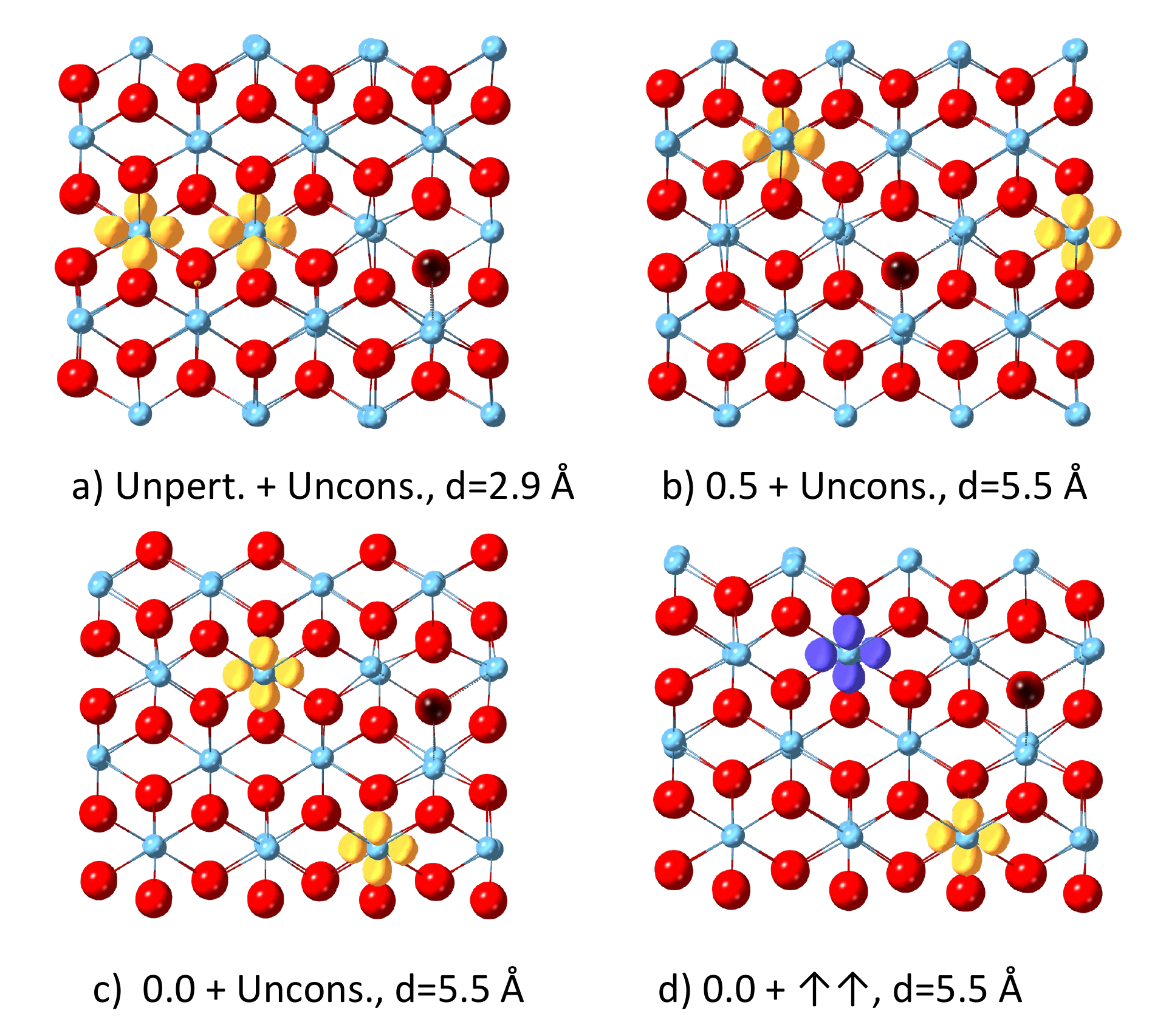}
	\caption{Lowest energy configurations identified for \kvc{V}{O}{0}\!\! in anatase \ce{TiO2}. The labels indicate the bond distortion leading to that state, magnetisation constraints and distance between Ti(III) sites. The spin density associated to the states introduced by \kvc{V}{O}{0}\!\! is shown, with different colours to differentiate spin channels. In all cases, the isosurfaces correspond to 25\% of the maximum charge density. \kv{V}{O} in black, Ti in blue and O in red.}
	\label{fig:vac_O_tio2_structures}
\end{figure}

The most favourable configurations correspond to the Ti(III) ions located in the second or third coordination shell (Fig. \ref{fig:vac_O_tio2_structures}), as reported by a previous HSE06 study\cite{Deak_2011}. This however differs from other computational studies which employed the Hubbard correction and found both Ti(III) to lie in the first coordination shell (N(1), N(1))\cite{Morgan_2010}. To verify our method was not missing this state, we performed additional geometry optimisations initialising the magnetic moments of two of the vacancy nearest neighbours. These configurations relaxed to a state with both Ti(III) ions located further away from the vacancy (first and second coordination shell), demonstrating that the N(1), N(1) state is not locally stable with this functional and supercell size, as previously reported by another HSE06 study\cite{Deak_2011}. \\
Finally, we note that all lowest energy states are successfully identified by the unconstrained run in this case, with comparable performance in the case of the spin-aligned ($\uparrow\uparrow$) constraint. 

\subsection{Neutral titanium interstitial in \ce{TiO2}}
\begin{figure}[ht]
	\includegraphics[width=1.0\textwidth]{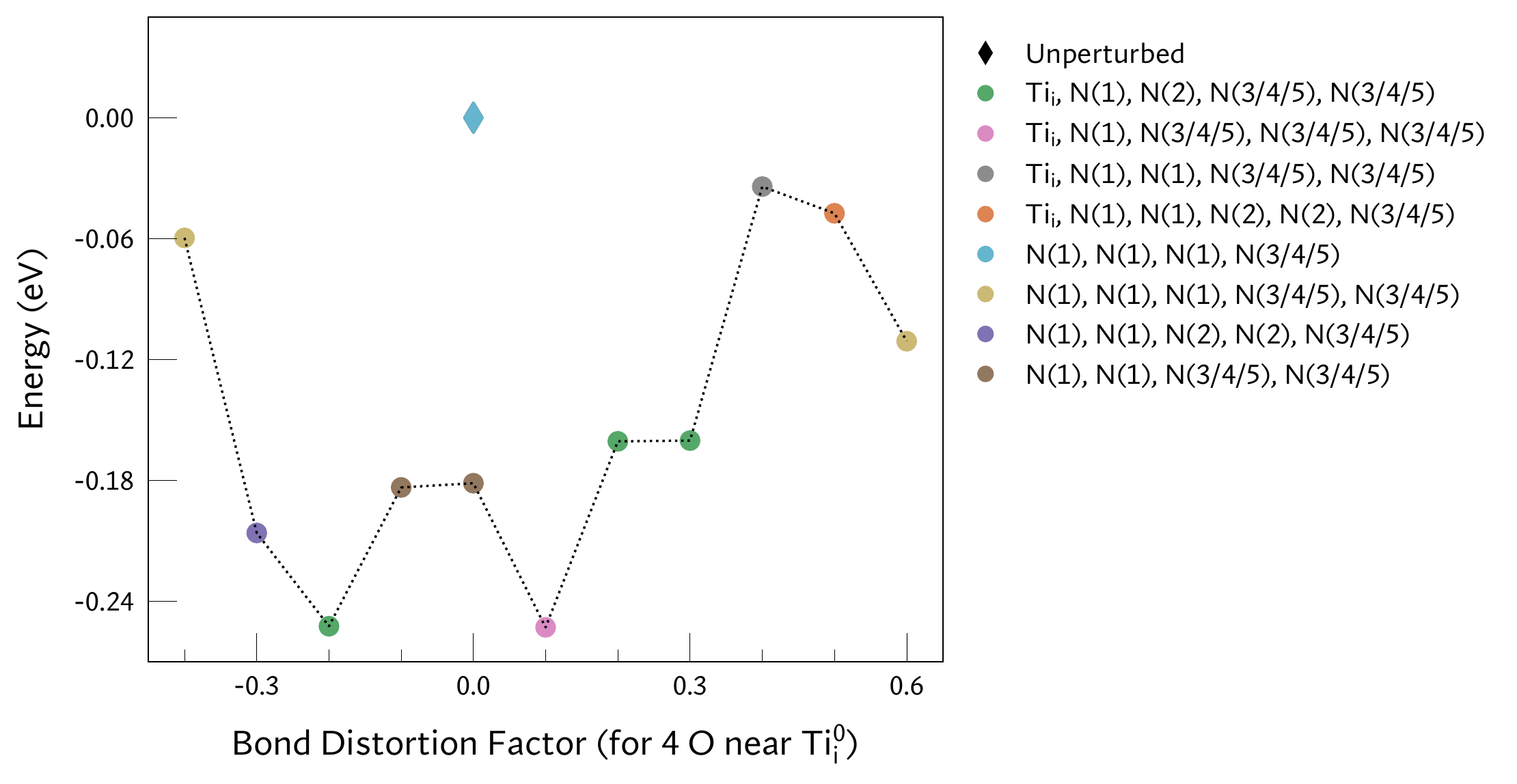}
	\caption{Relative energy (in eV) of the different configurations identified for the neutral titanium interstitial in rutile \ce{TiO2}. Standard relaxation from the high symmetry structure shown with a diamond (Unperturbed). The method was applied without magnetisation constraints. The different colours indicate the position of the Ti ions where the additional charge is localised, relative to $\rm Ti_{i}$: first coordination shell (N(1)), second coordination shell (N(2)) and so on. We find negligible energy difference between $\uparrow\downarrow$ and $\uparrow\uparrow$ (2 unpaired electrons) arrangements. We note that certain configurations with similar localisation patterns differ in their energy due to small structural differences and/or the degree of charge localisation.}
	\label{fig:int_ti_tio2}
\end{figure}

\begin{figure}[ht]
	\includegraphics[width=1.0\textwidth]{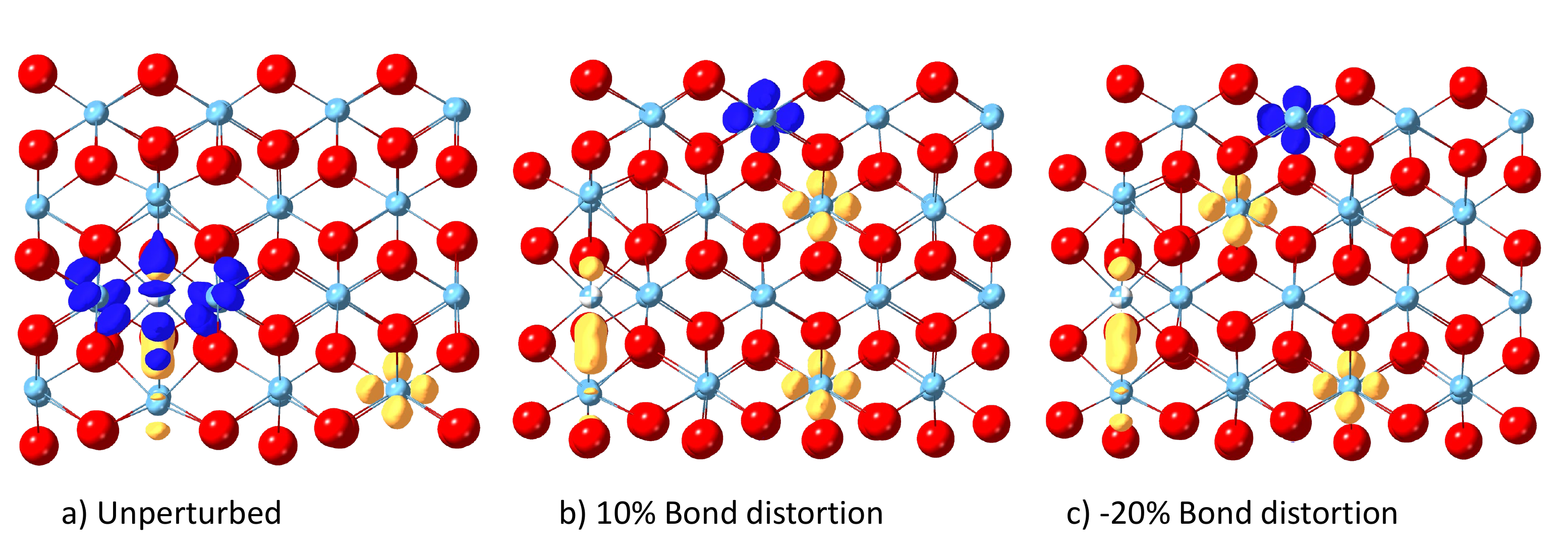}
	\caption{Low energy structures identified for the neutral titanium interstitial in rutile \ce{TiO2}. The spin density associated to the $\rm Ti_i$ states is shown, with different colors indicating different spin channels. In all cases, the isosurfaces correspond to 30\% of the maximum charge density.}\label{int_ti_tio2_structures}
\end{figure}

As with the oxygen vacancy, the titanium interstitial can lead to different states depending on where the donated electrons localise. The identified configurations are shown in Figure \ref{fig:int_ti_tio2} and \ref{int_ti_tio2_structures}. Performing a geometry optimisation from the high-symmetry structure results in most of the charge localised on the interstitial and its Ti neighbours (1 electron shared between $\rm Ti_i$ and a neighbouring Ti, 2 electrons localised on two NN and 1 electron localised further away, Fig. \ref{int_ti_tio2_structures} (a)). This state is similar to the configuration reported by previous hybrid\cite{Deak_2011,Finazzi_2008} and GGA+U studies\cite{Stausholm_2010}. In contrast, our approach finds lower energy states where three of the donated electrons localise on Ti ions further away from the defect while the remaining electron is shared between the interstitial and a neighbouring Ti.   

\subsection{Singly negative indium vacancy in \ce{In2O3}}
\begin{figure}[ht]
	\includegraphics[width=0.7\textwidth]{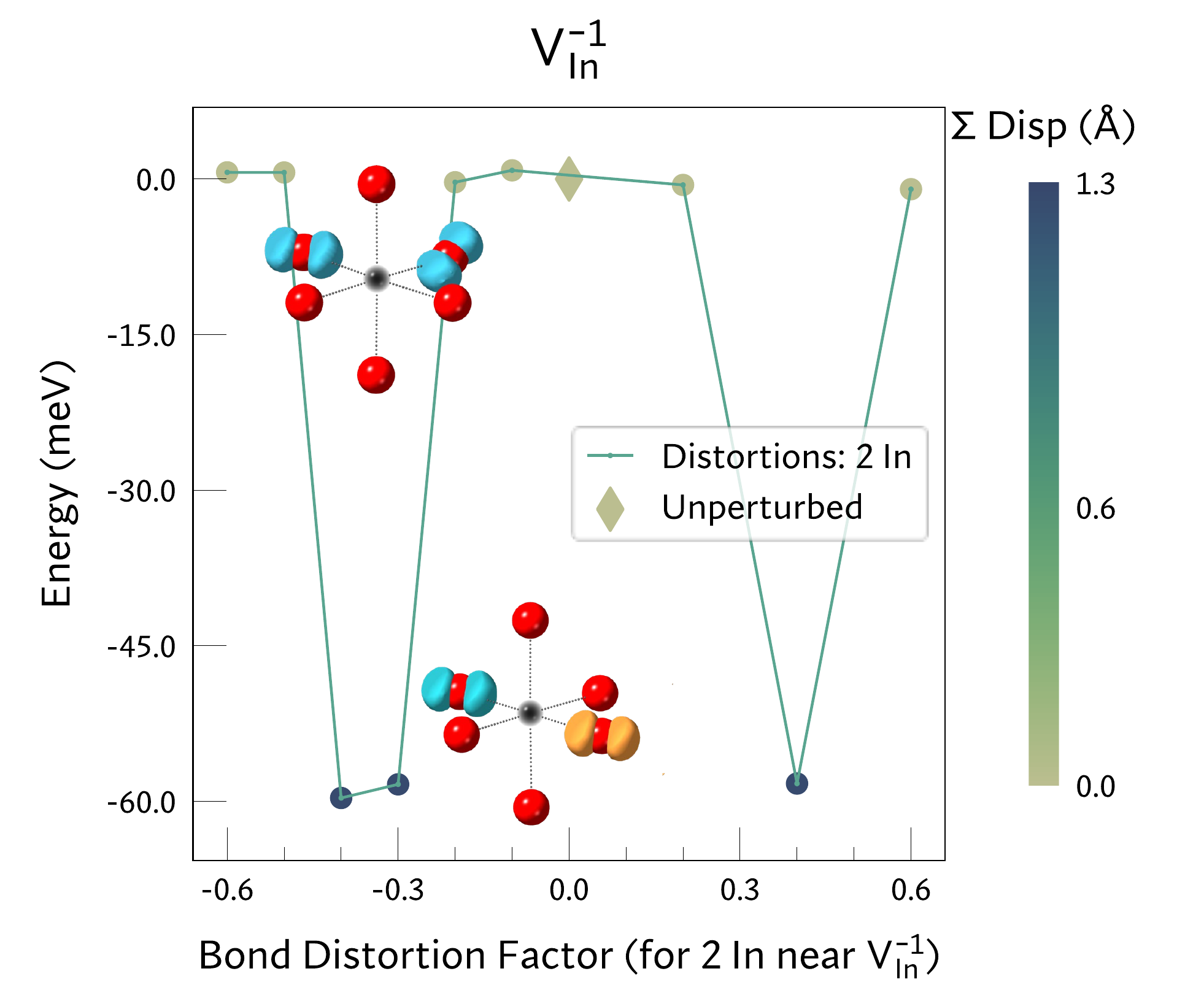}
	\caption{Relative energy (in meV) and structures of the different configurations identified for the singly negative indium vacancy in \ce{In2O3}. Standard relaxation from the high symmetry structure shown with a diamond (Unperturbed). The colourbar represents the structural difference between each structure and the `Unperturbed' one, measured as the sum of the atomic displacements between structures. \kv{V}{In} in black and O in red. The spin density of different channels is shown with different colours.}
	\label{fig:vac_in_in2o3}
\end{figure}

Finally, we also considered the singly negative charged indium vacancy in \ce{In2O3}, where the method was applied without magnetisation constraints. As shown in Figure \ref{fig:vac_in_in2o3}, a standard relaxation from the ideal structure  results in a spin-aligned state ($\uparrow\uparrow$), with two holes localised on two of the vacancy neighbours. In contrast, we find an \ce{anti-aligned} state ($\uparrow\downarrow$) to be more stable (by 60 meV). Compared to spin-aligned, in the anti-aligned case the two oxygens with a hole localised are displaced further away from each other (by 0.1 \AA). While for this simple PES relaxing with magnetisation constraints would also identify the $\uparrow\downarrow$ state, these results demonstrate that applying the method without magnetisation constraints can successfully identify both configurations. In general, these results exemplify the ability of the method to identify different polaronic states, even for defects with a complex PES with several low energy minima.  

\section{Impact on defect properties}
\begin{table}[ht]
    \centering
    \caption{Concentrations (in $\rm cm^{-3}$) of the stable charge states for the ground and metastable structures of \kv{V}{Sb} in \ce{Sb2S3}, for typical growth conditions (T = 550 K) and the Fermi level 0.97 eV above the VBM, as determined experimentally\cite{lian_revealing_2021}. Dashes (`-') signify unstable charge states.}
    \setlength{\tabcolsep}{8pt} 
    \label{tab:vac_sb_conc}
    \begin{tabular}{c|c|c||c|c|c}
         \hline \hline
         \multicolumn{3}{c}{\kv{V}{Sb,1}} &  \multicolumn{3}{c}{\kv{V}{Sb,2}}\\
         Charge & Ground state & Metastable & Charge & Ground state & Metastable \\
         \hline
         +2  & $ 2.1\times10^{-2} $ & $5.7\times10^{-17}$ &
         +2 & $4.4\times10^{-3} $  & $5.0\times10^{-22}$\\
         +1	& $ 4.2\times10^{2}$ & - & 
         +1 & $ 6.3\times10^{6}$ & $5.9\times10^{-15}$\\
         0	& - & - & 
         0 & - & $1.8\times10^{-7}$ \\
         -1 & $ 8.4 \times10^{8}$ & $2.3\times10^3$ & 
         -1 & - & $3.8\times10^{-1}$ \\
        -3	& $1.1\times10^{14}$ & $1.1\times10^{14}$ & 
        -3 & $1.1\times10^{12}$ & $1.1\times10^{12}$\\
        \hline \hline
    \end{tabular}
\end{table}

\begin{table}[ht]
    \centering
    \caption{Positions of thermodynamic charge transition levels (in eV above the VBM) for the ground and metastable structures of \kv{V}{Sb} in \ce{Sb2S3}. Dashes (`-') signify thermodynamically-unstable transition levels.}
    \setlength{\tabcolsep}{8pt} 
    \label{tab:vac_sb_charge_transitions}
    \begin{tabular}{c|c|c||c|c|c}
         \hline \hline
         \multicolumn{3}{c}{\kv{V}{Sb,1}} &  \multicolumn{3}{c}{\kv{V}{Sb,2}}\\
         Level & Ground state & Metastable & Level & Ground state & Metastable \\
         \hline
         (2/1)  & 0.50 & - & (2/1) & -0.03 & -0.14 \\
         (2/-1)	& -	& 0.26 & (1/0) & - & 0.15\\
         (1/-1)	& 0.63 & - & (0/-1) & - & 0.28 \\
         - & - & - & (1/-3) & 0.83 & - \\
        (-1/-3)	& 0.69 & 0.39& (-1/-3) & - & 0.29\\
        \hline \hline
    \end{tabular}
\end{table}

\section{Identifying metastable structures}
\begin{figure}[ht]
	\includegraphics[width=0.7\textwidth]{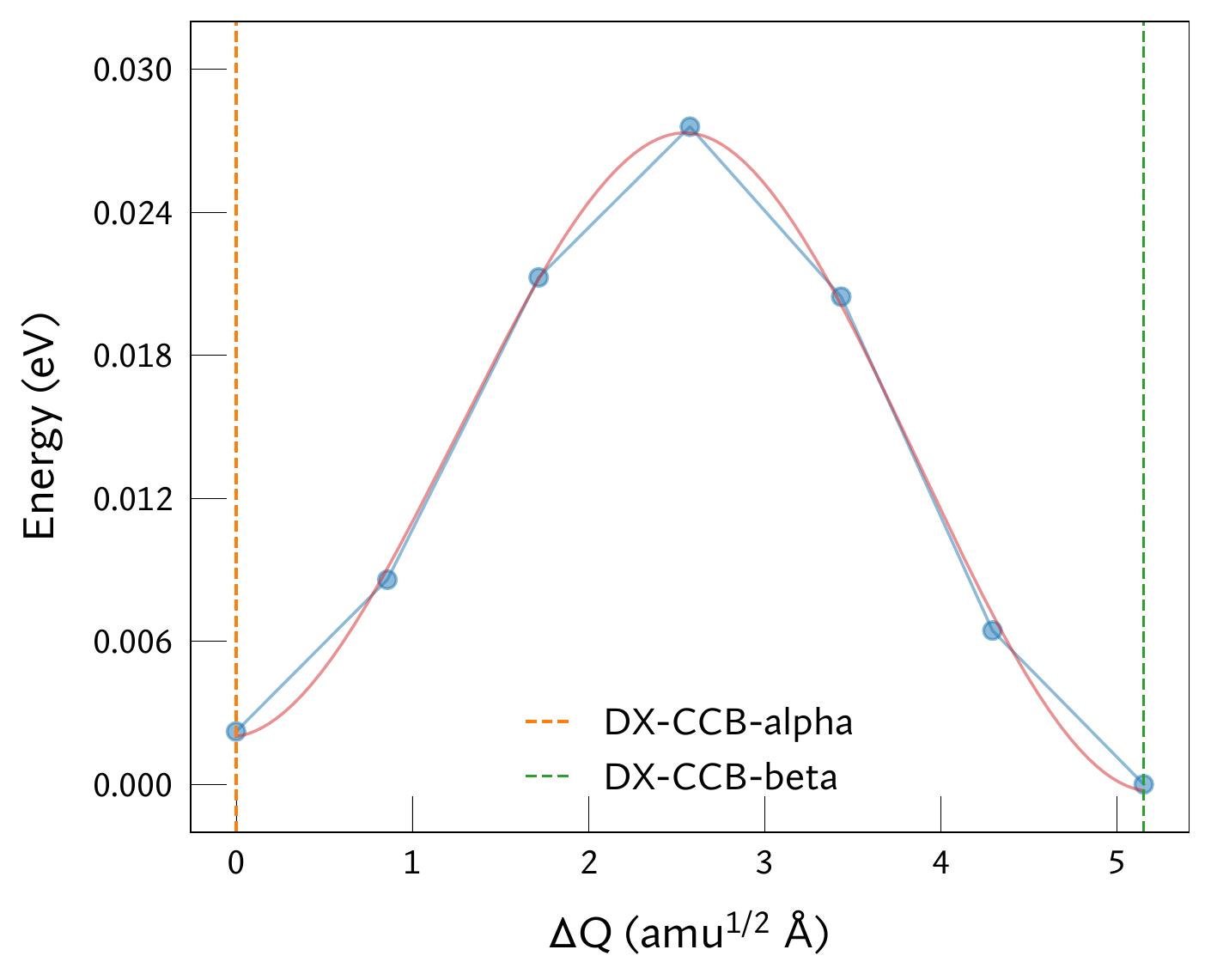}
	\caption{Energy barrier (in eV) between $\rm DX-CCB-\alpha $ and $\rm DX-CCB-\beta $ for $ \rm S_{As}^{-1}$ in \ce{GaAs}, calculated with the nudged elastic band method (5 images).}
	\label{fig:NEB_DX_CCB_alpha_to_beta}
\end{figure}

\section{References}
\bibliographystyle{naturemag}
\bibliography{SI_BDM}